\documentstyle[epsfig,a4p,12pt,rotating]{article}
\begin{document}

\def\toprule{\noalign{\hrule \medskip}}
\def\midrule{\noalign{\medskip\hrule }}
\def\botrule{\noalign{\medskip\hrule }}
\setlength{\parskip}{\medskipamount}


\newcommand{\xmax}{\mbox{$x_{max}$}}
\newcommand{\xmin}{\mbox{$x_{min}$}}
\newcommand{\xemu}{\mbox{$x_{e\mu}$}}
\newcommand{\xhad}{\mbox{$x_{had}$}}
\newcommand{\first}{\mbox{$1^{\mathrm{st}}$}}
\newcommand{\second}{\mbox{$2^{\mathrm{nd}}$}}

\newcommand{\epm}{\mbox{${\mathrm e}^\pm$}}
\newcommand{\mupm}{\mbox{$\mu^\pm$}}
\newcommand{\ee}{{\mathrm e}^+ {\mathrm e}^-}
\newcommand{\sq}{\tilde{\mathrm q}}
\newcommand{\seff}{\tilde{\mathrm f}}
\newcommand{\sele}{\tilde{\mathrm e}}
\newcommand{\sell}{\tilde{\ell}}
\newcommand{\snu}{\tilde{\nu}}
\newcommand{\smu}{\tilde{\mu}}
\newcommand{\stau}{\tilde{\tau}}
\newcommand{\chone}{\tilde{\chi}}
\newcommand{\ch}{\mbox{$\tilde{\chi}^\pm_1$}}
\newcommand{\chp}{\tilde{\chi}^+_1}
\newcommand{\chpm}{\tilde{\chi}^\pm_1}
\newcommand{\nt}{\tilde{\chi}^0}
\newcommand{\qq}{{\mathrm q}\bar{\mathrm q}}
\newcommand{\sleppair}{\sell^+ \sell^-}
\newcommand{\nunu}{\nu \bar{\nu}}
\newcommand{\mumu}{\mu^+ \mu^-}
\newcommand{\tautau}{\tau^+ \tau^-}
\newcommand{\ellell}{\ell^+ \ell^-}
\newcommand{\nulqq}{\nu \ell {\mathrm q} \bar{\mathrm q}'}
\newcommand{\MZ}{M_{\mathrm Z}}

\newcommand {\stopm}         {\tilde{\mathrm{t}}_{1}}
\newcommand {\stops}         {\tilde{\mathrm{t}}_{2}}
\newcommand {\stopbar}       {\bar{\tilde{\mathrm{t}}}_{1}}
\newcommand {\stopx}         {\tilde{\mathrm{t}}}
\newcommand {\sneutrino}     {\tilde{\nu}}
\newcommand {\slepton}       {\tilde{\ell}}
\newcommand {\stopl}         {\tilde{\mathrm{t}}_{\mathrm L}}
\newcommand {\stopr}         {\tilde{\mathrm{t}}_{\mathrm R}}
\newcommand {\stoppair}      {\tilde{\mathrm{t}}_{1}
\bar{\tilde{\mathrm{t}}}_{1}}
\newcommand {\gluino}        {\tilde{\mathrm g}}

\newcommand {\neutralino}    {\mbox{${\tilde{\chi }^{0}_{1}}$}}
\newcommand {\neutrala}      {\mbox{${\tilde{\chi }^{0}_{2}}$}}
\newcommand {\neutralb}      {\tilde{\chi }^{0}_{3}}
\newcommand {\neutralc}      {\tilde{\chi }^{0}_{4}}
\newcommand {\bino}          {\tilde{\mathrm B}^{0}}
\newcommand {\wino}          {\tilde{\mathrm W}^{0}}
\newcommand {\higginoa}      {\tilde{\rm H_{1}}^{0}}
\newcommand {\higginob}      {\tilde{\mathrm H_{1}}^{0}}
\newcommand {\chargino}      {\tilde{\chi }^{\pm}_{1}}
\newcommand {\charginop}     {\tilde{\chi }^{+}_{1}}
\newcommand {\KK}            {{\mathrm K}^{0}-\bar{\mathrm K}^{0}}
\newcommand {\ff}            {{\mathrm f} \bar{\mathrm f}}
\newcommand {\bstopm} {\mbox{$\boldmath {\tilde{\mathrm{t}}_{1}} $}}
\newcommand {\Mt}            {M_{\mathrm t}}
\newcommand {\mscalar}       {m_{0}}
\newcommand {\Mgaugino}      {M_{1/2}}
\newcommand {\rs}            {\sqrt{s}}
\newcommand {\WW}            {{\mathrm W}^+{\mathrm W}^-}
\newcommand {\MGUT}          {M_{\mathrm {GUT}}}
\newcommand {\Zboson}        {${\mathrm Z}^{0}$}
\newcommand {\Wpm}           {{\mathrm W}^{\pm}}
\newcommand {\allqq}         {\sum_{q \neq t} q \bar{q}}
\newcommand {\mixang}        {\theta _{\mathrm {mix}}}
\newcommand {\thacop}        {\theta _{\mathrm {Acop}}}
\newcommand {\cosjet}        {\cos\thejet}
\newcommand {\costhr}        {\cos\thethr}
\newcommand {\djoin}         {d_{\mathrm{join}}}
\newcommand {\mstop}         {m_{\stopm}}
\newcommand {\msell}         {m_{\sell}}
\newcommand {\mchi}          {m_{\neutralino}}
\newcommand {\pp}{p \bar{p}}

\newcommand{\epair}{\mbox{${\mathrm e}^+{\mathrm e}^-$}}
\newcommand{\mupair}{\mbox{$\mu^+\mu^-$}}
\newcommand{\taupair}{\mbox{$\tau^+\tau^-$}}
\newcommand{\qpair}{\mbox{${\mathrm q}\overline{\mathrm q}$}}
\newcommand{\eeee}{\mbox{\epair\epair}}
\newcommand{\eemumu}{\mbox{\epair\mupair}}
\newcommand{\eetautau}{\mbox{\epair\taupair}}
\newcommand{\eeqq}{\mbox{\epair\qpair}}
\newcommand{\fs}{ final states}
\newcommand{\epairf}{\mbox{\epair\fs}}
\newcommand{\mupairf}{\mbox{\mupair\fs}}
\newcommand{\taupairf}{\mbox{\taupair\fs}}
\newcommand{\qpairf}{\mbox{\qpair\fs}}
\newcommand{\eeeef}{\mbox{\eeee\fs}}
\newcommand{\eemumuf}{\mbox{\eemumu\fs}}
\newcommand{\eetautauf}{\mbox{\eetautau\fs}}
\newcommand{\eeqqf}{\mbox{\eeqq\fs}}
\newcommand{\ffff}{four fermion final states}
\newcommand{\llnunu}{\mbox{$\ell^+\nu\,\ell^-\nbar$}}
\newcommand{\wwllnunu}{\mbox{$\wpair\rightarrow\llnunu$}}
\newcommand{\lnuqq}{\mbox{\lept\nubar\qpair}}
\newcommand{\zee}{\mbox{\epair Z$^*/\gamma^*$}}
\newcommand{\zzg}{\mbox{ZZ/Z$\gamma$}}
\newcommand{\wenu}{\mbox{We$\nu$}}
\newcommand{\wtoenu}{\mbox{${\mathrm W}\rightarrow{\mathrm e}\nu$}}
\newcommand{\wtomunu}{\mbox{${\mathrm W}\rightarrow\mu\nu$}}
\newcommand{\wtotaunu}{\mbox{${\mathrm W}\rightarrow\tau\nu$}}

\newcommand{\el}{\mbox{${\mathrm e}^-$}}
\newcommand{\selem}{\mbox{$\tilde{\mathrm e}^-$}}
\newcommand{\smum}{\mbox{$\tilde\mu^-$}}
\newcommand{\staum}{\mbox{$\tilde\tau^-$}}
\newcommand{\slept}{\mbox{$\tilde{\ell}^\pm$}}
\newcommand{\sleptm}{\mbox{$\tilde{\ell}^-$}}
\newcommand{\lept}{\mbox{$\ell^-$}}
\newcommand{\Hl}{\mbox{$\mathrm{L}^\pm$}}
\newcommand{\Hm}{\mbox{$\mathrm{L}^-$}}
\newcommand{\Hnu}{\mbox{$\nu_{\mathrm{L}}$}}
\newcommand{\nul}{\mbox{$\nu_\ell$}}
\newcommand{\nubar}{\mbox{$\overline{\nu}_\ell$}}
\newcommand{\nbar}{\mbox{$\overline{\nu}$}}
\newcommand{\spair}{\mbox{$\tilde{\ell}^+\tilde{\ell}^-$}}
\newcommand{\lpair}{\mbox{$\ell^+\ell^-$}}
\newcommand{\staupair}{\mbox{$\tilde{\tau}^+\tilde{\tau}^-$}}
\newcommand{\smupair}{\mbox{$\tilde{\mu}^+\tilde{\mu}^-$}}
\newcommand{\selepair}{\mbox{$\tilde{\mathrm e}^+\tilde{\mathrm e}^-$}}
\newcommand{\Hpm}{\mbox{$\mathrm{H}^\pm$}}
\newcommand{\hpair}{\mbox{$\mathrm{H}^+\mathrm{H}^-$}}
\newcommand{\chpair}{\mbox{$\tilde{\chi}^+_1\tilde{\chi}^-_1$}}
\newcommand{\chm}{\mbox{$\tilde{\chi}^-_1$}}
\newcommand{\chmp}{\mbox{$\tilde{\chi}^\pm_1$}}
\newcommand{\chz}{\mbox{$\tilde{\chi}^0_1$}}
\newcommand{\dch}{\mbox{$\chm\rightarrow\lept\nubar\chz$}}
\newcommand{\dchtwo}{\mbox{$\chpm \rightarrow \ell^\pm {\tilde{\nu}_\ell}$}}
\newcommand{\dchthree}{\mbox{$\chmp\rightarrow{\mathrm W}^\pm\chz\rightarrow\ell^\pm\nubar\chz$}}
\newcommand{\dslept}{\mbox{$\sleptm\rightarrow\lept\chz$}}
\newcommand{\dL}{\mbox{$\Hm\rightarrow\lept\nubar\Hnu$}}
\newcommand{\dH}{\mbox{H$^-\rightarrow\tau^-\nbar_\tau$}}
\newcommand{\mch}{\mbox{$m_{\tilde{\chi}^\pm_1}$}}
\newcommand{\mslept}{\mbox{$m_{\tilde{\ell}}$}}
\newcommand{\mstau}{\mbox{$m_{\staum}$}}
\newcommand{\msmu}{\mbox{$m_{\smum}$}}
\newcommand{\msele}{\mbox{$m_{\selem}$}}
\newcommand{\msnu}{\mbox{$m_{\snu}$}}
\newcommand{\mchz}{\mbox{$m_{\tilde{\chi}^0_1}$}}
\newcommand{\mH}{\mbox{$m_{\mathrm{H}^-}$}}
\newcommand{\dm}{\mbox{$\Delta m$}}
\newcommand{\dmch}{\mbox{$\Delta m_{\ch-\chz}$}}
\newcommand{\dmslept}{\mbox{$\Delta m_{\slept-\chz}$}}
\newcommand{\dmhl}{\mbox{$\Delta m_{\Hl-\Hnu}$}}
\newcommand{\w}{\mbox{W$^\pm$}}

\newcommand{\acopc}{\mbox{$\phi_{\mathrm{acop}}$}}
\newcommand{\acolc}{\mbox{$\theta_{\mathrm{acol}}$}}
\newcommand{\acop}{\mbox{$\phi_{\mathrm{acop}}$}}
\newcommand{\acol}{\mbox{$\theta_{\mathrm{acol}}$}}
\newcommand{\ptpt}{\mbox{$p_{t}$}}
\newcommand{\pz}{\mbox{$p_{\mathrm{z}}^{\mathrm{miss}}$}}
\newcommand{\ptevt}{\mbox{$p_{t}^{\mathrm{miss}}$}}
\newcommand{\ptaxic}{\mbox{$a_{t}^{\mathrm{miss}}$}}
\newcommand{\stevt}{\mbox{$p_{t}^{\mathrm{miss}}$/\Ebeam}}
\newcommand{\staxic}{\mbox{$a_{t}^{\mathrm{miss}}$/\Ebeam}}
\newcommand{\dptaxic}{\mbox{missing $p_{t}$ wrt. event axis \ptaxic}}
\newcommand{\cosevt}{\mbox{$\mid\cos\theta_{\mathrm{p}}^{\mathrm{miss}}\mid$}}
\newcommand{\axicos}{\mbox{$\mid\cos\theta_{\mathrm{a}}^{\mathrm{miss}}\mid$}}
\newcommand{\pthet}{\mbox{$\theta_{\mathrm{p}}^{\mathrm{miss}}$}}
\newcommand{\athet}{\mbox{$\theta_{\mathrm{a}}^{\mathrm{miss}}$}}
\newcommand{\dcosevt}{\mbox{$\mid\cos\theta\mid$ of missing p$_{t}$}}
\newcommand{\daxicos}{\mbox{$\mid\cos\theta\mid$ of missing p$_{t}$ wrt. event
axis}}
\newcommand{\efdsw}{\mbox{$x_{\mathrm{FDSW}}$}}
\newcommand{\acopf}{\mbox{$\Delta\phi_{\mathrm{FDSW}}$}}
\newcommand{\acopm}{\mbox{$\Delta\phi_{\mathrm{MUON}}$}}
\newcommand{\acopt}{\mbox{$\Delta\phi_{\mathrm{trk}}$}}
\newcommand{\po}{\mbox{$E_{\mathrm{isol}}^\gamma$}}
\newcommand{\qprod}{\mbox{$q1$$*$$q2$}}
\newcommand{\lcode}{lepton identification code}
\newcommand{\nctro}{\mbox{$N_{\mathrm{trk}}^{\mathrm{out}}$}}
\newcommand{\necao}{\mbox{$N_{\mathrm{ecal}}^{\mathrm{out}}$}}
\newcommand{\mout}{\mbox{$m^{\mathrm{out}}$}}
\newcommand{\nctec}{\mbox{\nctro$+$\necao}}
\newcommand{\gfract}{\mbox{$F_{\mathrm{good}}$}}
\newcommand{\zz}       {\mbox{$|z_0|$}}
\newcommand{\dz}       {\mbox{$|d_0|$}}
\newcommand{\sint}      {\mbox{$\sin\theta$}}
\newcommand{\cost}      {\mbox{$\cos\theta$}}
\newcommand{\mcost}     {\mbox{$|\cos\theta|$}}
\newcommand{\dedx}     {\mbox{$\mathrm{d}E/\mathrm{d}x$}}
\newcommand{\wdedx}     {\mbox{$W_{\mathrm{d}E/\mathrm{d}x}$}}
\newcommand{\xe}     {\mbox{$x_E$}}

\newcommand{\signine}   {\mbox{$\sigma_{95}$}}
\newcommand{\Nnine}   {\mbox{$N_{95}$}}
\newcommand{\Nninef}   {\mbox{$N_{95}(N,\mu _B)$}}
\newcommand{\Ncand}   {\mbox{$N_{\mathrm{cand}}$}}
\newcommand{\Nsm}   {\mbox{$N_{\mathrm{SM}}$}}
\newcommand{\elw}   {\mbox{$\Sigma_{\varepsilon{\cal L}\omega}$}}

\newcommand{\ssix}     {\mbox{$\sqrt{s}$~=~161~GeV}}
\newcommand{\sseven}     {\mbox{$\sqrt{s}$~=~172~GeV}}
\newcommand{\sthree}     {\mbox{$\sqrt{s}$~=~130--136~GeV}}
\newcommand{\mrecoil}     {\mbox{$m_{\mathrm{recoil}}$}}
\newcommand{\llmass}     {\mbox{$m_{\ell\ell}$}}
\newcommand{\sml}{\mbox{Standard Model \llnunu\ events}}
\newcommand{\sme}{\mbox{Standard Model events}}
\newcommand{\sig}{events containing a lepton pair plus genuine missing transverse momentum}
\newcommand{\wpair}{\mbox{$\mathrm{W}^+\mathrm{W}^-$}}
\newcommand{\dW}{\mbox{W$^-\rightarrow\lept\nubar$}}
\newcommand{\dsele}{\mbox{$\selem\rightarrow\mathrm{e}^-\chz$}}
\newcommand{\dsmu}{\mbox{$\smum\rightarrow\mu^-\chz$}}
\newcommand{\dstau}{\mbox{$\staum\rightarrow\tau^-\chz$}}
\newcommand{\eeeell}{\mbox{$\epair\rightarrow\epair\lpair$}}
\newcommand{\eell}{\mbox{\epair\lpair}}
\newcommand{\llgam}{\mbox{$\ell^+\ell^-(\gamma)$}}
\newcommand{\nunugam}{\mbox{$\nu\bar{\nu}\gamma\gamma$}}
\newcommand{\acope}{\mbox{$\Delta\phi_{\mathrm{EE}}$}}
\newcommand{\nee}{\mbox{$N_{\mathrm{EE}}$}}
\newcommand{\eesum}{\mbox{$\Sigma_{\mathrm{EE}}$}}
\newcommand{\at}{\mbox{$a_{t}$}}
\newcommand{\spp}{\mbox{$p$/\Ebeam}}
\newcommand{\acoph}{\mbox{$\Delta\phi_{\mathrm{HCAL}}$}}
\newcommand{\Eovp}{\mbox{$E/p$}}
\newcommand{\ACOP}{\mbox{$\phi_{\mathrm{acop}}$}}
\newcommand{\XT}{\mbox{$p_{t}^{\mathrm{miss}}$/\Ebeam}}
\newcommand{\XONE}{\mbox{$x_{max}$}}
\newcommand{\XTWO}{\mbox{$x_{min}$}}  
\newcommand{\MLL}{\mbox{$m_{\ell\ell}$}}
\newcommand{\MRECOIL}{\mbox{$m_{\mathrm{recoil}}$}}
\newcommand {\mm}       {\mu^+ \mu^-}
\newcommand {\emu}         {\mbox{$\mathrm{e}^{\pm} \mu^{\mp}$}}
\newcommand {\etau}         {\mathrm{e}^{\pm} \tau^{\mp}}
\newcommand {\mutau}         {\mu^{\pm} \tau^{\mp}}
\newcommand {\et}         {\mathrm{e}^{\pm} \tau^{\mp}}
\newcommand {\mt}         {\mu^{\pm} \tau^{\mp}}
\newcommand {\lemu}       {\ell=\mathrm{e},\mu}
\newcommand{\Zz}{\mbox{${\mathrm{Z}^0}$}}
\newcommand{\mf}     {\mbox{$(m-15)/2$}}
\newcommand{\ZP}[3]    {Z. Phys. {\bf C#1} (#2) #3.}
\newcommand{\PL}[3]    {Phys. Lett. {\bf B#1} (#2) #3.}
\newcommand{\etal}     {{\it et al}.,\,\ }

\newcommand{\Ecm}{\mbox{$E_{\mathrm{cm}}$}}
\newcommand{\Ebeam}{\mbox{$E_{\mathrm{beam}}$}}
\newcommand{\ipb}{\mbox{pb$^{-1}$}}
\newcommand{\wrt}{with respect to}
\newcommand{\sm}{Standard Model}
\newcommand{\smb}{Standard Model background}
\newcommand{\smp}{Standard Model processes}
\newcommand{\smc}{Standard Model Monte Carlo}
\newcommand{\mc}{Monte Carlo}
\newcommand{\btb}{back-to-back}
\newcommand{\tp}{two-photon}
\newcommand{\tpb}{two-photon background}
\newcommand{\tpp}{two-photon processes}
\newcommand{\lp}{lepton pairs}
\newcommand{\vto}{\mbox{$\tau$ veto}}
\newcommand{\chargthree}{\chpair (3-body decays)}
\newcommand{\chargtwo}{\chpair (2-body decays)}
\newcommand{\chargthreee}{\chpair (3-body decays: \dchthree )}
\newcommand{\chargtwoo}{\chpair (2-body decays: \dchtwo )}
  
\newcommand{\gsim}{\;\raisebox{-0.9ex}
           {$\textstyle\stackrel{\textstyle >}{\sim}$}\;}
\newcommand{\lsim}{\;\raisebox{-0.9ex}{$\textstyle\stackrel{\textstyle<}
           {\sim}$}\;}

\newcommand{\degree}    {^\circ}
%
\newcommand{\roots}     {\sqrt{s}}
%
%
\newcommand{\thrust}    {T}
\newcommand{\nthrust}   {\hat{n}_{\mathrm{thrust}}}
\newcommand{\thethr}    {\theta_{\,\mathrm{thrust}}}
\newcommand{\phithr}    {\phi_{\mathrm{thrust}}}
\newcommand{\acosthr}   {|\cos\thethr|}
\newcommand{\thejet}    {\theta_{\,\mathrm{jet}}}
\newcommand{\acosjet}   {|\cos\thejet|}
\newcommand{\thmiss}    { \theta_{miss} }
\newcommand{\cosmiss}   {| \cos \thmiss |}
%
%
\newcommand{\Evis}      {E_{\mathrm{vis}}}
\newcommand{\Rvis}      {E_{\mathrm{vis}}\,/\roots}
\newcommand{\Mvis}      {M_{\mathrm{vis}}}
\newcommand{\Rbal}      {R_{\mathrm{bal}}}
%
%
\newcommand{\phiacop}   {\phi_{\mathrm{acop}}}
%
%
\newcommand{\PhysLett}  {Phys.~Lett.}
\newcommand{\PRL} {Phys.~Rev.~Lett.}
\newcommand{\PhysRep}   {Phys.~Rep.}
\newcommand{\PhysRev}   {Phys.~Rev.}
\newcommand{\NPhys}  {Nucl.~Phys.}
\newcommand{\NIM} {Nucl.~Instrum.~Methods}
\newcommand{\CPC} {Comp.~Phys.~Comm.}
\newcommand{\ZPhys}  {Z.~Phys.}
\newcommand{\IEEENS} {IEEE Trans.~Nucl.~Sci.}
%
%
\newcommand{\OPALColl}  {OPAL Collaboration}
%
\newcommand{\onecol}[2] {\multicolumn{1}{#1}{#2}}
\newcommand{\ra}        {\rightarrow}   


\begin{titlepage}
%
%
\begin{center}
    \large
    EUROPEAN LABORATORY FOR PARTICLE PHYSICS
\end{center}
\begin{flushright}
    \large
CERN-PPE/97-124 \\
12$^{\mathrm{th}}$ September 1997
\end{flushright}

%
%
\begin{center}
    \Large\bf\boldmath
    Search for Anomalous Production of 
    Di-lepton Events with Missing Transverse Momentum \\
    in e$^+$e$^-$ Collisions at $\sqrt{s} = 161$ and 172~GeV
\end{center}
\bigskip
\bigskip

\begin{center}
 \Large
 The OPAL Collaboration \\
\end{center}
\bigskip
\bigskip

%
%
\begin{abstract}

Events containing a pair of charged leptons and significant 
missing transverse momentum are selected from
a data sample corresponding to a total integrated luminosity of 20.6~pb$^{-1}$
at centre-of-mass energies of 161~GeV and 172~GeV.
The observed number of events, four at 161~GeV and nine at 172~GeV,
is consistent with the number expected from Standard Model processes, 
predominantly arising from \wpair\ production with each W decaying
leptonically. 
This topology is also an experimental
signature for the pair production of new particles that decay
to a charged lepton accompanied by one or more
invisible particles.
Further event selection criteria are described that optimise 
the sensitivity to particular new physics channels.
No evidence for new phenomena is observed and limits 
on the production
of scalar charged lepton pairs and other new particles are presented.

\end{abstract}

\bigskip
\bigskip
\begin{center}
(Submitted to Z.~Phys.~C.)
\end{center}
 
\end{titlepage}

\begin{center}{\Large        The OPAL Collaboration
}\end{center}\bigskip
\begin{center}{
K.\thinspace Ackerstaff$^{  8}$,
G.\thinspace Alexander$^{ 23}$,
J.\thinspace Allison$^{ 16}$,
N.\thinspace Altekamp$^{  5}$,
K.J.\thinspace Anderson$^{  9}$,
S.\thinspace Anderson$^{ 12}$,
S.\thinspace Arcelli$^{  2}$,
S.\thinspace Asai$^{ 24}$,
D.\thinspace Axen$^{ 29}$,
G.\thinspace Azuelos$^{ 18,  a}$
% ...
}\end{center}\bigskip
\bigskip
$^{  1}$School of Physics and Space Research, University of Birmingham,
Birmingham B15 2TT, UK
\newline
$^{  2}$Dipartimento di Fisica dell' Universit\`a di Bologna and INFN,
I-40126 Bologna, Italy
\newline
% ...
$^{  5}$Cavendish Laboratory, Cambridge CB3 0HE, UK
\newline
% ...
$^{  8}$CERN, European Organisation for Particle Physics,
CH-1211 Geneva 23, Switzerland
\newline
$^{  9}$Enrico Fermi Institute and Department of Physics,
University of Chicago, Chicago IL 60637, USA
\newline
% ...
$^{ 12}$Indiana University, Department of Physics,
Swain Hall West 117, Bloomington IN 47405, USA
\newline
% ...
$^{ 16}$Department of Physics, Schuster Laboratory, The University,
Manchester M13 9PL, UK
\newline
% ...
$^{ 18}$Laboratoire de Physique Nucl\'eaire, Universit\'e de Montr\'eal,
Montr\'eal, Quebec H3C 3J7, Canada
\newline
% ...
$^{ 23}$Department of Physics and Astronomy, Tel Aviv University,
Tel Aviv 69978, Israel
\newline
$^{ 24}$International Centre for Elementary Particle Physics and
Department of Physics, University of Tokyo, Tokyo 113, and
Kobe University, Kobe 657, Japan
\newline
% ...
$^{ 29}$University of British Columbia, Department of Physics,
Vancouver BC V6T 1Z1, Canada
\newline
% ...
$^{  a}$ and at TRIUMF, Vancouver, Canada V6T 2A3
\newline
% ...
\newline

\newpage

\section{Introduction}
\label{sec:intro}

We report on the selection of events containing a pair of charged leptons and 
significant missing transverse momentum.
We give the results of searches for several anomalous sources of 
such events at centre-of-mass energies of 161 and 172~GeV.

A number of \sm\ processes can lead to the final state \llnunu , the most 
common of which is expected to be \wpair\ production in which both W's
decay leptonically: \dW\ (with $\ell = \mathrm{e}, \mu, \tau $).

This topology is also an experimental
signature for the production of new particles that result in
final states with two charged leptons accompanied by one or more
invisible particles.
Such invisible particles can be
weakly interacting particles such as neutrinos or 
the hypothesised lightest stable supersymmetric~\cite{SUSY} particle (LSP), 
which may be the lightest neutralino, $\nt_1$, or the gravitino.
Experimentally, invisible particles may also be weakly interacting neutral
particles with long lifetimes, which decay outside the
detector volume.

One example of a new physics possibility for which this
topology is an experimental signature is the pair production of
charged scalar leptons (sleptons, $\sell^\pm$) in the framework of
supersymmetry: 
\begin{center}
$\ee \rightarrow \sell^+ \sell^-$,

$\sell^\pm \rightarrow  {\ell^\pm} \nt_1$,
\end{center}
where $\ell^\pm$ is the corresponding charged lepton.
We present searches for scalar electrons (selectrons, $\sele$ ), scalar
muons (smuons, $\smu$ ) and scalar taus (staus, $\stau$ ).
If the gravitino were the LSP and if the slepton were to decay to 
lepton-gravitino the same experimental signature would be produced. 

Models with extended Higgs sectors predict the 
existence of several new Higgs particles, including at least one pair
of charged Higgs particles.
The pair production of charged Higgs can lead to di-tau events
from the decay $\mathrm{H}^{\pm} \rightarrow \tau^\pm \nu_\tau$.

Di-lepton events with missing transverse momentum is a signature also 
for chargino pair production when both charginos decay  to produce a charged 
lepton and invisible particles:
\begin{center}
$\ee \rightarrow \chp \chm $,

$\chpm \rightarrow \ell^\pm \snu$ \ \ \   or \ \ \
$\chpm \rightarrow \ell^\pm \nu \chz$.
\end{center}

The $\chpm$ decay to the final state $\ell^\pm \nu \chz$ may take
place via an intermediate, real or virtual, W or charged slepton.
If the $\chpm$ decay via a W dominates then 90\% of the produced events
contain quarks in the final state;
for this case limits are given in~\cite{charg172,charg161}.
However,  if one or more sneutrinos or charged sleptons are light
then final states containing  a pair of charged 
leptons and invisible particles may dominate, in which case the search
presented in this paper is more sensitive.

The detailed properties of the expected events
(e.g., the type of leptons observed and their momenta) 
varies greatly depending on the type of new particles produced and on
free parameters within the models.
For example, in $\sell^+ \sell^-$ production, if the mass 
difference \dm\ between the $\sell$ and $\nt_1$ is small,
the visible energy and transverse momentum will be small.
The main experimental challenge in this case is to detect such ``small \dm ''
events
in the presence of the large potential background from \tpp , in
particular  \eeeell .
Conversely, if \dm\ is large the visible energy and transverse 
momentum will be large and the new physics events may be difficult to
distinguish from \wpair\ events. 

The event selection is performed in two stages.
The first stage consists of a general selection for all the possible
\sig\ (section~\ref{general}).
In this context the \sm\ \llnunu\ events are considered as signal
in addition to the possible new physics sources.
All \smp\ that do not lead to \llnunu\ final states --- e.g., \eell\ and
\llgam\ --- are considered as background and are reduced to a
rather low level in this  selection.
In the second stage the detailed properties of
the events are used to separate as far as  possible the events
consistent with
potential new physics sources from  \wpair\ and other \smp\
  (section~\ref{sec:addkin}).

The number and properties of the observed events are found to be
consistent with the expectations for  \smp\  (section~\ref{efficiency}).
We present limits on the production
of charged scalar leptons, leptonically decaying charged Higgs 
and charginos that decay to produce a charged lepton and invisible particles
 (section~\ref{results}).
In a companion paper~\cite{ref:wwpaper} we use the same event 
selections to measure the production of  \wwllnunu\ events.

\section{OPAL Detector and Monte Carlo Simulation}
\label{opaldet}

A complete description of the  OPAL detector\footnote{
A right-handed coordinate system is adopted,
in which the $x$-axis points to the centre of the LEP ring,
and positive $z$ is along  the electron beam direction.
The angles $\theta$ and $\phi$ are the polar and azimuthal angles,
respectively.} 
can be found 
elsewhere~\cite{ref:OPAL-detector}.
Subdetectors relevant 
to the current analyses are described very briefly here. 

The central detector consists of
a system of tracking chambers
providing charged particle tracking
over 96\% of the full solid 
angle
inside a 0.435~T uniform magnetic field parallel to the beam axis. 
It consists of a two-layer
silicon microstrip vertex detector, a high precision drift chamber (CV),
a large volume jet chamber (CJ) and a set of $z$ chambers (CZ) that measure 
the track coordinates along the beam direction. 

A lead-glass electromagnetic
calorimeter (ECAL) located outside the magnet coil
covers the full azimuthal range with excellent hermeticity
in the polar angle range of $|\cos \theta |<0.82$ for the barrel
region (EB)  and $0.81<|\cos \theta |<0.984$ for the endcap region (EE).
Electromagnetic calorimeters close to the beam axis 
complete the geometrical acceptance down to approximately 25 mrad.
These include 
the forward detectors (FD) which are
lead-scintillator sandwich calorimeters and, at smaller angles,
silicon tungsten calorimeters (SW)
located on both sides of the interaction point.
The gap between the EE and FD calorimeters
is instrumented with an additional lead-scintillator 
electromagnetic calorimeter,
called the gamma-catcher (GC).

The magnet return yoke is instrumented for hadron calorimetry (HCAL)
and consists of barrel and endcap sections along with pole tip detectors that
together cover the region $|\cos \theta |<0.99$.
Outside the hadron calorimeter, four layers of muon chambers 
cover the polar angle range of $|\cos \theta |<0.98$. 

The following \smp\ are simulated.
4-fermion production is generated using grc4f~\cite{grc4f},
{\sc Pythia}~\cite{pythia} and {\sc Excalibur}~\cite{excalibur}.
Two-photon processes are generated using the program of
Vermaseren~\cite{vermaseren}
for \eell , and 
using {\sc Phojet}~\cite{phojet} and 
{\sc Pythia} for \eeqq .
The production of lepton pairs is generated using 
{\sc Bhwide}~\cite{bhwide} and {\sc Teegg}~\cite{teegg} for $\ee(\gamma)$, 
 and using {\sc
  Koralz}~\cite{koralz} for $\mumu(\gamma)$ and $\tautau(\gamma)$.
The production of quark pairs, \qpair($g$), is generated using {\sc Pythia} 
and the final state \nunugam\ is generated using 
{\sc Nunugpv}~\cite{NUNUGPV} and {\sc Koralz}.

The following new physics processes are simulated.
Slepton pair production is generated using {\sc Susygen}~\cite{SUSYGEN}. 
Charged Higgs pair production is generated using  {\sc Pythia}.
Chargino pair production is generated using  {\sc Susygen}
 and {\sc Pythia}.
All \sm\ and new physics \mc\ samples are processed with a full simulation  
of the OPAL detector \cite{gopal} and subjected to the same
reconstruction and analysis programs as used for the OPAL data.

\section{General Selection of Di-lepton Events with Significant Missing Momentum}
\label{general}

The first stage in the analysis
consists of a general selection for all the possible
\sig .
At this stage \sm\ \llnunu\ events are considered as signal
in addition to the possible new physics sources.
In this section we give a general overview of the selection; details
are given in the appendices.
 
The event selection requires evidence that a pair of leptons is
produced in association with an invisible system that carries away
significant missing energy and momentum.
A number of additional cuts are applied to distinguish events with
genuine prompt missing momentum (\sm\ \llnunu\ events and potential
new physics) from the background, in which the signature of 
missing momentum is faked by \sm\ processes containing, e.g.,
secondary neutrinos from tau decays or
particles that strike regions of the detector where they are
undetected or poorly measured.

Leptons may be identified as electrons, muons or hadronically decaying
taus\footnote{
Leptonically decaying taus are normally identified as isolated
electrons or muons.}.
The flavour of the two lepton candidates is not required to be the same.

A number of \sm\ processes can lead to high energy particles
travelling down the beam pipe, thus being undetected and giving rise
to missing momentum along the beam axis.
Therefore, in selecting candidate signal events we require a significant 
missing momentum in the plane perpendicular
to the beam axis (\ptevt ) and require that the total missing momentum
vector points away from the beam axis. 
Nevertheless, \sm\ events
containing neutrinos (in particular from tau decay) 
or poorly measured particles represent a
potential background to this signature. 
In these events the value of \ptevt\ may be large and the missing momentum
vector may point at a large angle to the beam axis; such events may
thus survive cuts on these quantities.
However, in the majority of  \sm\ events the two leptons tend to be
approximately \btb\  in
the plane perpendicular to the beam axis (coplanar).
In coplanar events the component of \ptevt\
that is perpendicular to the event thrust axis in the 
transverse plane, which is referred to as \ptaxic , 
is much less sensitive than \ptevt\ to the presence of 
neutrinos from tau decays or poorly measured particles.
A cut on \ptaxic\ is also more
performant than a cut on the acoplanarity angle\footnote{
The acoplanarity angle (\acop ) is defined  as 
180$^{\circ}$ minus  
the angle
between the two lepton candidates in the plane transverse to the 
beam direction.}, \acop .
Consider, for example, electrons produced in tau decay.
Low momentum electrons from this source 
may be produced at very large angles to the original
tau direction (thus causing a large \acop ), but their momentum transverse
to the original tau direction (and thus their contribution to \ptaxic
) is small.
In events that are relatively coplanar
a cut on  \ptaxic\ is applied and 
the direction of the missing momentum vector is
calculated using  \ptaxic\ rather than \ptevt .
At large acoplanarity cuts using \ptaxic\ 
no longer discriminate sufficiently between
signal and background and a hard cut 
on \ptevt\ and a conventional cut on the direction of the missing
momentum are made.

The dominant background that survives cuts on missing
momentum arises from \tp\ \lp\
in which one of the initial state
beam particles is scattered at an appreciable angle to
the beam direction.
Events that may have arisen from such processes
 are suppressed by requiring no significant energy 
to be present in the GC, FD and SW detectors.

Muons and hadrons produced at small angles to the beam direction
($\theta < 0.2$~rad) are likely to be poorly measured in the detector. 
In order to reduce the \sm\ background from events containing such
particles,  the muon chambers, hadron calorimeters
and central detector are  examined for any evidence of particles
 escaping in the very forward region that might fake
 the signature of 
missing momentum.

Two independently developed selections are used in the first stage
general selection of acoplanar di-lepton events.
In one selection particular emphasis has been placed on retaining efficiency 
for events with very low visible energy, but nevertheless significant 
missing transverse momentum.
It is referred to below as ``selection I'' and
is described in detail in  appendix~\ref{sec-selI} of this paper.
An earlier version of this selection has been used to set limits on new
particle production at \sthree~\cite{ref:paperppe}.
A second selection
has been  optimised to select and measure the rate of
high visible energy events such as those from \wpair\ events in which both
 W's decay leptonically. 
It is referred to below as ``selection II''
and is described in detail in  appendix~\ref{sec-selII}.
An earlier version of this selection has been used at \ssix\ to
measure the production of  \wwllnunu\ events~\cite{Wmass}.

For events with high visible energy such as \wpair\ or in new physics
scenarios with large mass differences betweeen the parent and the visible
daughter particles, the two selections have similar efficiency and 
background acceptance. However, a significant fraction of the Monte Carlo
signal events are selected exclusively by only one of the two selections.
The general selection of acoplanar di-lepton events is, therefore, 
defined as the logical ``.or.'' of the  selections ``I'' and ``II''.

The general selection of acoplanar di-lepton events 
has been applied to
the data at centre-of-mass energies of 161 and 172~GeV.
Four candidate events are selected at 161~GeV.
The total number of selected events predicted by the \smc\ for the
integrated luminosity\footnote{
The overall error on the luminosity amounts to 0.53\% at 161~GeV and
0.55\% at 172~GeV~\protect\cite{fpair}.}
 of 10.3~\ipb\  is 
$4.7 \pm 0.3$~(stat)
events, of which 3.5 events arise from 4-fermion processes with
genuine prompt missing energy and momentum (\llnunu ) and 1.2 events are
background coming mainly from processes with four charged leptons in
the final state (of which only two are observed in the detector).
About half of the \eell\ background arises from two-photon (multiperipheral)
processes and about half from other four-fermion processes, most
notably \zee~$\rightarrow$~\eemumu.

Nine candidate events are selected at 172~GeV.
The total number of selected events predicted by the \smc\ for the
integrated luminosity of 10.3~\ipb\   is 
$12.1 \pm 0.3$~(stat)
events, of which 11.0 events arise from \llnunu\ and 1.1 events do not
have genuine missing momentum.

\begin{figure}[htbp]
 \epsfxsize=\textwidth 
 \epsffile[0 0 580 600]{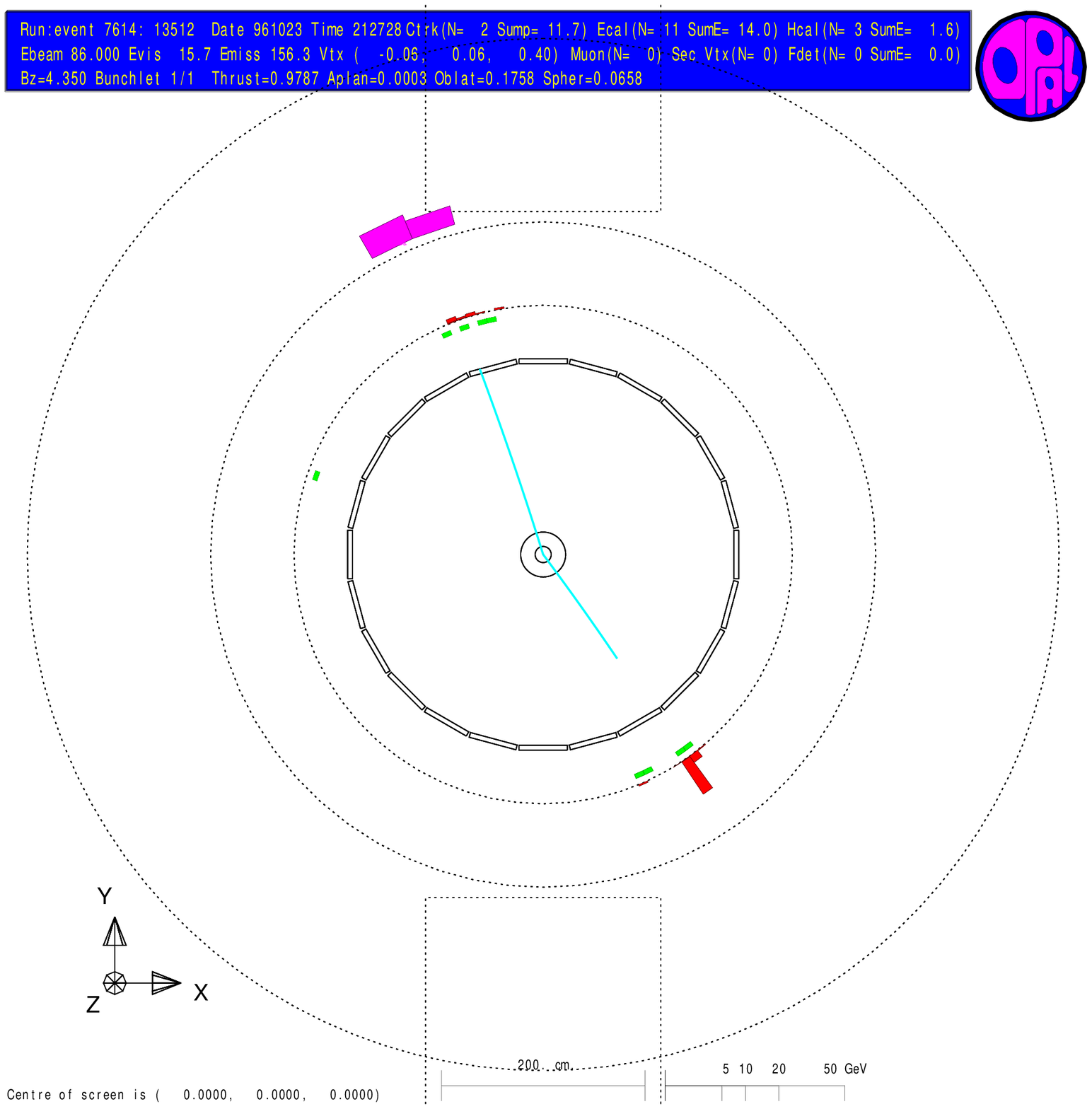}
 \caption{
Acoplanar di-lepton candidate number 3 at 172~GeV.
This event is selected as a \wpair\ candidate and is considered as a
candidate in the searches for
stau pair, charged Higgs pair and chargino pair production.
} 
\label{epic3}
\end{figure}

\begin{figure}[htbp]
 \epsfxsize=\textwidth 
 \epsffile[0 0 580 600]{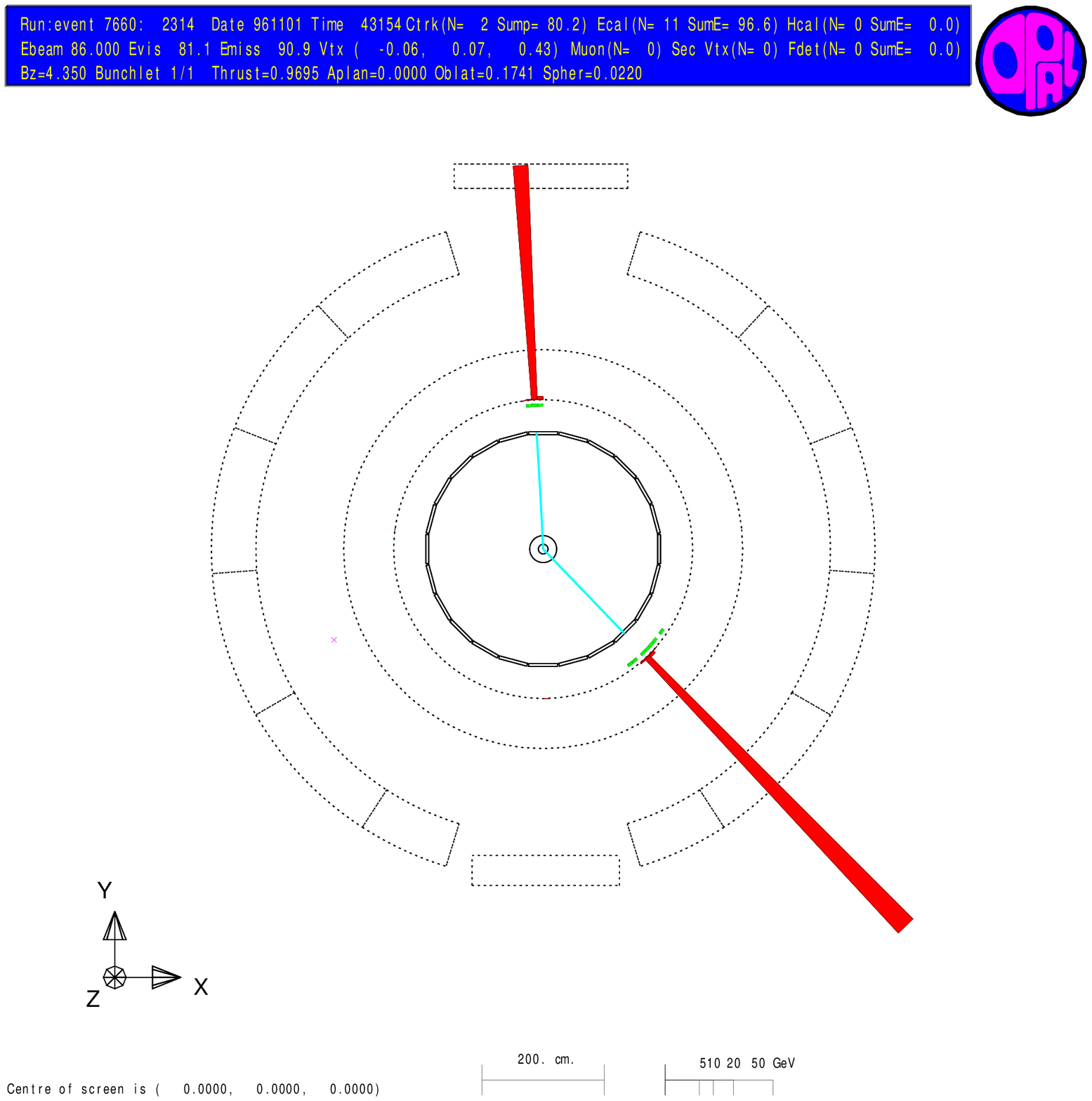}
 \caption{
Acoplanar di-lepton candidate number 6 at 172~GeV.
This event is selected as a \wpair\ candidate and is considered as a
candidate in the search for
selectron pair production.
} 
\label{epic6}
\end{figure}

The properties of the four events selected
 at \ssix\ are given in table~\ref{tab-lnln-events}.
The properties of the nine events selected
 at \sseven\ are given in table~\ref{tab-lnln-events2}.
Event displays of two of the
candidates at \sseven\
are shown in figures~\ref{epic3} and~\ref{epic6}.

\begin{table}[htbp]
\centering
\begin{tabular}{||l||c|c|c|c||} 
\hline\hline
 & \multicolumn{4}{c||}{Candidate} \\ \cline{2-5}
              & 1   & 2  &  3 &  4     \\ \hline\hline
Run Number & 7270 &  7274 & 7402  &  7463  \\
Event Number & 9719 & 61080 & 203918 &  62162 \\ 
Selection  & I &  I,II    &  I,II  & I \\
Lepton id. & $h^{+}h^{-}$ & $\mu^+\mu^-$ & e$^-\mu^+$ & e$^+h^-$ \\
\wpair\ classification & Y  &   Y    &  Y    & N   \\
Candidate for: & $\stau$,H,$\chone$ & $\smu$,$\chone$ & -- & $\sele$,$\stau$,$\chone$ \\
\acop\ (rad)      & 2.31   & 0.50   & 0.43  & 3.06    \\
\stevt            & 0.14   & 0.27   & 0.19  & 0.05   \\
$x_{+}$           & 0.15   & 0.63   & 0.55  & 0.07   \\
$x_{-}$           & 0.21   & 0.28   & 0.56  & 0.01   \\
\MLL\  (GeV)      & 22.1   & 64.9   & 87.7  & 2.0   \\
\MRECOIL\ (GeV)   & 136.1  & 81.2   & 69.1  & 154.4   \\
\hline\hline
\end{tabular}
\caption{\sl Properties of the selected events 
at 161~GeV.
The row labelled ``Lepton id.''
gives the results of the lepton identification;  ``$h$'' means
that the lepton is identified neither as an electron nor muon and
so is probably the product of a hadronic tau decay.
Leptonic decays of taus are usually classified
as electron or muon.
In the row labelled ``\wpair\ classification'' a ``Y'' indicates 
that an event is selected as a \wpair\ candidate and
an ``N'' indicates that it is not.
Events are considered as \wpair\ candidates: (i)
if they pass selection I and the
additional \wpair\ selection
criteria designed to suppress two-photon background and
non-\wpair\ \sm\ \llnunu\ events as
described in appendix~\protect\ref{sec-wwcuts}; 
or (ii) if they pass selection II.
The row labelled ``Candidate for:'' is relevant to the new particle
searches described in section~\protect\ref{sec:addkin} below.
It gives the new physics categories for which a given event is
selected as a candidate.
A dash ``--'' indicates that an event is not consistent with any new
physics category according to the cuts of section~\protect\ref{sec:addkin}.
$x_{+}$ and $x_{-}$ are the momenta of the positively charged lepton
and negatively charged lepton, respectively, as a fraction of the beam
energy.
\MLL\ and \MRECOIL\ are the invariant masses of the lepton pair and of the
invisible particles, respectively.
}
\label{tab-lnln-events}
\end{table}

\begin{table}[htbp]
\centering
\begin{tabular}{||l||c|c|c|c|c|c|c|c|c||} 
\hline\hline
 & \multicolumn{9}{c||}{Candidate} \\ \cline{2-10}
                     & 1     &    2   &    3  &    4   &    5  &    6 & 7 & 8 & 9         \\ \hline\hline
Run Number           &  7596 &  7596  & 7614  & 7631   & 7631  & 7660  &  7718  & 7729  & 7757    \\
Event Number         &  7824 &  34581 & 13512 & 58714  & 59299 & 2314  &  48005 & 32594 & 20446   \\ 
Selection            &   I,II &   I,II  & I     &   I,II  & I     &  I,II   &   I,II   &   I,II  &  I,II     \\
Lepton id.  & $h^{+}h^{-}$ & e$^{-}h^+$ & $h^{+}h^{-}$ & e$^+\mu^-$
& $\mu^-h^+$ & e$^+$e$^-$ & 
e$^+\mu^-$ & $\mu^+\mu^-$ & e$^+$e$^-$ \\
\wpair\ classification     &   Y   &   Y    &  Y    &   Y    &  N    &   Y    &   Y &
Y & Y \\
Candidate for:        &$\stau$,H,$\chone$&$\stau$,$\chone$&$\stau$,H,$\chone$
 &  $\chone$   &
$\smu$,$\stau$,$\chone$    &  $\sele$ & -- &
-- & $\sele$,$\chone$ \\
\acop\ (rad)         & 0.58  &  0.34  & 0.34   & 0.65   & 3.13  & 0.71 & 2.82  & 2.16  &  0.25
 \\
\stevt               & 0.27  &  0.23  & 0.041  & 0.38   & 0.047 & 0.51 &  0.71 &  0.29 &  0.12
 \\
$x_{+}$              & 0.17  &  0.15  & 0.17   & 0.84   & 0.01 & 0.79 & 0.38 &  0.61 &  0.55
\\
$x_{-}$              & 0.42  &  0.38  & 0.09   & 0.66   & 0.04 & 0.43 & 0.40 &  0.04 &  0.44
\\
\MLL\    (GeV)       & 44.2  &  32.9  & 17.2  & 117.0  & 1.0 &  95.1 & 26.2 & 16.1  &  76.3
\\
\MRECOIL\  (GeV)     & 118.5 & 127.5  & 154.2 & --     & 167.7 &  50.0 & 84.2 & 106.4 &  77.6
 \\
\hline\hline
\end{tabular}
\caption{\sl Properties of the selected events at 172~GeV.
}
\label{tab-lnln-events2}
\end{table}

\begin{figure}[htbp]
 \epsfxsize=\textwidth 
 \epsffile[0 0 580 600]{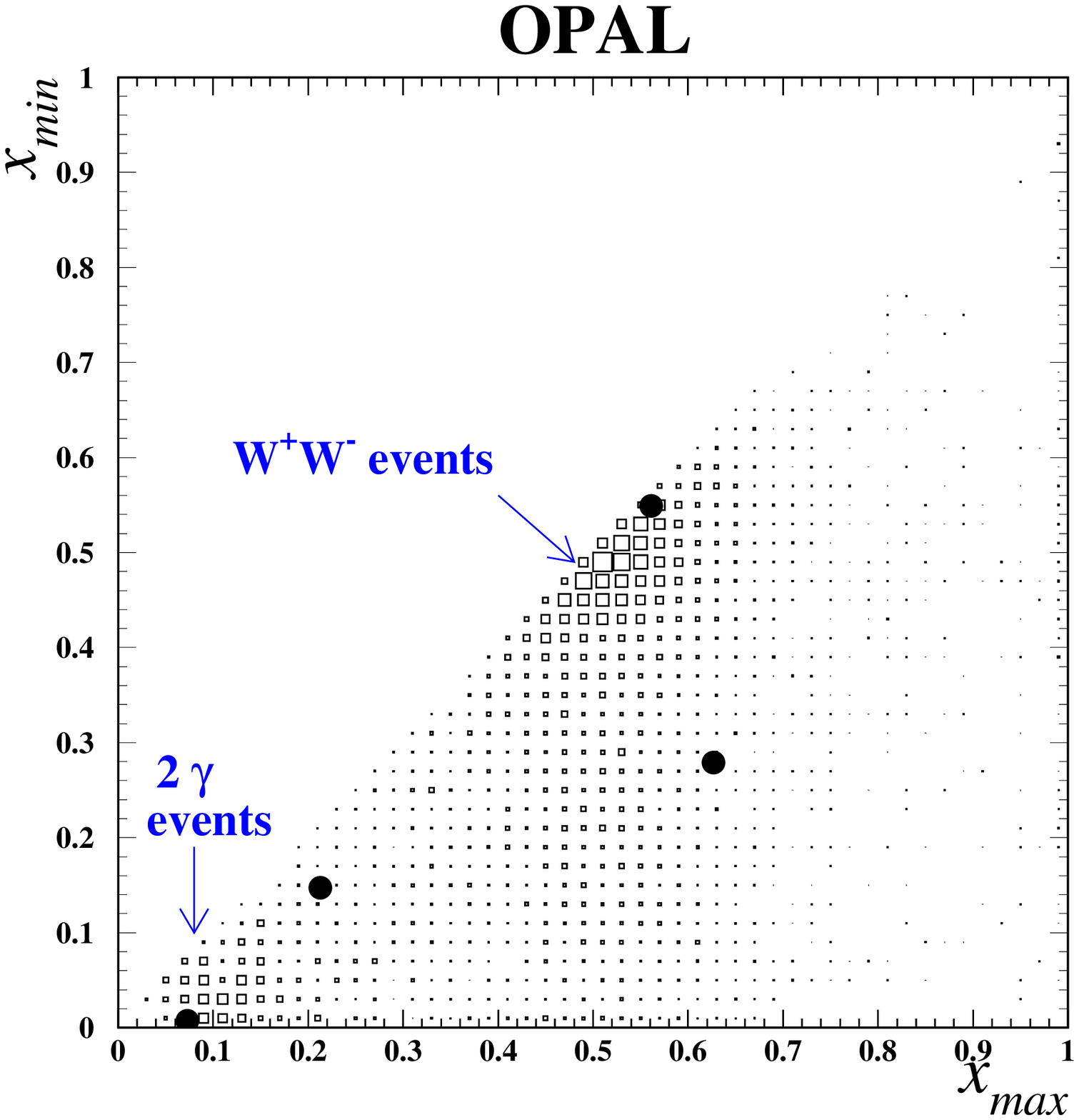}
 \caption{
Distribution in the (\xmax, \xmin) plane of events passing the general
selection of acoplanar di-leptons at $\protect\sqrt{s}$~=~161~GeV.
The four data events are shown as the circular points.
The \smc\ distribution is shown as the squares.
The regions that correspond to \wpair\
production and to two-photon processes are indicated.
} 
\label{ebkd161}
\end{figure}

\begin{figure}[htbp]
 \epsfxsize=\textwidth 
 \epsffile[0 0 580 600]{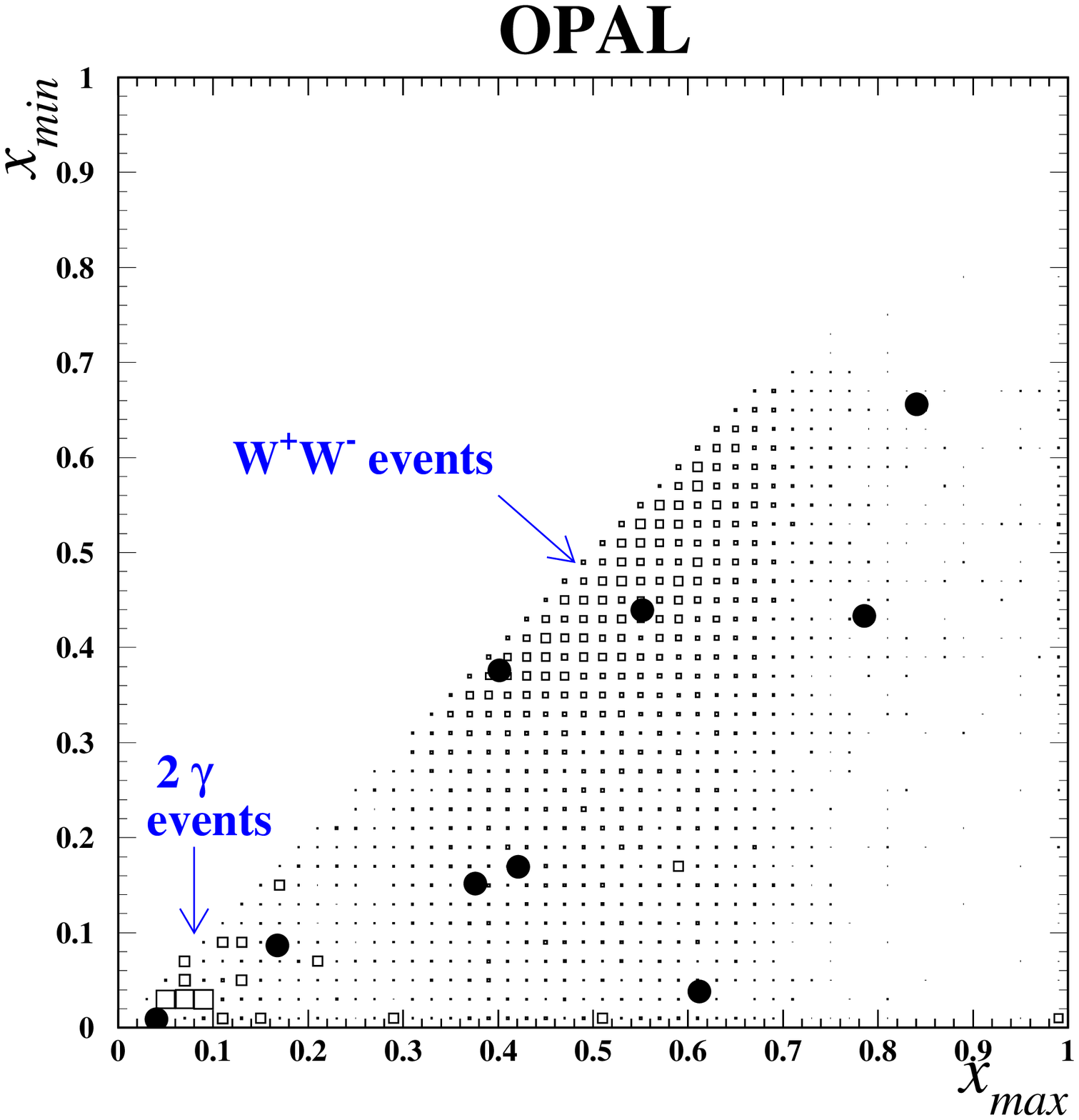}
 \caption{
Distribution in the (\xmax, \xmin) plane of events passing the general
selection of acoplanar di-leptons at $\protect\sqrt{s}$~=~172~GeV.
The nine data events are shown as the circular points.
The \smc\ distribution is shown as the squares.
The regions that correspond to \wpair\
production and to two-photon processes are indicated.
} 
\label{ebkd172}
\end{figure}

In figure~\ref{ebkd161} the selected candidates at \ssix\  are plotted 
 in the (\xmax, \xmin) plane. 
 The scaled
momenta of the lepton candidates are defined by:
\begin{center}
$\xmax = \frac{p_{max}}{E_{beam}},$\ \ \ \ \ \ \ \ 
$\xmin = \frac{p_{min}}{E_{beam}},$
\end{center}
where $p_{max}$ is the momentum of the higher momentum lepton, and $p_{min}$
is the momentum of the lower momentum lepton.
The \sm\ Monte Carlo distribution in the (\xmax, \xmin) plane is also
shown in figure~\ref{ebkd161}.
Because the W's are produced
approximately at rest, electrons and muons
from W decays, \wtoenu\ and \wtomunu , have a momentum close to 
half the beam energy.
This is seen in the clustering of events with \xmin\ and \xmax\ close
to 0.5.
In general, the process \wtotaunu\ produces lepton 
candidates with lower values of
scaled momentum than those produced in \wtoenu\ and \wtomunu\ decays.
The clustering of events at very low \xmax\ and \xmin\ results 
mainly from the \eell\ background.  
The \mc\ at \sseven\ (figure~\ref{ebkd172}) shows a
\wpair\ peak that is broader than at \ssix , 
because the W's are no longer at rest.

The lepton identification information in the selected data events is
compared with the \sm\ expectation in table~\ref{tab-mi}.
The numbers of events observed in the data are consistent
with \sm\ expectations.
Of particular interest are the numbers of
events in which the flavour of the two observed leptons is the same.
In \wwllnunu\ events there is no correlation between the flavour of the
two leptons.
Some new physics sources of acoplanar di-leptons would produce events with
correlated lepton flavour, for example, 
the pair production of sleptons or charged Higgs.

The event classes \epair , \mupair\  and \emu\ receive a large
contribution from \wpair .
In order to increase the potential sensitivity to new physics sources 
of events in these categories whilst
minimising the assumptions we have to make about the kinematics of any
new physics events, we can remove events in which both leptons
have momentum in the region expected for prompt
\wtoenu\ and \wtomunu\  decays.
We remove events in which both leptons have $0.40<p/\Ebeam<0.65$ at
\ssix\ and $0.36<p/\Ebeam<0.68$ at \sseven .
As can be seen in figures~\ref{ebkd161} and~\ref{ebkd172} these
regions contain a large fraction of the expected \wpair\ events.
The results are shown in the final three
rows of  table~\ref{tab-mi}.
There is no indication of an excess in the data \wrt\ the \sm\
expectations in any category.
In the following sections we will apply more sophisticated selections
in order to optimise the sensitivity to various potential new physics
sources of acoplanar di-lepton events.

\begin{table}[htb]
\centering
\begin{tabular}{||l||c|c|c|c||} 
\hline\hline
Event & \multicolumn{2}{c|}{\ssix}  & \multicolumn{2}{c||}{\sseven} \\ 
\cline{2-5}
Class                          & data &  SM  & data &  SM   \\ 
\hline
\epair\                        & 0    & 0.6  & 2    & 1.7   \\
\mupair\                       & 1    & 0.8  & 1    & 1.9   \\
\emu\                          & 1    & 1.5  & 2    & 4.0   \\
$\mathrm{e}^{\pm} h^{\mp}$     & 1    & 0.8  & 1    & 2.2   \\
$\mu^{\pm} h^{\mp}$            & 0    & 0.8  & 1    & 1.9   \\
$h^{\pm} h^{\mp}$              & 1    & 0.2  & 2    & 0.4   \\
\hline
\multicolumn{5}{||l||}{Apply veto on \wtoenu\ and \wtomunu\  decays } \\
\hline
\epair\                        & 0    & 0.4  & 1    & 0.8   \\
\mupair\                       & 1    & 0.5  & 1    & 1.0   \\
\emu\                          & 0    & 1.0  & 1    & 2.1   \\
\hline\hline
\end{tabular}
\caption{\sl Lepton identification information in the selected data events 
compared with the \sm\ expectation.
For further details see the text.
}
\label{tab-mi}
\end{table}

\begin{figure}[htbp]
 \epsfxsize=\textwidth 
 \epsffile{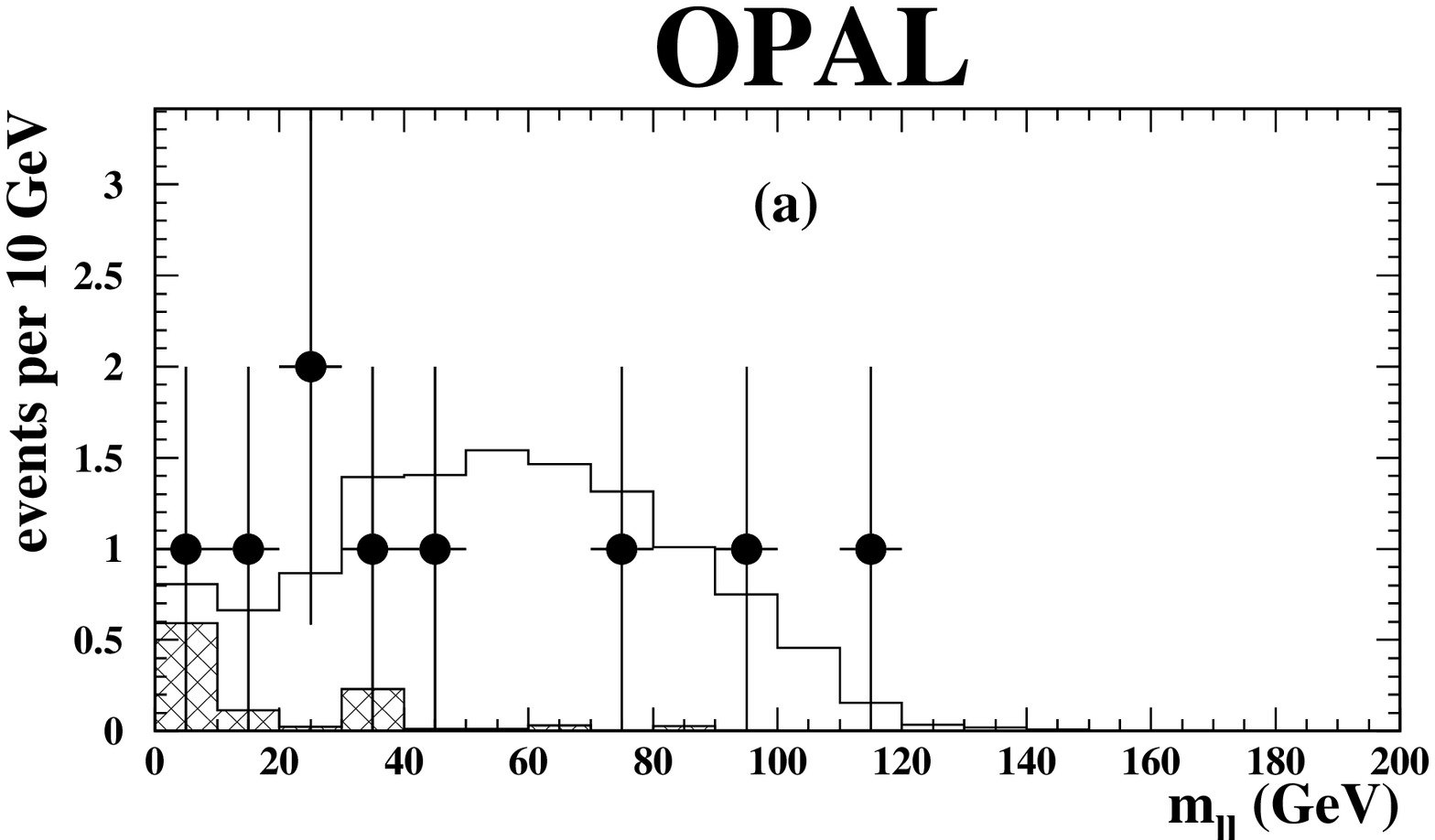}
 \epsfxsize=\textwidth 
 \epsffile{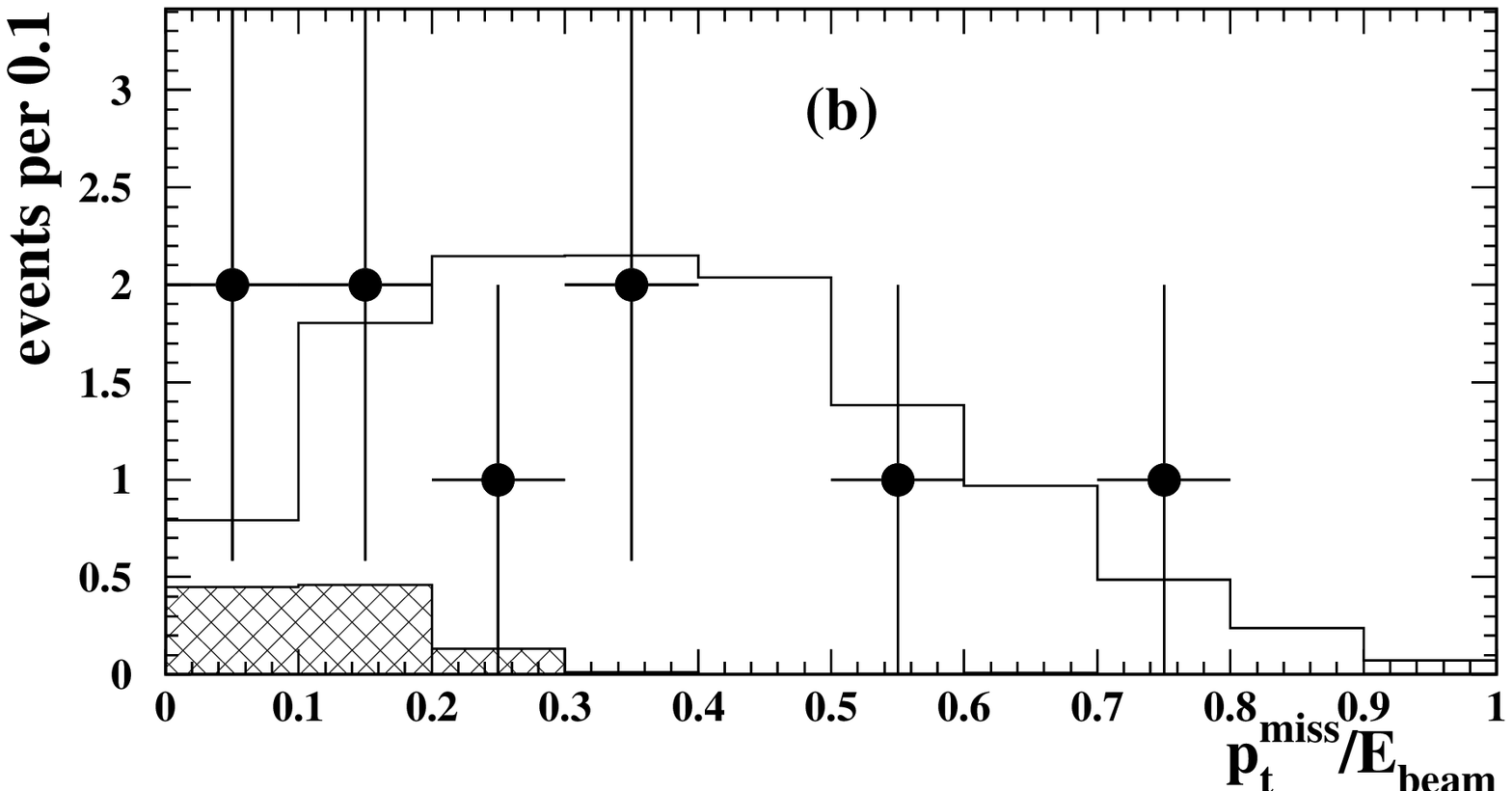}
 \caption{
Distributions at  $\protect\sqrt{s}$~=~172~GeV  of
(a) \llmass\ and (b) \stevt\
 of the observed events
compared with the \smc .
The data are shown as the points with error bars.
The \mc\ prediction for  4-fermion processes with
genuine prompt missing energy and momentum (\llnunu ) is shown as the
open histogram and the
background, coming mainly from processes with four charged leptons in
the final state, is shown as the shaded histogram.
} 
\label{fig-llmass}
\end{figure}

Figure~\ref{fig-llmass} shows the distributions of
(a) \llmass\ and (b) \stevt\ of the observed events at \sseven\ 
compared with the \smc .
The \smc\ and data are consistent.
Additional comparisons between the data and \smc\ expectations are
given in appendix~\ref{sec-selI}.

\section{Additional Selection Criteria for Specific Search Channels}
\label{sec:addkin}

\subsection{Introduction}

Starting from the general selection of acoplanar di-lepton events
described in section~\ref{general}, we search for the production of
new particles by applying
additional cuts
to suppress \sm\ sources of such events, the most important of which
are \llnunu\ and \eell .

The \sml\ from \wpair\ are characterised by the production of two leptons,
both with \spp\ around 0.5.
Equal numbers of \epm , \mupm\ and $\tau^\pm$ are produced and there is no
correlation between the flavours of the two charged leptons in the
event.
In the \sm\ \eell\ events the two observed leptons both tend to 
have low momentum. 

In the signal events the momentum distribution of the expected 
leptons varies strongly
as a function of the mass difference, \dm\ between the parent particle (e.g.,
selectron) and the invisible daughter particle (e.g., lightest neutralino),
and, to a lesser extent, $m$, the mass of the parent particle.
When performing a search for a particular new particle at a
particular point in $m$ and \dm , an event is considered as a potential
candidate only if the results of the lepton identification and the 
momenta of the observed leptons are
consistent with expectations.

In order to maximise the  {\it a priori}\, potential to discover new
physics the additional selection cuts must be tuned to give the
optimal balance between signal selection efficiency and the number of
selected background events.
There is no unique prescription for how to achieve this, particularly
when the expected production cross-section and/or branching
ratio for new physics channels are unknown, as is the case here.
In this analysis we use the following prescription.
The precise values of the additional selection cuts  are chosen in such
a way
that, in the hypothesis of no signal being present, the {\it a priori}\,
average value
of the 95\% CL upper limit on the cross-section for new physics is minimised.
This is achieved by an
automated optimisation procedure that makes use of Monte Carlo
samples of signal and Standard Model backgrounds, but not the
experimental data.

\subsection{Optimisation of the Cut Values}
\label{sec:opt}

For a given set of selection cuts one can estimate $\mu _B$, the mean
number of selected \sm\ events expected from \mc . 
If $N$ candidate events are actually observed in an experiment, 
then it is possible to 
calculate the 95\% CL upper limit \Nninef\ on the expected number of
signal events, using the method advocated by the PDG~\cite{PDG}.

The upper limit at 95\% CL on the cross-section for new 
particle production is then given by:
$$ \signine = \frac{\Nninef}{\varepsilon {\cal L}}, $$
where $\varepsilon$ is the selection efficiency for the particular
type of new 
particle production and decay being considered and ${\cal L}$ is the
integrated luminosity of the experiment.

In the absence of any new physics sources, $N$ is expected to follow 
a  Poisson distribution, $P(N,\mu _B)$, with mean $\mu_B$.
We can, therefore, calculate the expectation value of \Nnine\ for an
ensemble of experiments:
$$\langle\Nnine (\mu _B)\rangle = \sum_{N=0}^{\infty} P(N,\mu _B)\Nninef.$$
This in turn gives us the expectation value of the 95\% CL upper
limit on the cross-section for new particle production:
$$\langle\signine\rangle = \frac{\langle\Nnine (\mu _B)\rangle}{\varepsilon {\cal L}}. $$
A modification to the selection cuts changes both $\langle\Nnine (\mu_B)\rangle$ and
$\varepsilon$.
An optimised set
of cuts is found by minimising $\langle\signine\rangle$ using an iterative procedure.

The above can be generalised to include data from more than one
$\roots$ value.
The 95\% CL upper
limit on the cross-section for new particle production  at \sseven ,
obtained by combining the data at the two centre-of-mass 
energies \ssix\ and \sseven\ is given by:
$$\langle\signine\rangle = \frac{\langle\Nnine (\sum_i \mu _{B_i})\rangle}
{\sum_i \varepsilon_i {\cal L}_i \omega_i},$$
where the sums run over the two centre-of-mass energies.
$\omega_i$ is a weight factor which takes into account that
the expected production cross-section varies with $\roots$,
but the limit on the observed cross-section is quoted at \sseven.
$$ \omega_i = \frac{\sigma_i}{\sigma_{172}}, $$
where $\sigma_{172}$ is the expected cross-section for \sseven\
and $\sigma_i$ is the expected cross-section for the $i$'th value of
$\roots$.  
For scalar particles, for example sleptons, 
we assume that the expected cross-section
varies as $\beta^3/s$.
For spin 1/2 particles, for example charginos, 
we assume that the expected cross-section
varies as $\beta/s$.

We can expect the optimal selection cuts to be different at the  two 
centre-of-mass energies, given that we expect different
cross-sections at the  two centre-of-mass energies, both 
for signal and background.
For example, the \wpair\ cross-section is significantly higher at
\sseven\ than at \ssix .
In regions of ($m$, \dm ) where  \wpair\ is the dominant source of
background we can expect the optimised selection cuts to be tighter 
at \sseven\ than at \ssix . 
The selection cuts at the  two centre-of-mass energies
 are therefore optimised simultaneously.
Note that this optimisation of the cut values uses only \mc\ signal
and background distributions, and thus is not biased by the real data.

\subsection{Selection Criteria for Selectrons}
\label{sec:sele}

In this section we describe the additional selection criteria in the
search for selectron pair production and decay: 
\begin{center}
$\ee \rightarrow \sele^+ \sele^-,$

$\sele^\pm \rightarrow  {\mathrm e}^\pm \nt_1.$
\end{center}
In the region of high mass difference, \dm , between $\sele^\pm$ and $\nt_1$
this process would lead to events containing two energetic electrons
and the dominant background arises from W pair production:
\begin{center}
$\ee \rightarrow \wpair,$

${\mathrm{\Wpm}} \rightarrow \ell^\pm \nu.$
\end{center}

There are two important distinctions between signal and background:

(i)  Selectrons always decay to produce electrons, whereas a 
leptonically decaying 
     W pair  decays to $\ee\nu_{\mathrm e}\nbar_{\mathrm e}$ with
     probability of only 1/9.

(ii) $\nt_1$ may be massive whereas the neutrino is massless.  
     Therefore electrons from selectron decay will in general be softer
     than those from W decay.  Also the mass of the W is 80~GeV,
     whereas the mass of the selectron accessible at \sseven\ could be
     anywhere between\footnote{
       Masses below 45~GeV are not considered given the negative results of
       searches at LEP1.
     } 45~GeV and~85~GeV, again resulting in different
     kinematic properties of the observed electrons.

In the region of low \dm ,  selectron pair production would lead to
events containing two low energy electrons and the dominant background
arises from \eell\ events.

In order to discriminate between selectron pair production and the \smb\
 we make requirements on the results of the lepton identification 
and on the momenta of the observed leptons.

\subsubsection{Kinematic Cuts for Selectrons}
\label{sec:kin}

\begin{figure}[htbp]
 \epsfxsize=\textwidth 
 \epsffile[0 0 580 600]{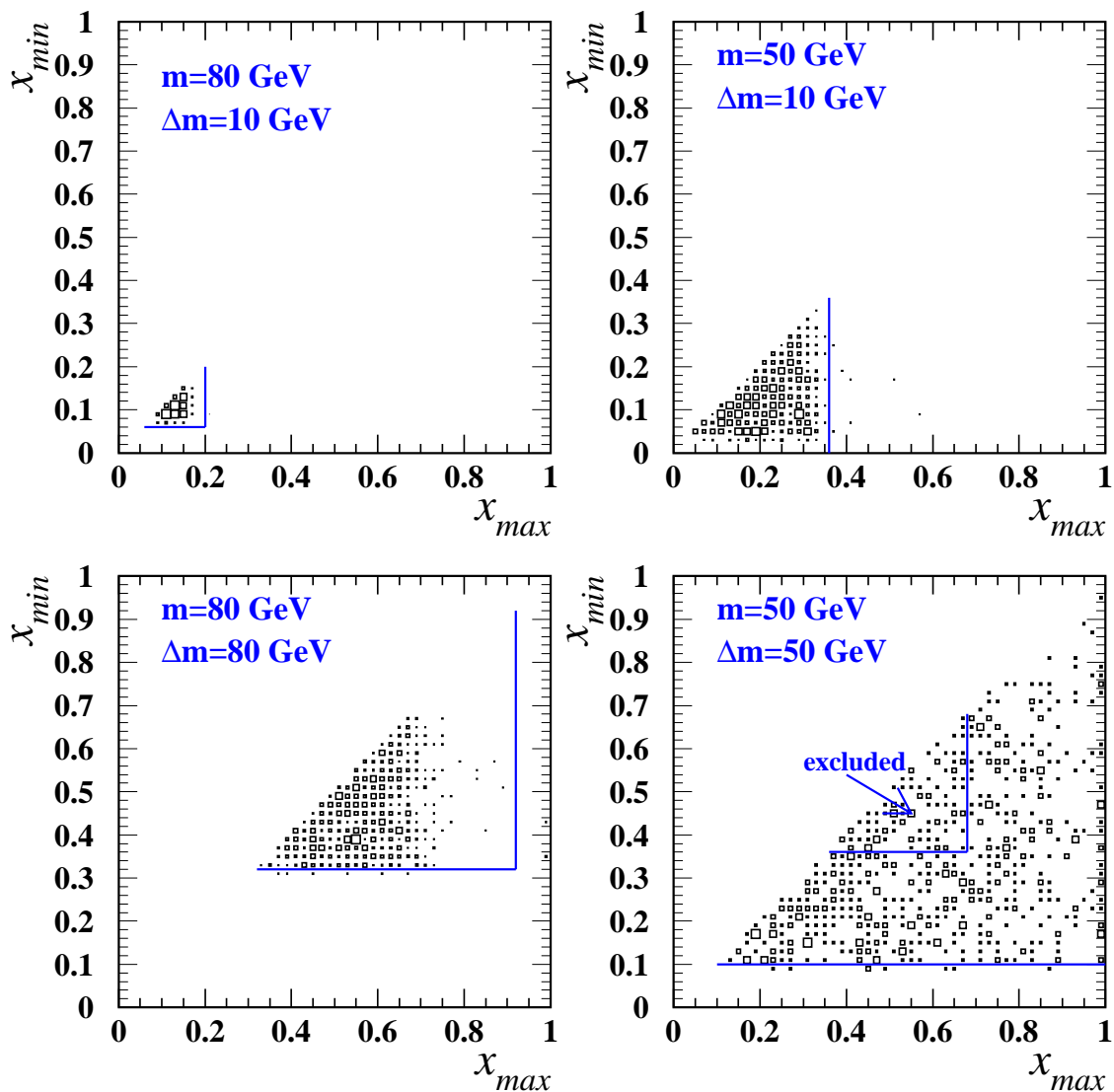}
 \caption{
Selectron \mc\ signal distributions in the (\xmax, \xmin) plane at 
$\protect\sqrt{s}$~=~172~GeV  
for four different combinations of $m = \msele$ and $\dm = \msele
- \mchi$.  
The kinematic cuts that are applied for these values of $m$ and
\dm\ are indicated. 
} 
\label{esig}
\end{figure}

The (\xmax, \xmin) distribution expected for the selectron
 signal varies considerably with the 
values of $m$ and \dm.
The signal \mc\ distribution at \sseven\ is shown for four different
$m$ and \dm\ values in figure~\ref{esig}.  As \dm\ increases, so does
the amount of energy available to the electron.
As the selectron mass $m$ decreases,
the Lorentz boost given to the electrons increases, spreading out the
distribution.  The signal behaves similarly at \ssix.

Comparison of the \sm\ background (figure~\ref{ebkd172}) 
and new physics signal 
(figures~\ref{esig}) distributions suggests that
two sets of kinematic cuts can be made:

(i)  {\bf First stage cuts}:  The signal is contained within 
a triangular region in the (\xmax, \xmin) plane.
Therefore an upper cut on \xmax\ and a lower cut on \xmin\ are made.

(ii) {\bf Second stage cuts}:  In order to remove the large peak
in the background due to \wpair\ production, events within the triangular 
region containing a large fraction of the \wpair\  are excluded.

The cuts on \xmax, \xmin\ are indicated in figure~\ref{esig}.
The ``second stage'' cuts are applied only for those values of  $m$ and
\dm\ for which they result in a decrease in $\langle\signine\rangle$,
calculated using the \mc\ events as described in section~\ref{sec:opt}.

Selectron \mc\ samples 
are available for about 50 different combinations of $m$ and \dm\ at 
each centre-of-mass energy.  
The values of $m$ range from $m = 45$~GeV up to $m \approx \Ebeam$ in
5~GeV steps.
For each value of $m$, the values of \dm\  range from $\dm = 1.5$~GeV up
to $\dm = m$.
There are a total of eight kinematic cut values to be optimised 
for a given $m$ and \dm\ (four at
each centre-of-mass energy) .  
The available \mc\ samples are used to calculate the optimum values of the
kinematic cuts for a discrete set of values of $m$ and \dm.  The cut
values are then parameterised to give cuts that vary continuously as
functions of $m$ and \dm.  This procedure smoothes out the effect of
statistical fluctuations in the individual signal \mc\ samples.
The
validity of the fit is established by verifying that the value of
$\langle\signine\rangle$ corresponding to the parameterised cuts does not differ
significantly from the value corresponding to the optimised cuts.

\subsubsection{Lepton Identification Requirements for Selectrons}

Although a selectron pair always decays to produce a pair of electrons, it is
not necessarily optimal to require that two electrons are identified, as
this will result in a loss in efficiency.
At each value of $\roots$,  $m$ and \dm, two possibilities are considered:

(i)  require two identified electrons,

or

(ii) require at least one identified electron and no identified 
muons\footnote{
If the lepton identification fails then the electron track is likely to be
identified as a candidate hadronic tau decay.
An electron is very unlikely to be identified as a muon candidate.
However, a significant fraction of the \smb\ to the search at low \dm , 
in particular \eemumu ,
contain one observed electron and one observed muon.}.

We use the combination of lepton type requirements at the two centre-of-mass
energies that, after optimisation of the kinematic cuts, gives the
lowest value of $\langle\signine\rangle$.

It is found that at high values of \dm, where the expected background
from \wpair\ is at its highest, $\langle\signine\rangle$ is minimised if two
electrons are required.
Conversely, for low values of \dm, where the efficiency of the 
general selection is already lower and the expected background
from \smp\ is also lower,  $\langle\signine\rangle$ is minimised by 
requiring one electron and no muons.

\subsection{Selection Criteria for Smuons}
\label{sec:smuon}

The procedure to optimise the selection cuts in the search for smuons
is similar to that described for selectrons
 in section~\ref{sec:sele}, except that the two lepton type
 requirements considered are:

(i)  At low \dm:  require no identified electrons and at least one
identified muon

(ii) At high \dm: require two identified muons.

\subsection{Selection Criteria for Staus}
\label{sec:stau}

The search for staus is complicated by the fact that the observed
particles are the decay products of the taus, rather than the taus
themselves:

\begin{center}

$\ee \rightarrow \stau^+ \stau^-,$

$\stau^\pm \rightarrow  {\tau^\pm} \nt_1,$

$\tau^\pm \rightarrow \mathrm{e}^\pm \nu \bar{\nu}$\ \ \  or 
\ \ \   $\mu^\pm \nu \bar{\nu}$\ \ \  
or \ \ \   hadrons+$\nu_{\tau}$.
\end{center}

The events are first classified according to the number of identified
electrons or muons in the final state. Both \wpair\ and \staupair\ can decay
to final states containing two, one or zero electrons or muons.
However,  it is important to note that the relative fractions of events
in these three classes is very different for \wpair\ and \staupair , as
is shown in table~\ref{tab:br}.

\begin{table}[htbp]
\centering
\begin{tabular}{||l||c|c||c|c|c|c||} 
\hline\hline
            & \multicolumn{2}{c||}{Fraction} & \multicolumn{4}{c||}
                                                {Number of Events} \\ 
\cline{2-7}
Final State & $\wpair\rightarrow$ & $\staupair\rightarrow$ 
& \multicolumn{2}{c|}{\ssix}
                  & \multicolumn{2}{c||}{\sseven} \\
\cline{4-7}
              &  \ \llnunu\  & $\ \tau^+\nt_1\tau^-\nt_1$ 
 & data &  SM  bgrd   & data &  SM  bgrd  \\
\hline\hline
\epair\  or \mupair\ or \emu\                       
 & 0.61   & 0.12 &   2  &  2.9$\pm$0.2  &   5  &  7.6$\pm$0.2 \\
$\mathrm{e}^{\pm} h^{\mp}$ or $\mu^{\pm} h^{\mp}$   
 & 0.34   & 0.46 &   1  &  1.6$\pm$0.1  &   2  &  4.1$\pm$0.1 \\
$h^{\pm} h^{\mp}$                                   
 & 0.05   & 0.42 &   1  &  0.2$\pm$0.0  &   2  &  0.4$\pm$0.0 \\
\hline\hline
\end{tabular}
\caption{\sl 
Columns 2 and 3 show the relative  fractions of leptonically decaying 
\wpair\ and \staupair\ pairs containing
two, one or zero \epm\  or \mupm\  (before selection cuts).  
Note that in the case of \wpair\, the branching 
ratio to \epm\  or \mupm\ includes decays to \epm\  or \mupm\ via a 
$\tau^\pm$.
Columns 4 and 5 show the number of
events in the general selection at  $\protect\sqrt{s}$~=~161~GeV
in these three classes compared with the  SM expectation before the cut
on the lepton momenta. Columns 6 and 7 show the same information at
$\protect\sqrt{s}$~=~170 -- 172~GeV.
}
\label{tab:br}
\end{table}

\begin{figure}[htbp]
 \epsfxsize=\textwidth 
 \epsffile[0 0 580 600]{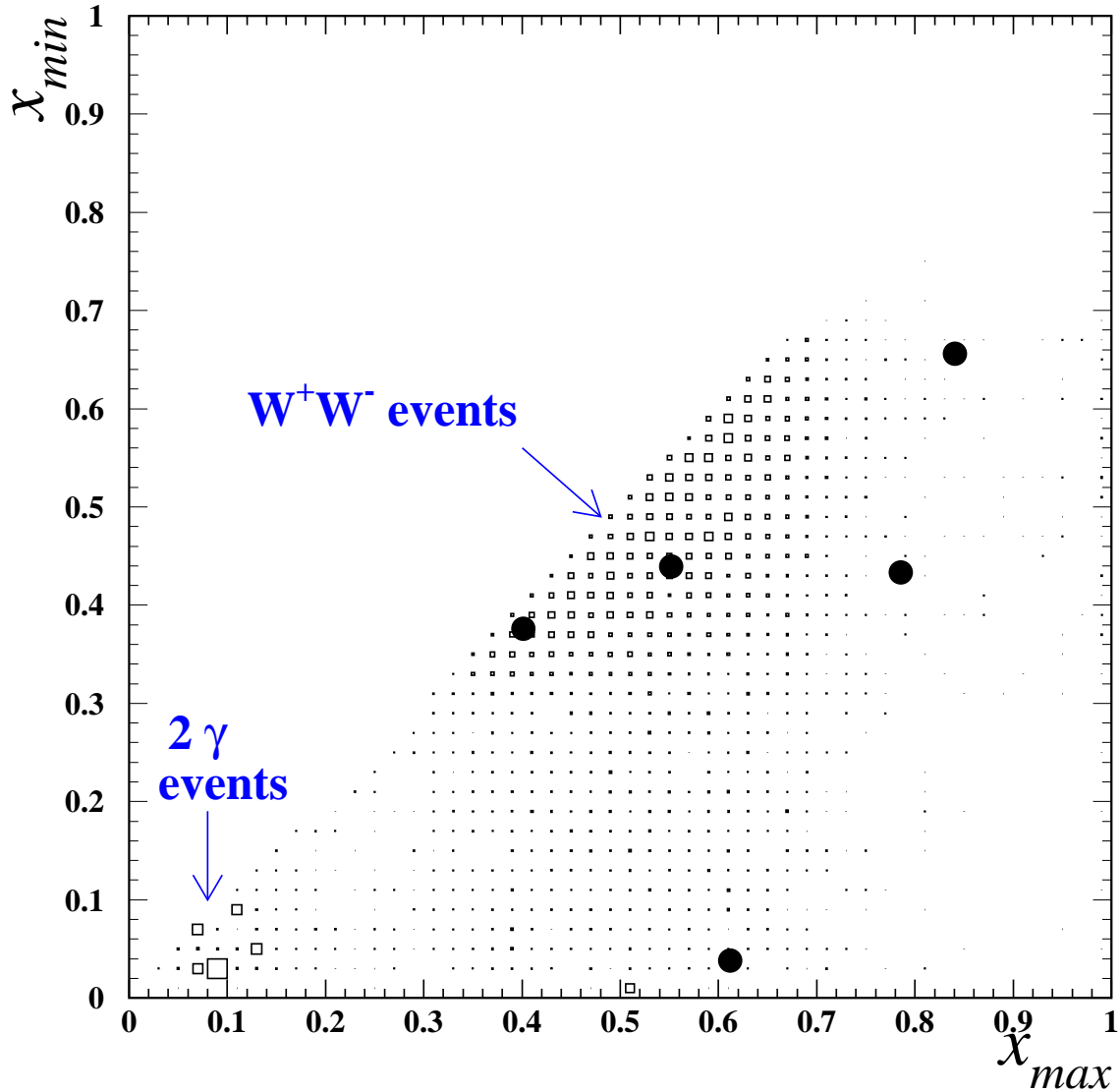}
 \caption{
Distribution in the (\xmax, \xmin) plane at 
$\protect\sqrt{s}$~=~172~GeV, with
the requirement that both the observed leptons are identified as either
electrons or muons.  
The data events are shown as the circular points.
The \smc\ distribution is shown as the squares.
The regions that correspond to \wpair\
production and to two-photon processes are indicated.
} 
\label{fig:tbkdem}
\end{figure}

\begin{figure}[htbp]
 \epsfxsize=\textwidth 
 \epsffile[0 0 580 600]{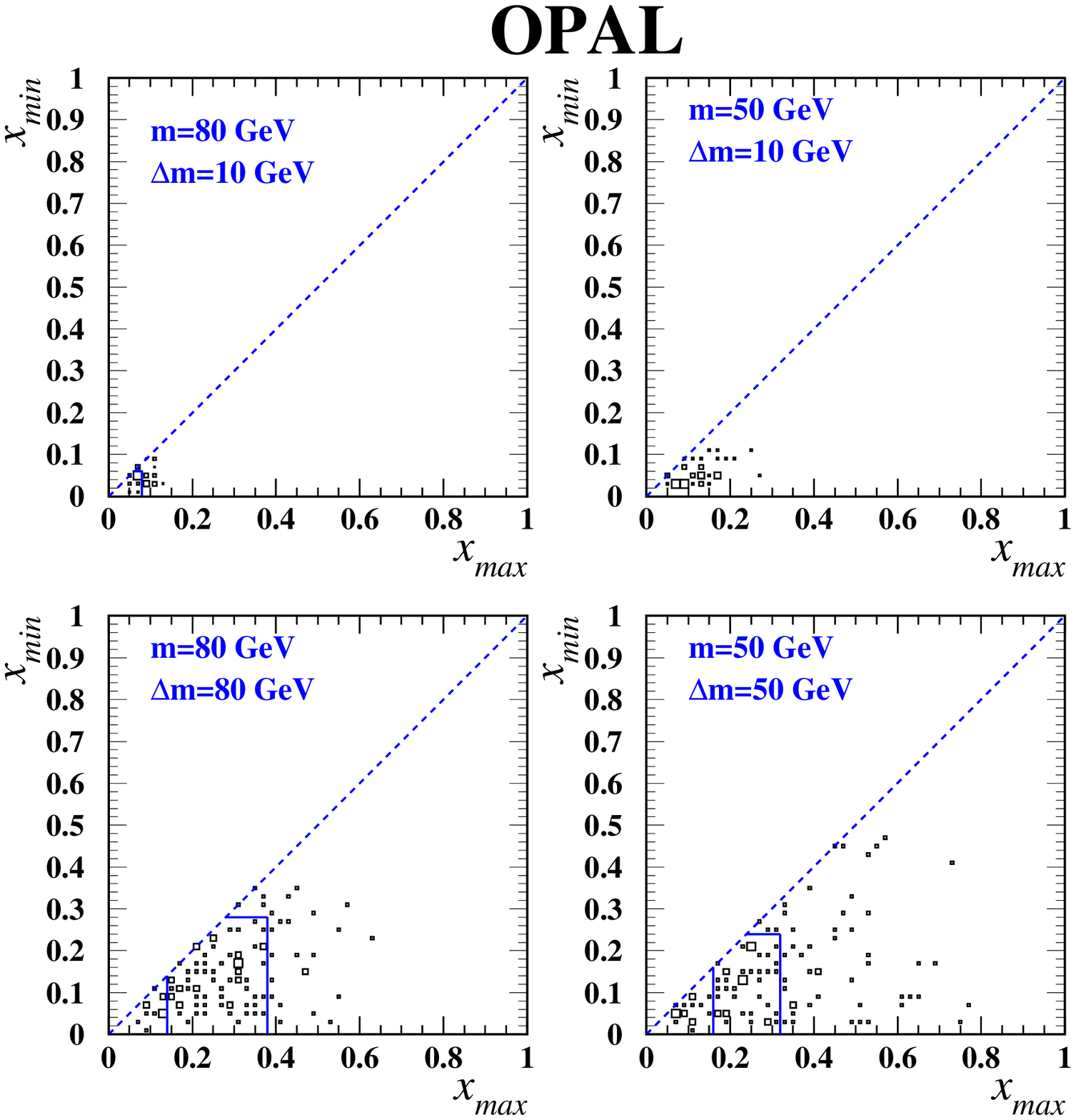}
 \caption{
Stau \mc\ signal distribution in the (\xmax, \xmin) plane at 
$\protect\sqrt{s}$~=~172~GeV,
for four different combinations of $m = \mstau$ and $\dm = \mstau-\mchi$,
 with the requirement that both the observed leptons are
identified as either electrons or muons.
The kinematic cuts that are applied
for these values of $m$ and
\dm\ are illustrated.  Note that for $m$=50~GeV,
\dm=10~GeV, all events in this class are
rejected by the automated cut optimisation procedure; 
therefore no acceptance box is shown on the figure.
For $m$=80~GeV,
\dm=10~GeV, events in the region $\xmax < 0.08, \xmin < 0.06$ are
accepted, although the  acceptance box is difficult to see in the
figure.
For further discussion see section~\protect\ref{sec:stau} and 
table~\protect\ref{tab:br}.
} 
\label{fig:tsigem}
\end{figure}

Events containing two identified  electrons or muons form a large
fraction of the \wpair\ background, but only a  small fraction of the
\taupair\ signal. Therefore  rather tight cuts on lepton
momentum are necessary in this case. This is illustrated in
figures~\ref{fig:tbkdem} and~\ref{fig:tsigem}, which show the distributions 
in the (\xmin , \xmax) plane for events containing two identified
electrons or muons. 
Figure~\ref{fig:tbkdem} shows the Standard Model
expectation and  figure~\ref{fig:tsigem} the
simulated \staupair\ signal for four example $m$, \dm\ combinations.
The cuts in \xmin\ and \xmax\ are shown
 in figure~\ref{fig:tsigem}. 
It can be seen that the region dominated by \wpair\ events is excluded
in all cases.
The region in which both observed leptons are soft is dominated by
\eell\ events.
It can be seen in figure~\ref{fig:tsigem} that this region is excluded
by the \staupair\ cuts employed at high \dm .
At low \dm\ the \staupair\ signal occupies a region of
 the (\xmin , \xmax) plane  very similar to the \eell\ background;
events with two identified electrons or muons must be completely or
almost completely excluded from the search (see figure~\ref{fig:tsigem}).

\begin{figure}[htbp]
 \epsfxsize=\textwidth 
 \epsffile[0 0 580 600]{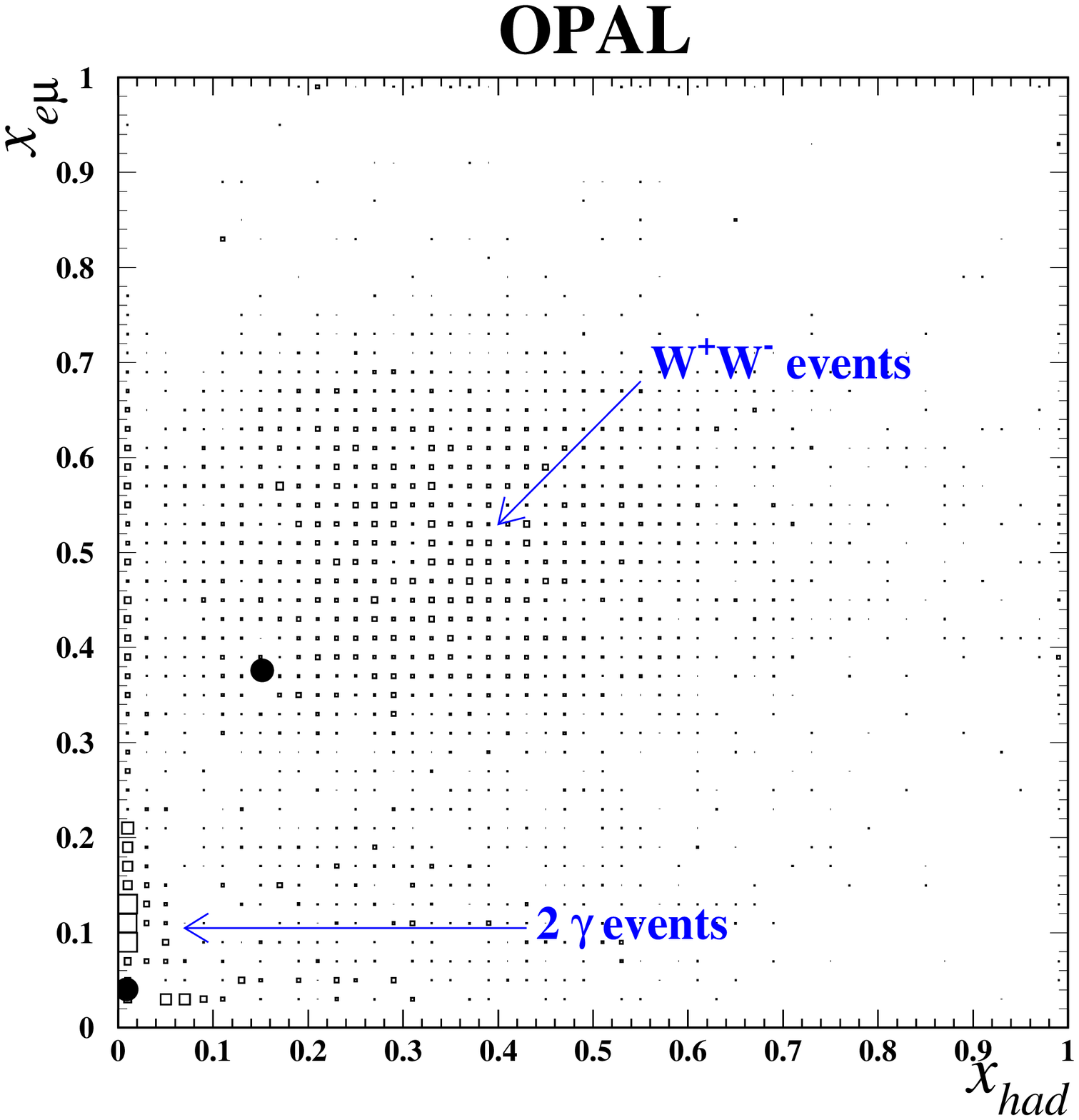}
 \caption{
Distribution in the (\xemu, \xhad) plane at
$\protect\sqrt{s}$~=~172~GeV, with
the requirement that one of the observed leptons is identified as \epm\ 
or \mupm , and the other is not.  
The data events are shown as the circular points.
The \smc\ distribution is shown as the squares.
The regions  that correspond to \wpair\
production and to two-photon processes are indicated.
} 
\label{fig:tbkdemt}
\end{figure}

\begin{figure}[htbp]
 \epsfxsize=\textwidth 
 \epsffile[0 0 580 600]{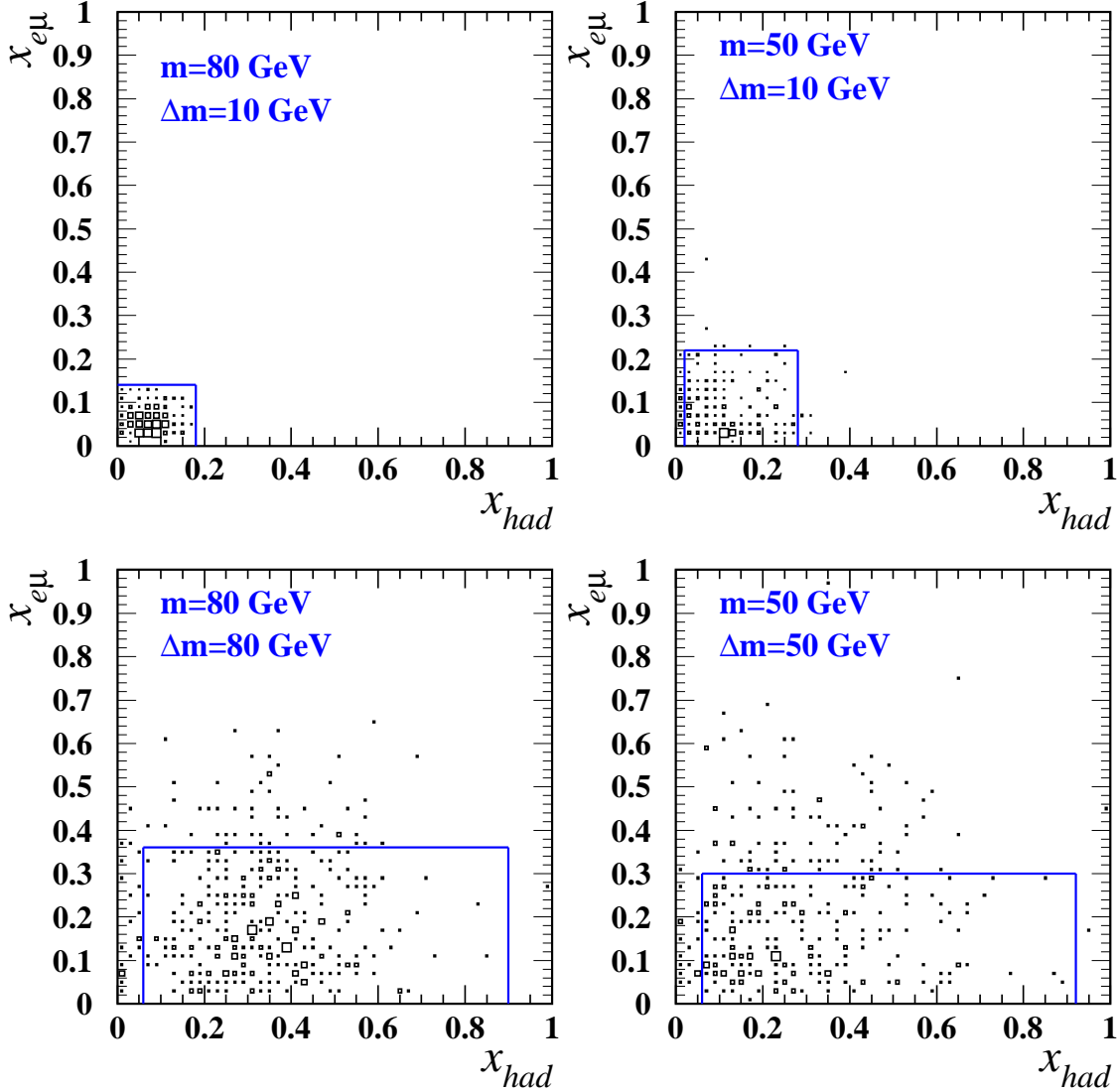}
 \caption{
Stau \mc\ signal distribution in the (\xemu, \xhad) plane at 
$\protect\sqrt{s}$~=~172~GeV,
for four different combinations of $m = \mstau$ and $\dm = \mstau-\mchi$,
with the requirement that one of the observed leptons is
identified as \epm\ 
or \mupm , and the other is not.
The kinematic cuts that are applied
for these values of $m$ and
\dm\ are illustrated.  
}
\label{fig:tsigemt}
\end{figure}

For the intermediate case, in which one lepton is identified as \epm\  or
\mupm\ and the other as a  hadronically decaying
tau, figures~\ref{fig:tbkdemt} 
and~\ref{fig:tsigemt} show the
distributions in the  (\xemu , \xhad) plane for \smc\ and
\staupair\ signal events, respectively.
\xemu\ and \xhad\ are the scaled momentum of the  \epm\  or
\mupm\ and the hadronically decaying tau, respectively.
The cuts in \xmin\ and \xmax\ are shown
 in figure~\ref{fig:tsigemt}.

At the other extreme, events without identified \epm\  or \mupm\ form a large
fraction of the \taupair\ signal, but only a small fraction of the
\wpair\ background.
Very loose cuts on momentum are appropriate in this case.
Events in which both tau candidates have very low momentum are
rejected in order to
suppress the background from \eell\ events, but
 essentially the
entire region in momentum space occupied by the signal is accepted.

\subsection{Selection Criteria for Charged Higgs}

Charged Higgs pair production can lead to acoplanar di-tau events
from the decay $\mathrm{H}^{\pm} \rightarrow \tau \nu_\tau$.
These events would have the same properties as stau pair events for
the case of massless neutralinos.
The optimised kinematic cuts obtained for staus with
massless neutralinos are therefore
parameterised and used in the search for charged Higgs pair production.
The kinematic cuts are parameterised as  a 1-dimensional
function of \mH .

\subsection{Selection Criteria for Charginos}

Charginos can decay to a final state containing a charged lepton and
one or more invisible particles by three possible processes:

\begin{center}

(i)  $\chpm \rightarrow \ell^\pm {\tilde{\nu}_\ell}$,

possibly followed by the decay:\ \ \ \ 
${\tilde{\nu}_\ell} \rightarrow \chz \nu_\ell$

(ii)  $\chpm \rightarrow {\mathrm{\Wpm}} \chz$,

      $\Wpm \rightarrow \ell^\pm \nu_\ell$

(iii)  $\chpm \rightarrow \slepton^\pm \nu_\ell$,

      $\slepton^\pm \rightarrow \ell^\pm \chz$

\end{center}

\subsubsection{Charginos Undergoing Two-Body Decay}

The first case, in which the chargino decays via a sneutrino, \dchtwo, is 
essentially a two-body decay.
The observed charged lepton is produced directly in the decay of the
chargino.
If ${\tilde{\nu}_\ell}$ is lighter than $\chpm$ then
for a given choice of the two parameters
\mch\ and \msnu\ the lepton is mono-energetic in the chargino rest-frame.
(The possible sneutrino decay is to two
invisible particles and hence  is unobserved.)

Efficiencies have been calculated for the 
case where the three sneutrino 
generations are mass degenerate.
In this case the same di-lepton mixture is expected as 
for \wpair\ and for chargino decays via a W.

\subsubsection{Charginos Undergoing Three-Body Decay}

\begin{figure}[htbp]
 \epsfxsize=\textwidth 
 \epsffile[0 0 580 600]{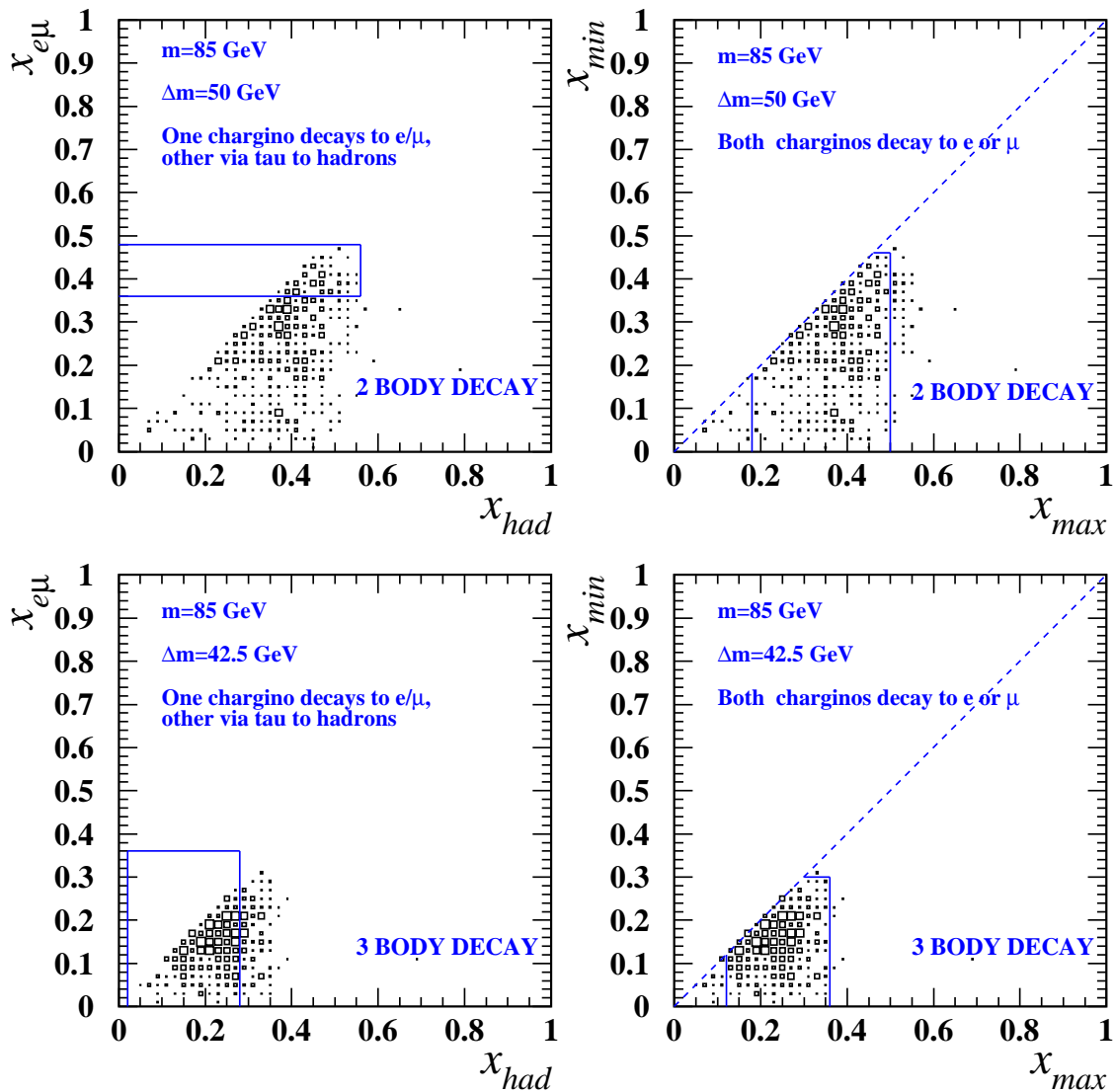}
 \caption{
Comparison between the 2- and 3-body decay hypotheses of the
scaled momentum distributions of the leptons in \chpair\ events.
Note the need for a minimum cut on \xemu\ for 2-body decay
in the case of there being one identified electron or muon.  The kinematic 
cuts applied for these values of $m$ and \dm\ are illustrated. 
}
\label{figchar}
\end{figure}

The second case, in which the chargino decays via a virtual or real W,
\dchthree , we refer  to as three-body decay.
The observed charged lepton is produced in the second stage of the chargino
decay and its momentum spectrum is determined by the two parameters
\mch\ and \mchz .
Some kinematic distributions for two-body and three-body decay are compared 
in figure~\ref{figchar}.  It can be seen that for a given value of $m$ and 
\dm , the 
momenta tend to be higher for two-body decay, as would be expected because 
the lepton comes directly from the decay of the chargino. 

The analysis applied to the three-body channel is completely 
analogous to the analysis 
applied to staus, with different kinematic cuts being applied for the three 
classes of two, one or zero electrons or muons being identified in the final 
state.
Since both signal and background now involve W decay, the relative fractions 
of events in each of the three classes is the same for both signal and 
background.  This was not the case for staus, and the advantage of the 
different fractions discussed in section~\ref{sec:stau} is lost.
The cuts on \xmax, \xmin\ are indicated in figure~\ref{figchar}.

In the case of two-body chargino decays,
 the top left 
plot in figure~\ref{figchar} shows a clear need for a minimum cut on
\xemu .
This was 
not necessary for three-body chargino decays or staus.  The application 
of a minimum cut on \xemu\ (or \xmin) is the only important difference between 
the analyses for two- and three-body chargino decays.

\subsubsection{Charginos Decaying via a Charged Slepton}

Specific results for the case in which the chargino decays via a
charged slepton, $\chpm \rightarrow \slepton^\pm \nu_\ell$, are not given 
because the kinematic distributions for the
signal will depend on $m_{\chargino}$, $m_{\tilde{\chi}^0_1}$ 
and the masses of the three charged sleptons, all of which are
unknown.
In general, we expect our search to be sensitive to this process.
However, for particularly unfavourable choices of these masses, 
for example, if $\sell^\pm$ and $\chz$ are close in 
mass, then the final state lepton will be very soft, and events from such 
a decay would not be selected.
 Note that direct searches for charged slepton pair production
 are described in sections~\ref{sec:sele}--\ref{sec:stau}.

\section{Selection Efficiencies, Number of Candidates, Expected Backgrounds}
\label{efficiency}

Tables~\ref{tab-eff1}--\ref{tab-eff4} give the 
selection efficiency, the number of selected events and 
number of events expected from \smp\  of the selections
for  \selepair , \smupair ,
\staupair , \hpair , \chargtwoo\ and \chargthreee , respectively.
It should be noted that an individual candidate event may be
consistent with a given new physics hypothesis over a range of $m$ and
\dm\ values.
Similarly, an individual candidate event may be
consistent with more than one new physics hypothesis and may,
therefore, give entries in more than one of the above tables (for
illustration see the rows labelled ``Candidate for:'' in 
tables~\ref{tab-lnln-events} and~\ref{tab-lnln-events2}).

It can be seen that sizable selection efficiencies have been obtained
for \selepair , \smupair\ and  \chargtwo , even when \dm\ is as low as
2.5~GeV.
Non-zero  efficiencies have been obtained down to  $\dm = 1.5$~GeV in
these channels.
However, in the case of \staupair\ and \chargthree\ there are additional
invisible particles (neutrinos) in the final state.
 The visible
leptons are therefore less energetic and the selection efficiencies at
low \dm\ values are reduced.

Slepton pair efficiencies are evaluated for right-handed sleptons, both 
decaying to lepton and lightest 
neutralino $\sell^\pm \rightarrow  {\ell^\pm} \nt_1$. 
The selectron pair events were generated at $\mu = -200$~GeV and
$\tan{\beta} = 1.5$.
The selection efficiency depends on
the angular distribution of the produced selectrons
and this will depend on the size of the neutralino-mediated t-channel
contribution to the cross-section. 
We have found by varying $\mu$ and $\tan{\beta}$ that the above choice
gives a conservative estimate of the selection efficiency. 

Given the limit of 37.1~GeV on the mass of the lightest sneutrino 
from LEP1~\cite{ref-snu},
\chargtwo\ events are not generated for sneutrino masses less than
35~GeV.

Inefficiencies arising from random detector occupancy 
and other effects in the data that are not modelled in the \mc\
have been studied using random triggers.
The most important influence of such un-modelled effects is to activate
one of the  ``2-photon veto cuts'' (cuts 6, 7  and 11  of
selection I --- see appendix~\ref{sec-selI}).
These vetoes are  applied only if the event has low missing
transverse momentum and so this un-modelled inefficiency is relevant
mainly for events at low \dm .  
The selection efficiencies given in tables~\ref{tab-eff1}--\ref{tab-eff4} 
contain a correction for this effect, which amounts to a 3\% reduction
in efficiency for low values of \dm .

At \sthree\ we use our previously published~\cite{ref:paperppe} 
selection efficiencies, numbers of selected events and 
numbers of events expected from \smp\  in the search 
for slepton and chargino pair production. 
For charged Higgs pair production at \sthree\ the results given 
in~\cite{ref:paperppe} for staus in the case of massless neutralinos
are used.

\begin{table}[htb]
\centering
\begin{tabular}{||l||c|c|c||c|c|c|c||}
\hline\hline
       & \multicolumn{3}{c||}{\ssix\ }  
       & \multicolumn{4}{c||}{\sseven\ } \\
\cline{2-8}
  \dm\ & \multicolumn{3}{c||}{\msele\ (GeV)}
       & \multicolumn{4}{c||}{\msele\ (GeV)} \\
\cline{2-8}
 (GeV) & 50 & 65 & 75  & 50 & 65 & 75 & 85 \\
\hline\hline
  \multicolumn{8}{||l||}{selection efficiency (\%)} \\
\hline
 1.5  
 &   5 $\pm$ 1
 &   1 $\pm$ 0
 &   0 $\pm$ 0
 &   3 $\pm$ 1
 &   1 $\pm$ 0
 &   1 $\pm$ 0
 &   0 $\pm$ 0
 \\
 2.5  
 &  32 $\pm$ 1
 &  29 $\pm$ 1
 &  26 $\pm$ 1
 &  30 $\pm$ 1
 &  27 $\pm$ 1
 &  23 $\pm$ 1
 &  18 $\pm$ 1
 \\
 5    
 &  57 $\pm$ 1
 &  60 $\pm$ 1
 &  62 $\pm$ 1
 &  50 $\pm$ 2
 &  52 $\pm$ 2
 &  60 $\pm$ 2
 &  57 $\pm$ 2
 \\
 10   
 &  67 $\pm$ 1
 &  72 $\pm$ 1
 &  73 $\pm$ 1
 &  63 $\pm$ 2
 &  65 $\pm$ 2
 &  68 $\pm$ 1
 &  68 $\pm$ 1
 \\
$m/2$  
 &  66 $\pm$ 1
 &  69 $\pm$ 1
 &  67 $\pm$ 1
 &  52 $\pm$ 2
 &  53 $\pm$ 2
 &  68 $\pm$ 1
 &  68 $\pm$ 1
 \\
 $m$    
 &  62 $\pm$ 1
 &  71 $\pm$ 1
 &  75 $\pm$ 1
 &  54 $\pm$ 2
 &  55 $\pm$ 2
 &  74 $\pm$ 1
 &  68 $\pm$ 1
 \\

\hline\hline
  \multicolumn{8}{||l||}{number of selected events} \\
\hline
 1.5  
 &          1
 &          0
 &          0
 &          0
 &          0
 &          0
 &          0
 \\
 2.5  
 &          1
 &          1
 &          0
 &          0
 &          0
 &          0
 &          0
 \\
 5    
 &          0
 &          0
 &          0
 &          0
 &          0
 &          0
 &          0
 \\
 10   
 &          0
 &          0
 &          0
 &          0
 &          0
 &          0
 &          0
 \\
 $m/2$  
 &          0
 &          0
 &          0
 &          0
 &          0
 &          1
 &          0
 \\
 $m$    
 &          0
 &          0
 &          0
 &          1
 &          1
 &          2
 &          1
 \\
\hline\hline
\multicolumn{ 8}{||l||}{number of events expected from \smp} \\
\hline
 1.5  
 &        0.0 $\pm$       0.0
 &        0.0 $\pm$       0.0
 &        0.0 $\pm$       0.0
 &        0.0 $\pm$       0.0
 &        0.0 $\pm$       0.0
 &        0.0 $\pm$       0.0
 &        0.0 $\pm$       0.0
 \\
 2.5  
 &        0.1 $\pm$       0.0
 &        0.1 $\pm$       0.0
 &        0.1 $\pm$       0.0
 &        0.2 $\pm$       0.1
 &        0.0 $\pm$       0.0
 &        0.0 $\pm$       0.0
 &        0.0 $\pm$       0.0
 \\
 5    
 &        0.3 $\pm$       0.2
 &        0.2 $\pm$       0.1
 &        0.0 $\pm$       0.0
 &        0.0 $\pm$       0.0
 &        0.0 $\pm$       0.0
 &        0.2 $\pm$       0.1
 &        0.1 $\pm$       0.1
 \\
 10   
 &        0.3 $\pm$       0.1
 &        0.0 $\pm$       0.0
 &        0.0 $\pm$       0.0
 &        0.1 $\pm$       0.0
 &        0.1 $\pm$       0.0
 &        0.0 $\pm$       0.0
 &        0.0 $\pm$       0.0
 \\
 $m/2$  
 &        0.4 $\pm$       0.0
 &        0.2 $\pm$       0.0
 &        0.1 $\pm$       0.0
 &        0.6 $\pm$       0.0
 &        0.4 $\pm$       0.0
 &        0.9 $\pm$       0.0
 &        0.2 $\pm$       0.0
 \\
 $m$    
 &        0.3 $\pm$       0.0
 &        0.2 $\pm$       0.0
 &        0.2 $\pm$       0.0
 &        0.6 $\pm$       0.0
 &        0.5 $\pm$       0.0
 &        1.3 $\pm$       0.1
 &        0.3 $\pm$       0.0
 \\

\hline\hline
\end{tabular}
\caption[]{\sl
  \protect{\parbox[t]{15cm}{
Selection efficiency, number of selected events and 
number of events expected from \smp\ in the search for \selepair\ 
production at \ssix\ and \sseven\ for different values of \msele\ and \dm .
The errors are statistical only.
}} }
\label{tab-eff1}
\end{table}

\begin{table}[htb]
\centering
\begin{tabular}{||l||c|c|c||c|c|c|c||}
\hline\hline
       & \multicolumn{3}{c||}{\ssix\ }  
       & \multicolumn{4}{c||}{\sseven\ } \\
\cline{2-8}
  \dm\ & \multicolumn{3}{c||}{\msmu\ (GeV)}
       & \multicolumn{4}{c||}{\msmu\ (GeV)} \\
\cline{2-8}
 (GeV) & 50 & 65 & 75  & 50 & 65 & 75 & 85 \\
\hline\hline
  \multicolumn{8}{||l||}{selection efficiency (\%)} \\
\hline
 1.5  
 &   5 $\pm$ 1
 &   1 $\pm$ 0
 &   0 $\pm$ 0
 &   4 $\pm$ 1
 &   1 $\pm$ 0
 &   0 $\pm$ 0
 &   0 $\pm$ 0
 \\
 2.5  
 &  44 $\pm$ 2
 &  45 $\pm$ 2
 &  37 $\pm$ 2
 &  31 $\pm$ 1
 &  29 $\pm$ 1
 &  22 $\pm$ 1
 &  16 $\pm$ 1
 \\
 5    
 &  62 $\pm$ 2
 &  63 $\pm$ 2
 &  62 $\pm$ 2
 &  58 $\pm$ 2
 &  62 $\pm$ 2
 &  61 $\pm$ 2
 &  60 $\pm$ 2
 \\
 10   
 &  72 $\pm$ 1
 &  71 $\pm$ 1
 &  73 $\pm$ 1
 &  71 $\pm$ 1
 &  70 $\pm$ 1
 &  72 $\pm$ 1
 &  73 $\pm$ 1
 \\
 $m/2$  
 &  79 $\pm$ 1
 &  80 $\pm$ 1
 &  71 $\pm$ 1
 &  62 $\pm$ 2
 &  67 $\pm$ 1
 &  76 $\pm$ 1
 &  77 $\pm$ 1
 \\
 $m$    
 &  78 $\pm$ 1
 &  83 $\pm$ 1
 &  77 $\pm$ 1
 &  63 $\pm$ 2
 &  68 $\pm$ 1
 &  83 $\pm$ 1
 &  74 $\pm$ 1
 \\

\hline\hline
  \multicolumn{8}{||l||}{number of selected events} \\
\hline
 1.5  
 &          0
 &          0
 &          0
 &          1
 &          1
 &          1
 &          0
 \\
 2.5  
 &          0
 &          0
 &          0
 &          1
 &          1
 &          1
 &          0
 \\
 5    
 &          0
 &          0
 &          0
 &          0
 &          0
 &          0
 &          0
 \\
 10   
 &          0
 &          0
 &          0
 &          0
 &          0
 &          0
 &          0
 \\
 $m/2$  
 &          1
 &          1
 &          0
 &          0
 &          0
 &          0
 &          0
 \\
 $m$    
 &          1
 &          1
 &          0
 &          0
 &          0
 &          0
 &          0
 \\
\hline\hline
\multicolumn{ 8}{||l||}{number of events expected from \smp} \\
\hline
 1.5  
 &        0.0 $\pm$       0.0
 &        0.0 $\pm$       0.0
 &        0.0 $\pm$       0.0
 &        0.0 $\pm$       0.0
 &        0.0 $\pm$       0.0
 &        0.0 $\pm$       0.0
 &        0.0 $\pm$       0.0
 \\
 2.5  
 &        0.0 $\pm$       0.0
 &        0.0 $\pm$       0.0
 &        0.0 $\pm$       0.0
 &        0.0 $\pm$       0.0
 &        0.0 $\pm$       0.0
 &        0.0 $\pm$       0.0
 &        0.0 $\pm$       0.0
 \\
 5    
 &        0.1 $\pm$       0.0
 &        0.1 $\pm$       0.0
 &        0.0 $\pm$       0.0
 &        0.0 $\pm$       0.0
 &        0.0 $\pm$       0.0
 &        0.0 $\pm$       0.0
 &        0.0 $\pm$       0.0
 \\
 10   
 &        0.2 $\pm$       0.1
 &        0.0 $\pm$       0.0
 &        0.0 $\pm$       0.0
 &        0.1 $\pm$       0.0
 &        0.0 $\pm$       0.0
 &        0.0 $\pm$       0.0
 &        0.0 $\pm$       0.0
 \\
 $m/2$  
 &        0.5 $\pm$       0.0
 &        0.4 $\pm$       0.0
 &        0.1 $\pm$       0.0
 &        0.7 $\pm$       0.0
 &        0.7 $\pm$       0.0
 &        0.7 $\pm$       0.0
 &        0.2 $\pm$       0.0
 \\
 $m$    
 &        0.4 $\pm$       0.0
 &        0.4 $\pm$       0.0
 &        0.4 $\pm$       0.0
 &        0.6 $\pm$       0.0
 &        0.8 $\pm$       0.0
 &        1.5 $\pm$       0.1
 &        0.5 $\pm$       0.0
 \\

\hline\hline
\end{tabular}
\caption[]{\sl
  \protect{\parbox[t]{15cm}{
Selection efficiency, number of selected events and 
number of events expected from \smp\ in the search for \smupair\ 
production at \ssix\ and \sseven\ for different values of \msmu\ and \dm .
The errors are statistical only.
}} }
\label{tab-eff2}
\end{table}

\begin{table}[htb]
\centering
\begin{tabular}{||l||c|c|c||c|c|c|c||}
\hline\hline
       & \multicolumn{3}{c||}{\ssix\ }  
       & \multicolumn{4}{c||}{\sseven\ } \\
\cline{2-8}
  \dm\ & \multicolumn{3}{c||}{\mstau\ (GeV)}
       & \multicolumn{4}{c||}{\mstau\ (GeV)} \\
\cline{2-8}
 (GeV) & 50 & 65 & 75  & 50 & 65 & 75 & 85 \\
\hline\hline
  \multicolumn{8}{||l||}{selection efficiency (\%)} \\
\hline
 2.5  
 &   3 $\pm$ 1
 &   1 $\pm$ 0
 &   0 $\pm$ 0
 &   1 $\pm$ 0
 &   0 $\pm$ 0
 &   0 $\pm$ 0
 &   0 $\pm$ 0
 \\
 5    
 &  12 $\pm$ 1
 &  12 $\pm$ 1
 &  11 $\pm$ 1
 &  13 $\pm$ 1
 &  12 $\pm$ 1
 &  10 $\pm$ 1
 &  10 $\pm$ 1
 \\
 10   
 &  31 $\pm$ 1
 &  31 $\pm$ 1
 &  23 $\pm$ 1
 &  28 $\pm$ 1
 &  28 $\pm$ 1
 &  29 $\pm$ 1
 &  29 $\pm$ 1
 \\
 $m/2$  
 &  44 $\pm$ 2
 &  51 $\pm$ 2
 &  51 $\pm$ 2
 &  43 $\pm$ 2
 &  49 $\pm$ 2
 &  51 $\pm$ 2
 &  59 $\pm$ 2
 \\
 $m$    
 &  50 $\pm$ 2
 &  55 $\pm$ 2
 &  53 $\pm$ 2
 &  44 $\pm$ 2
 &  47 $\pm$ 2
 &  50 $\pm$ 2
 &  62 $\pm$ 2
 \\

\hline\hline
  \multicolumn{8}{||l||}{number of selected events} \\
\hline
 2.5  
 &          1
 &          0
 &          0
 &          1
 &          1
 &          0
 &          0
 \\
 5    
 &          1
 &          1
 &          0
 &          2
 &          1
 &          1
 &          1
 \\
 10   
 &          2
 &          2
 &          0
 &          1
 &          1
 &          1
 &          2
 \\
 $m/2$  
 &          1
 &          1
 &          1
 &          2
 &          2
 &          2
 &          2
 \\
 $m$    
 &          1
 &          1
 &          1
 &          2
 &          2
 &          2
 &          3
 \\
\hline\hline
\multicolumn{ 8}{||l||}{number of events expected from \smp} \\
\hline
 2.5  
 &        0.2 $\pm$       0.1
 &        0.1 $\pm$       0.0
 &        0.0 $\pm$       0.0
 &        0.2 $\pm$       0.1
 &        0.0 $\pm$       0.0
 &        0.0 $\pm$       0.0
 &        0.0 $\pm$       0.0
 \\
 5    
 &        0.2 $\pm$       0.1
 &        0.1 $\pm$       0.0
 &        0.1 $\pm$       0.0
 &        0.2 $\pm$       0.1
 &        0.2 $\pm$       0.1
 &        0.0 $\pm$       0.0
 &        0.1 $\pm$       0.0
 \\
 10   
 &        0.6 $\pm$       0.2
 &        0.5 $\pm$       0.2
 &        0.2 $\pm$       0.1
 &        0.4 $\pm$       0.1
 &        0.3 $\pm$       0.1
 &        0.2 $\pm$       0.1
 &        0.4 $\pm$       0.2
 \\
 $m/2$  
 &        1.0 $\pm$       0.2
 &        1.0 $\pm$       0.2
 &        0.6 $\pm$       0.1
 &        1.2 $\pm$       0.2
 &        1.4 $\pm$       0.2
 &        1.2 $\pm$       0.2
 &        1.9 $\pm$       0.2
 \\
 $m$    
 &        1.0 $\pm$       0.2
 &        1.1 $\pm$       0.2
 &        0.7 $\pm$       0.1
 &        1.3 $\pm$       0.2
 &        1.4 $\pm$       0.2
 &        1.5 $\pm$       0.2
 &        2.6 $\pm$       0.2
 \\

\hline\hline
\end{tabular}
\caption[]{\sl
  \protect{\parbox[t]{15cm}{
Selection efficiency, number of selected events and 
number of events expected from \smp\ in the search for \staupair\ 
production at \ssix\ and \sseven\ for different values of \mstau\ and \dm .
The errors are statistical only.
}} }
\label{tab-eff3}
\end{table}

\begin{table}[htb]
\centering
\begin{tabular}{||c|c|c||c|c|c|c||}
\hline\hline
        \multicolumn{3}{||c||}{\ssix\ }  
      & \multicolumn{4}{c||}{\sseven\ } \\
\hline
        \multicolumn{3}{||c||}{\mH\ (GeV)}
      & \multicolumn{4}{c||}{\mH\ (GeV)} \\
\hline
  45 & 55 & 65  & 45 & 55 & 65 & 75 \\
\hline\hline
  \multicolumn{7}{||l||}{selection efficiency (\%)} \\
\hline
    47 $\pm$ 2
 &  51 $\pm$ 2
 &  56 $\pm$ 2
 &  35 $\pm$ 2
 &  45 $\pm$ 2
 &  47 $\pm$ 2
 &  51 $\pm$ 2
 \\

\hline\hline
  \multicolumn{7}{||l||}{number of selected events} \\
\hline
      
            1
 &          1
 &          1
 &          2
 &          2
 &          2
 &          2
 \\
\hline\hline
\multicolumn{ 7}{||l||}{number of events expected from \smp} \\
\hline
      
          0.8 $\pm$       0.1
 &        1.0 $\pm$       0.2
 &        1.1 $\pm$       0.2
 &        1.2 $\pm$       0.2
 &        1.3 $\pm$       0.2
 &        1.3 $\pm$       0.2
 &        1.5 $\pm$       0.2
 \\

\hline\hline
\end{tabular}
\caption[]{\sl
  \protect{\parbox[t]{15cm}{
Selection efficiency, number of selected events and 
number of events expected from \smp\ in the search for \hpair\ in
which both Higgs particles undergo the decay \dH\
at \ssix\ and \sseven\ for different values of \mH .
The errors are statistical only.
}} }
\label{tab-eff5}
\end{table}

\begin{table}[htb]
\centering
\begin{tabular}{||l||c|c|c||c|c|c|c||}
\hline\hline
       & \multicolumn{3}{c||}{\ssix\ }  
       & \multicolumn{4}{c||}{\sseven\ } \\
\cline{2-8}
  \dm\ & \multicolumn{3}{c||}{\mch\ (GeV)}
       & \multicolumn{4}{c||}{\mch\ (GeV)} \\
\cline{2-8}
 (GeV) & 55 & 65 & 75  & 55 & 65 & 75 & 85 \\
\hline\hline
  \multicolumn{8}{||l||}{selection efficiency (\%)} \\
\hline
 1.5  
 &   3 $\pm$ 1
 &   1 $\pm$ 0
 &   0 $\pm$ 0
 &   3 $\pm$ 1
 &   1 $\pm$ 0
 &   0 $\pm$ 0
 &   0 $\pm$ 0
 \\
 2.5  
 &  17 $\pm$ 1
 &  17 $\pm$ 1
 &  11 $\pm$ 1
 &  17 $\pm$ 1
 &  18 $\pm$ 1
 &  11 $\pm$ 1
 &   8 $\pm$ 1
 \\
 5    
 &  43 $\pm$ 2
 &  44 $\pm$ 2
 &  46 $\pm$ 2
 &  42 $\pm$ 2
 &  45 $\pm$ 2
 &  45 $\pm$ 2
 &  46 $\pm$ 2
 \\
 10   
 &  50 $\pm$ 2
 &  50 $\pm$ 2
 &  53 $\pm$ 2
 &  52 $\pm$ 2
 &  53 $\pm$ 2
 &  50 $\pm$ 2
 &  54 $\pm$ 2
 \\
 20   
 &  66 $\pm$ 1
 &  68 $\pm$ 1
 &  61 $\pm$ 2
 &  48 $\pm$ 2
 &  52 $\pm$ 2
 &  58 $\pm$ 2
 &  66 $\pm$ 2
 \\
 \mf\ 
 &  \dm = 20
 &   --
 &  68 $\pm$ 1
 &  \dm = 20
 &   --
 &  56 $\pm$ 2
 &  67 $\pm$ 1
 \\
 $m-35$ 
 &  \dm = 20
 &  65 $\pm$ 2
 &  72 $\pm$ 1
 &  \dm = 20
 &  44 $\pm$ 2
 &  57 $\pm$ 2
 &  71 $\pm$ 1
 \\

\hline\hline
  \multicolumn{8}{||l||}{number of selected events} \\
\hline
 1.5  
 &          0
 &          0
 &          0
 &          1
 &          1
 &          1
 &          0
 \\
 2.5  
 &          1
 &          1
 &          0
 &          1
 &          1
 &          1
 &          1
 \\
 5    
 &          1
 &          1
 &          1
 &          1
 &          1
 &          1
 &          1
 \\
 10   
 &          0
 &          0
 &          0
 &          1
 &          1
 &          1
 &          0
 \\
 20   
 &          1
 &          1
 &          0
 &          0
 &          1
 &          1
 &          0
 \\
 \mf\ 
 &   \dm = 20
 &   --  
 &          1
 &   \dm = 20
 &   --  
 &          2
 &          1
 \\
 $m-35$ 
 &   \dm = 20  
 &          1
 &          2
 &   \dm = 20
 &          0
 &          3
 &          2
 \\
\hline\hline
\multicolumn{ 8}{||l||}{number of events expected from \smp} \\
\hline
 1.5  
 &        0.1 $\pm$       0.0
 &        0.0 $\pm$       0.0
 &        0.0 $\pm$       0.0
 &        0.0 $\pm$       0.0
 &        0.0 $\pm$       0.0
 &        0.0 $\pm$       0.0
 &        0.0 $\pm$       0.0
 \\
 2.5  
 &        0.3 $\pm$       0.1
 &        0.1 $\pm$       0.0
 &        0.0 $\pm$       0.0
 &        0.3 $\pm$       0.2
 &        0.1 $\pm$       0.0
 &        0.0 $\pm$       0.0
 &        0.0 $\pm$       0.0
 \\
 5    
 &        0.5 $\pm$       0.1
 &        0.6 $\pm$       0.2
 &        0.3 $\pm$       0.1
 &        0.6 $\pm$       0.2
 &        0.4 $\pm$       0.2
 &        0.2 $\pm$       0.1
 &        0.2 $\pm$       0.1
 \\
 10   
 &        0.6 $\pm$       0.2
 &        0.3 $\pm$       0.1
 &        0.4 $\pm$       0.2
 &        0.9 $\pm$       0.2
 &        0.6 $\pm$       0.2
 &        0.4 $\pm$       0.2
 &        0.3 $\pm$       0.1
 \\
 20   
 &        1.5 $\pm$       0.1
 &        0.7 $\pm$       0.1
 &        0.2 $\pm$       0.0
 &        1.7 $\pm$       0.2
 &        1.1 $\pm$       0.2
 &        0.5 $\pm$       0.1
 &        0.3 $\pm$       0.1
 \\
 \mf\ 
 &   \dm = 20
 &   --  
 &        0.6 $\pm$       0.1
 &   \dm = 20
 &   --  
 &        1.7 $\pm$       0.2
 &        1.2 $\pm$       0.2
 \\
 $m-35$ 
 &   \dm = 20
 &        1.8 $\pm$       0.1
 &        2.0 $\pm$       0.1
 &   \dm = 20
 &        2.1 $\pm$       0.2
 &        3.7 $\pm$       0.2
 &        3.3 $\pm$       0.2
 \\

\hline\hline
\end{tabular}
\caption[]{\sl
  \protect{\parbox[t]{15cm}{
Selection efficiency, number of selected events and 
number of events expected from \smp\ in the search for \chargtwoo\
at \ssix\ and \sseven\ for different values of \mch\ and \dm .
The entries indicated by ``--'' correspond to a combination of \mch\
and \dm\ for which no signal \mc\ samples were generated.
Efficiencies have been calculated for the 
case where the three sneutrino 
generations are mass degenerate.
The errors are statistical only.
}} }
\label{tab-eff8}
\end{table}

\begin{table}[htb]
\centering
\begin{tabular}{||l||c|c|c||c|c|c|c||}
\hline\hline
       & \multicolumn{3}{c||}{\ssix\ }  
       & \multicolumn{4}{c||}{\sseven\ } \\
\cline{2-8}
  \dm\ & \multicolumn{3}{c||}{\mch\ (GeV)}
       & \multicolumn{4}{c||}{\mch\ (GeV)} \\
\cline{2-8}
 (GeV) & 50 & 65 & 75  & 50 & 65 & 75 & 85 \\
\hline\hline
  \multicolumn{8}{||l||}{selection efficiency (\%)} \\
\hline
 3    
 &   4 $\pm$ 1
 &   3 $\pm$ 1
 &   1 $\pm$ 0
 &   7 $\pm$ 1
 &   4 $\pm$ 1
 &   3 $\pm$ 1
 &   1 $\pm$ 0
 \\
 5    
 &  16 $\pm$ 1
 &  17 $\pm$ 1
 &  16 $\pm$ 1
 &  20 $\pm$ 1
 &  18 $\pm$ 1
 &  18 $\pm$ 1
 &  14 $\pm$ 1
 \\
 10   
 &  38 $\pm$ 2
 &  40 $\pm$ 2
 &  43 $\pm$ 2
 &  36 $\pm$ 2
 &  39 $\pm$ 2
 &  41 $\pm$ 2
 &  42 $\pm$ 2
 \\
 20   
 &  53 $\pm$ 2
 &  52 $\pm$ 2
 &  52 $\pm$ 2
 &  45 $\pm$ 2
 &  48 $\pm$ 2
 &  51 $\pm$ 2
 &  56 $\pm$ 2
 \\
 $m/2$  
 &  54 $\pm$ 2
 &  61 $\pm$ 2
 &  59 $\pm$ 2
 &  41 $\pm$ 2
 &  53 $\pm$ 2
 &  53 $\pm$ 2
 &  62 $\pm$ 2
 \\
 $m-20$ 
 &  56 $\pm$ 2
 &  64 $\pm$ 2
 &  65 $\pm$ 2
 &  44 $\pm$ 2
 &  50 $\pm$ 2
 &  49 $\pm$ 2
 &  66 $\pm$ 1
 \\
 $m-10$ 
 &  59 $\pm$ 2
 &  61 $\pm$ 2
 &  64 $\pm$ 2
 &  43 $\pm$ 2
 &  45 $\pm$ 2
 &  49 $\pm$ 2
 &  70 $\pm$ 1
 \\
 $m$    
 &  56 $\pm$ 2
 &  60 $\pm$ 2
 &  68 $\pm$ 1
 &  39 $\pm$ 2
 &  42 $\pm$ 2
 &  52 $\pm$ 2
 &  74 $\pm$ 1
 \\

\hline\hline
  \multicolumn{8}{||l||}{number of selected events} \\
\hline
 3    
 &          1
 &          0
 &          0
 &          1
 &          1
 &          1
 &          1
 \\
 5    
 &          1
 &          1
 &          1
 &          1
 &          1
 &          1
 &          1
 \\
 10   
 &          1
 &          1
 &          1
 &          1
 &          1
 &          1
 &          1
 \\
 20   
 &          1
 &          0
 &          0
 &          1
 &          0
 &          0
 &          0
 \\
 $m/2$  
 &          1
 &          1
 &          1
 &          1
 &          1
 &          1
 &          1
 \\
 $m-20$ 
 &          1
 &          1
 &          1
 &          1
 &          1
 &          1
 &          2
 \\
 $m-10$ 
 &          2
 &          2
 &          1
 &          1
 &          1
 &          1
 &          4
 \\
 $m$    
 &          2
 &          2
 &          2
 &          1
 &          1
 &          3
 &          5
 \\
\hline\hline
\multicolumn{ 8}{||l||}{number of events expected from \smp} \\
\hline
 3    
 &        0.1 $\pm$       0.0
 &        0.0 $\pm$       0.0
 &        0.0 $\pm$       0.0
 &        0.2 $\pm$       0.1
 &        0.0 $\pm$       0.0
 &        0.0 $\pm$       0.0
 &        0.0 $\pm$       0.0
 \\
 5    
 &        0.5 $\pm$       0.2
 &        0.3 $\pm$       0.1
 &        0.2 $\pm$       0.1
 &        0.4 $\pm$       0.2
 &        0.2 $\pm$       0.1
 &        0.1 $\pm$       0.0
 &        0.1 $\pm$       0.0
 \\
 10   
 &        0.9 $\pm$       0.2
 &        0.7 $\pm$       0.2
 &        0.6 $\pm$       0.2
 &        0.7 $\pm$       0.2
 &        0.6 $\pm$       0.2
 &        0.6 $\pm$       0.2
 &        0.4 $\pm$       0.2
 \\
 20   
 &        1.3 $\pm$       0.2
 &        1.0 $\pm$       0.2
 &        0.6 $\pm$       0.2
 &        1.2 $\pm$       0.2
 &        0.9 $\pm$       0.2
 &        0.8 $\pm$       0.2
 &        0.7 $\pm$       0.2
 \\
 $m/2$  
 &        1.5 $\pm$       0.2
 &        1.0 $\pm$       0.2
 &        0.7 $\pm$       0.2
 &        1.3 $\pm$       0.2
 &        1.2 $\pm$       0.2
 &        1.1 $\pm$       0.2
 &        1.1 $\pm$       0.2
 \\
 $m-20$ 
 &        1.8 $\pm$       0.2
 &        1.6 $\pm$       0.2
 &        1.3 $\pm$       0.2
 &        1.4 $\pm$       0.2
 &        1.5 $\pm$       0.2
 &        1.5 $\pm$       0.2
 &        2.3 $\pm$       0.2
 \\
 $m-10$ 
 &        1.9 $\pm$       0.2
 &        1.8 $\pm$       0.2
 &        1.7 $\pm$       0.2
 &        1.7 $\pm$       0.2
 &        1.7 $\pm$       0.2
 &        2.1 $\pm$       0.2
 &        3.9 $\pm$       0.2
 \\
 $m$    
 &        1.9 $\pm$       0.2
 &        1.9 $\pm$       0.2
 &        1.9 $\pm$       0.2
 &        1.9 $\pm$       0.2
 &        2.2 $\pm$       0.2
 &        3.5 $\pm$       0.2
 &        6.1 $\pm$       0.2
 \\

\hline\hline
\end{tabular}
\caption[]{\sl
  \protect{\parbox[t]{15cm}{
Selection efficiency, number of selected events and 
number of events expected from \smp\ in the search for \chargthreee\
at \ssix\ and \sseven\ for different values of \mch\ and \dm .
The errors are statistical only.
}} }
\label{tab-eff4}
\end{table}

\clearpage

\section{New Particle Search Results}
\label{results}

The number of observed candidate events 
and their kinematic properties are
compatible with the 
expected background from Standard Model processes. 
We present limits on the pair production
of charged scalar leptons, leptonically decaying charged Higgs 
and charginos that decay to produce a charged lepton and invisible particles.
The limits are computed combining  
data collected at $\sqrt{s} = 161$ and $172$~GeV, described in
this paper, together 
with data previously collected at 
$\sqrt{s} = 130-136$~GeV~\cite{ref:paperppe}. 

As described in section~\ref{sec:addkin}, 
the additional event selection cuts 
for a given search channel vary as a function of  $m$ and \dm .
The number of selected candidate events ($N$) is calculated at each
kinematically allowed point on a 0.2~GeV by 0.2~GeV grid of $m$ and \dm .
The data at the different centre-of-mass energies are combined by
simply adding the number of observed candidates.
The number of expected \sm\ events ($\mu _B$) 
is calculated in a similar fashion.

The 95\% CL upper
limit on  new particle production  at \sseven ,
obtained by combining the data at the three centre-of-mass 
energies \sthree , \ssix\ and \sseven\ is given by:
$$\signine\cdot B^2 = \frac{\Nninef}
{\sum_i \varepsilon_i {\cal L}_i \omega_i},$$
where the sum runs over the three centre-of-mass energies.
$B$ is the branching ratio for the decay mode studied
and the other terms are as defined\footnote{
In the background subtraction we conservatively do not take into
account the expected background at $\sqrt{s} = 130-136$~GeV.}
in section~\ref{sec:opt}.

Monte Carlo signal events are available only at certain particular
 values of $m$ and \dm .
The values of $m$ range typically from $m = 45$~GeV up to $m \approx \Ebeam$ in
5~GeV steps.
The values of \dm\ correspond to those
given in tables~\ref{tab-eff1}--\ref{tab-eff4}.
Signal efficiencies at intermediate values of $m$ and \dm\ are
obtained by a linear 2-dimensional interpolation from the values in the tables.
In addition to the \mc\ statistical error, we assign a 5\% 
systematic error on the estimated selection efficiency to take into
account uncertainties in the: trigger efficiency, detector
occupancy, lepton identification efficiency, luminosity measurement, 
interpolation procedure, and 
deficiencies in the \mc\ generators and the detector simulation.

At high values of 
\dm\ the dominant background results from \wpair\ production, for
which high statistics \mc\ samples are available that describe well
the OPAL data~\cite{ref:wwpaper,Wmass}.
In addition to the \mc\ statistical error, we assign a 5\% 
systematic error on the estimated background to take into account the
uncertainty in the expected \wpair\ cross-section at \ssix\ (arising from the
uncertainty in the measured W mass) and deficiencies in the \mc\ detector
simulation.
At low values of \dm\ the dominant background results from \eell\
events.
Additional checks of the degree to which the \smc\ describes these events
are given at the end of appendix~\ref{sec-gen}.
The background uncertainty at low \dm\ is dominated by the limited \mc\
statistics. 
(As can be seen in tables~\ref{tab-eff1}--\ref{tab-eff4} the
uncertainty is typically 20--60\% at low \dm ). 
In setting limits the \mc\ statistical errors and other
systematics are taken into account according to the 
method described in~\cite{cousins}.

\begin{figure}[htbp]
 \epsfxsize=\textwidth 
 \epsffile[0 0 580 600]{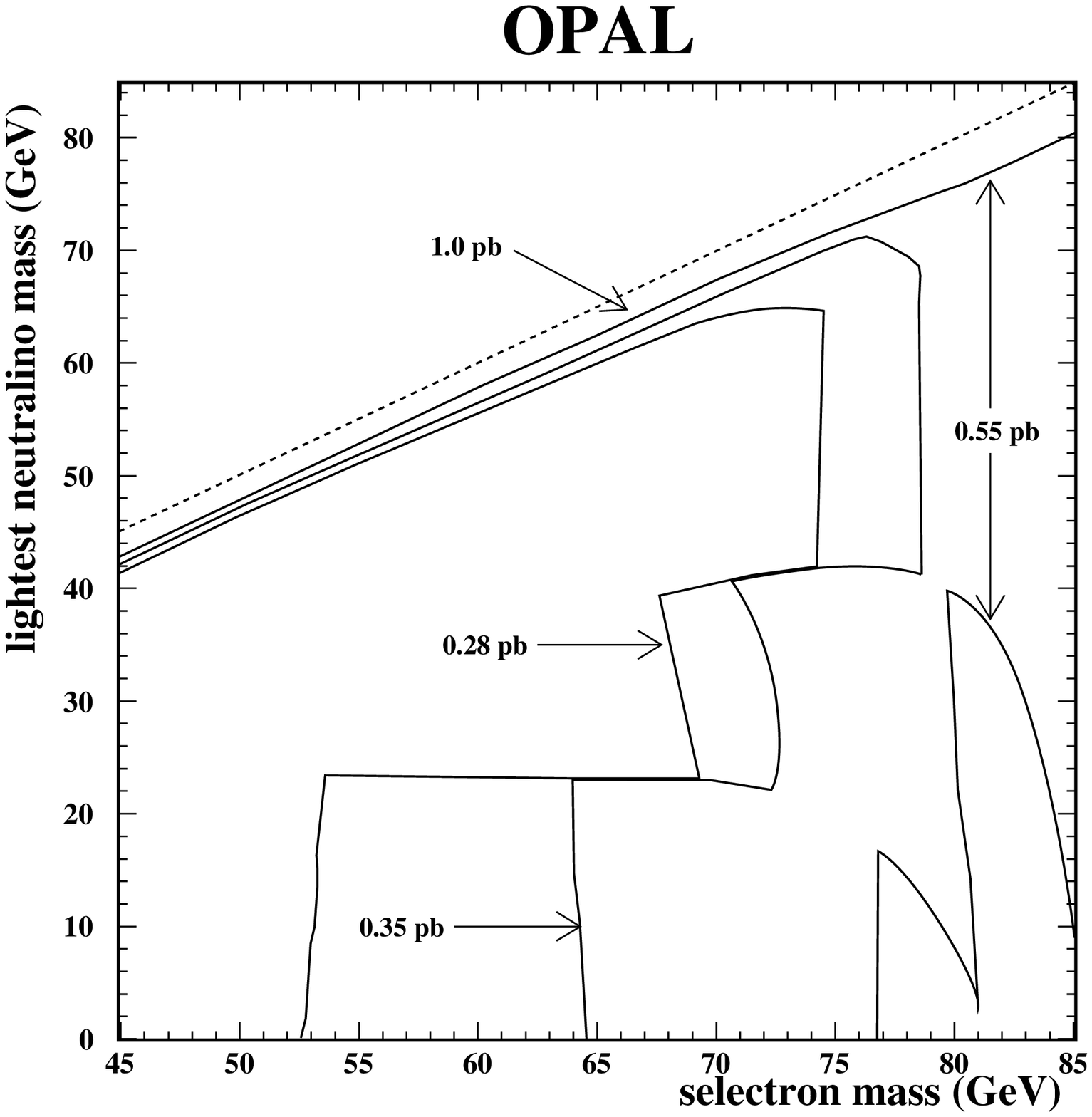}
 \caption{
Contours of the 95\% CL upper limits on the selectron pair
cross-section times branching ratio squared for 
$\sele \rightarrow \mathrm{e} \nt_1$
at $\protect\sqrt{s}$~=~172~GeV.
The limit is obtained by combining the 130--172~GeV data-sets 
assuming a $\beta^3/s$ dependence of the cross-section.
The kinematically allowed region is indicated by the dashed lines.
} 
\label{fig:limit_1}
\end{figure}

\begin{figure}[htbp]
 \epsfxsize=\textwidth 
 \epsffile[0 0 580 600]{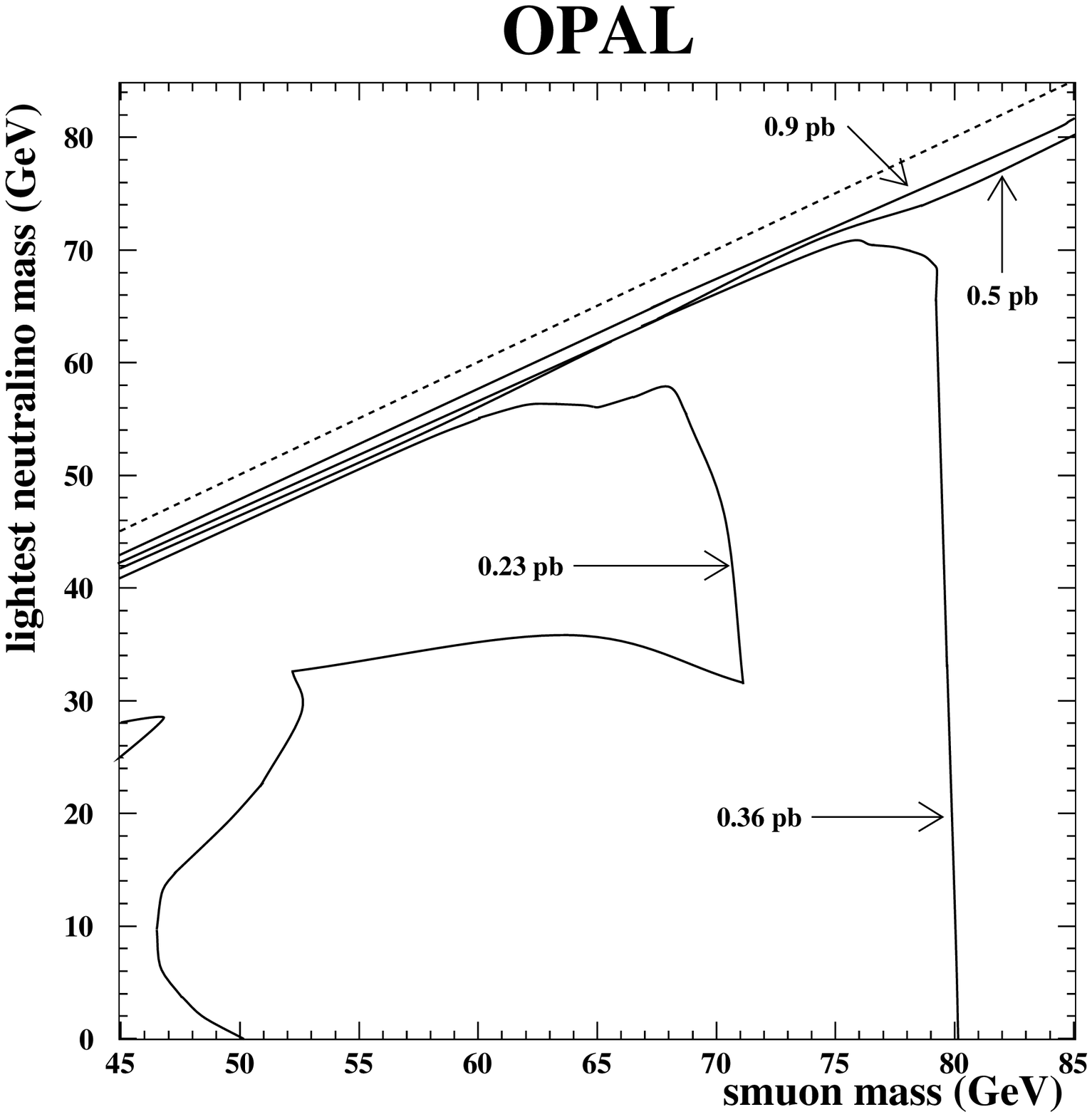}
 \caption{
Contours of the 95\% CL upper limits on the smuon pair
cross-section times branching ratio squared for 
$\smu \rightarrow \mu \nt_1$
at 172~GeV.
The limit is obtained by combining the 130--172~GeV data-sets 
assuming a $\beta^3/s$ dependence of the cross-section.
The kinematically allowed region is indicated by the dashed lines.
} 
\label{fig:limit_2}
\end{figure}

\begin{figure}[htbp]
 \epsfxsize=\textwidth 
 \epsffile[55 230 486 673]{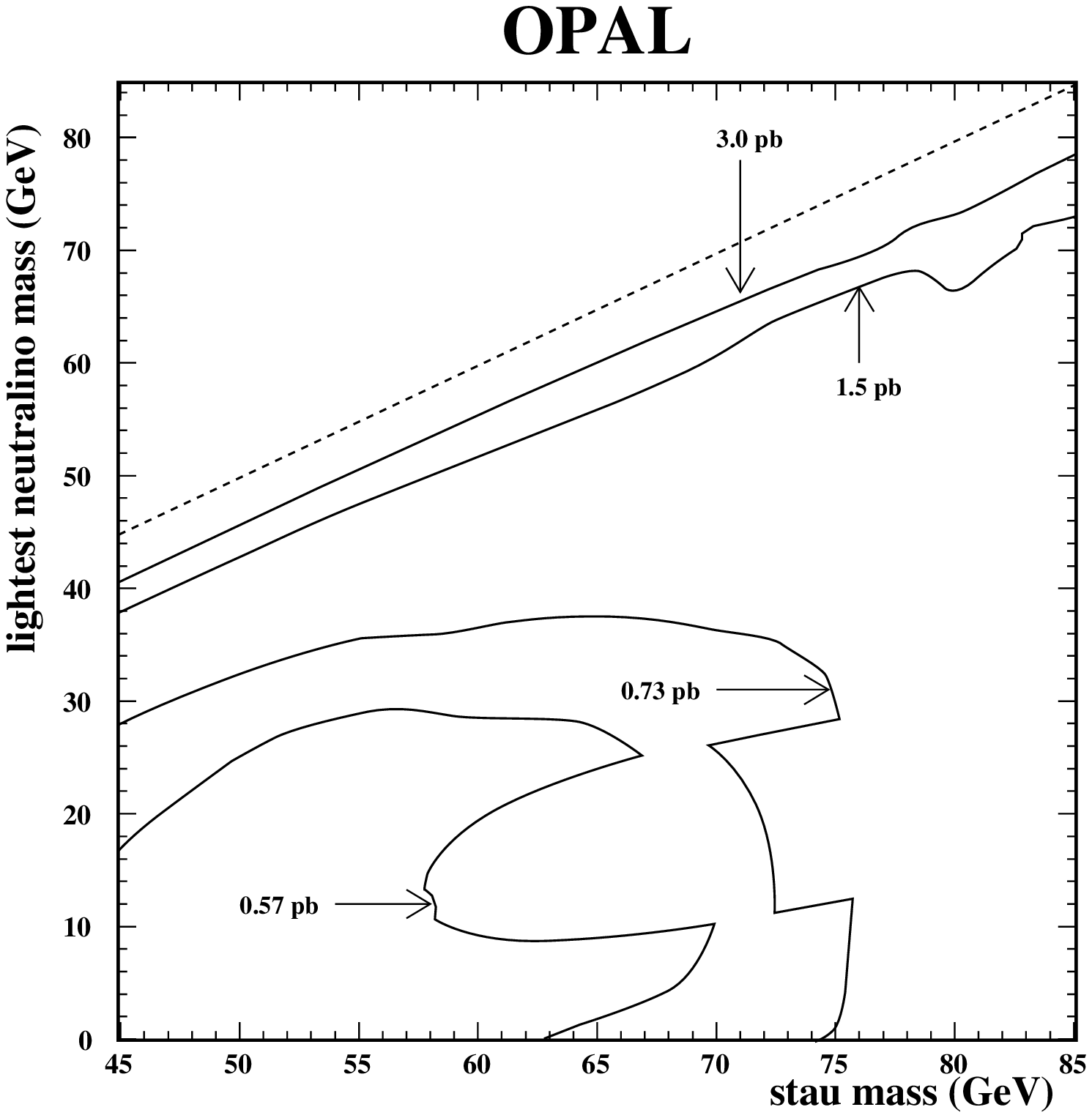}
 \caption{
Contours of the 95\% CL upper limits on the stau pair
cross-section times branching ratio squared for 
$\stau \rightarrow \tau \nt_1$
at $\protect\sqrt{s}$~=~172~GeV.
The limit is obtained by combining the 130--172~GeV data-sets 
assuming a $\beta^3/s$ dependence of the cross-section.
The kinematically allowed region is indicated by the dashed lines.
} 
\label{fig:limit_3}
\end{figure}

Limits on production cross-section times branching ratio squared for new
physics processes are now presented.
Upper limits at  95\% CL on the selectron pair cross-section 
at 172~GeV times 
branching ratio squared for the decay \dsele\
are shown in figure~\ref{fig:limit_1} as a function of selectron mass 
and lightest neutralino mass.
These limits are valid for
$\tilde{\mathrm e}^+_{\mathrm L}\tilde{\mathrm e}^-_{\mathrm L}$ and 
$\tilde{\mathrm e}^+_{\mathrm R}\tilde{\mathrm e}^-_{\mathrm R}$ production.
The corresponding plots for  the smuon and stau pair searches are shown in  
figures~\ref{fig:limit_2} and~\ref{fig:limit_3}, respectively.
The data at the different centre-of-mass energies have been combined 
by weighting the integrated luminosity according to
$\beta^3/s$, which corresponds to the approximate dependence
on $\beta$ and $s$ of the expected production cross-section for scalar
particles.

\begin{figure}[htbp]
 \epsfxsize=\textwidth 
 \epsffile[0 0 580 600]{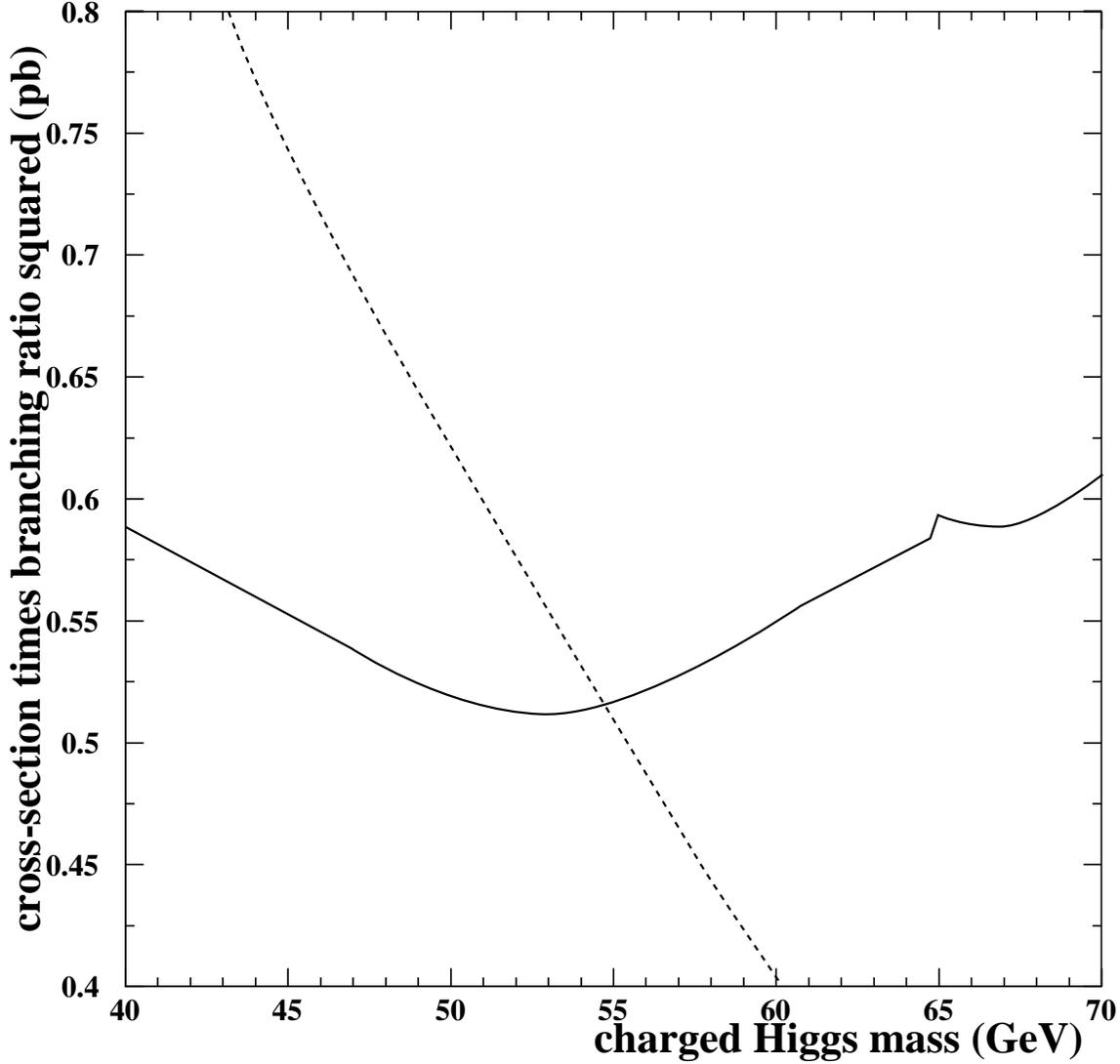}
 \caption{
The solid line shows the 95\% CL upper limit on the 
charged Higgs pair production 
cross-section times branching ratio squared for the decay \dH\
at $\protect\sqrt{s}$~=~172~GeV.
The limit is obtained by combining the 130--172~GeV data-sets 
assuming the \mH\ and $\protect\sqrt{s}$ 
dependence of the cross-section predicted
by {\sc Pythia}.
For comparison, the dashed curve shows the prediction from {\sc Pythia}
 at $\protect\sqrt{s}$~=~172~GeV
assuming a 100\% branching ratio for the decay \dH .
} 
\label{fig:limit_5}
\end{figure}

The upper limit at 95\% CL on the charged Higgs pair production 
cross-section times branching ratio squared for the decay \dH\
is shown as a function of \mH\ as the solid line in figure \ref{fig:limit_5}. 
The dashed  line in figure \ref{fig:limit_5} 
shows the prediction from {\sc Pythia}
 at $\protect\sqrt{s}$~=~172~GeV
for a 100\% branching ratio for the decay \dH .
With this assumption we set a lower limit  at 95\% CL on \mH\ of 54.8~GeV.
In a forthcoming paper~\cite{higgs} the search described here
for acoplanar di-tau events will be combined with searches in
the final states $\tau\nbar\qpair$ and \qpair\qpair\ to set limits on
charged Higgs pair production for arbitrary \dH\ branching ratio. 

\begin{figure}[htbp]
 \epsfxsize=\textwidth 
 \epsffile[0 0 580 600]{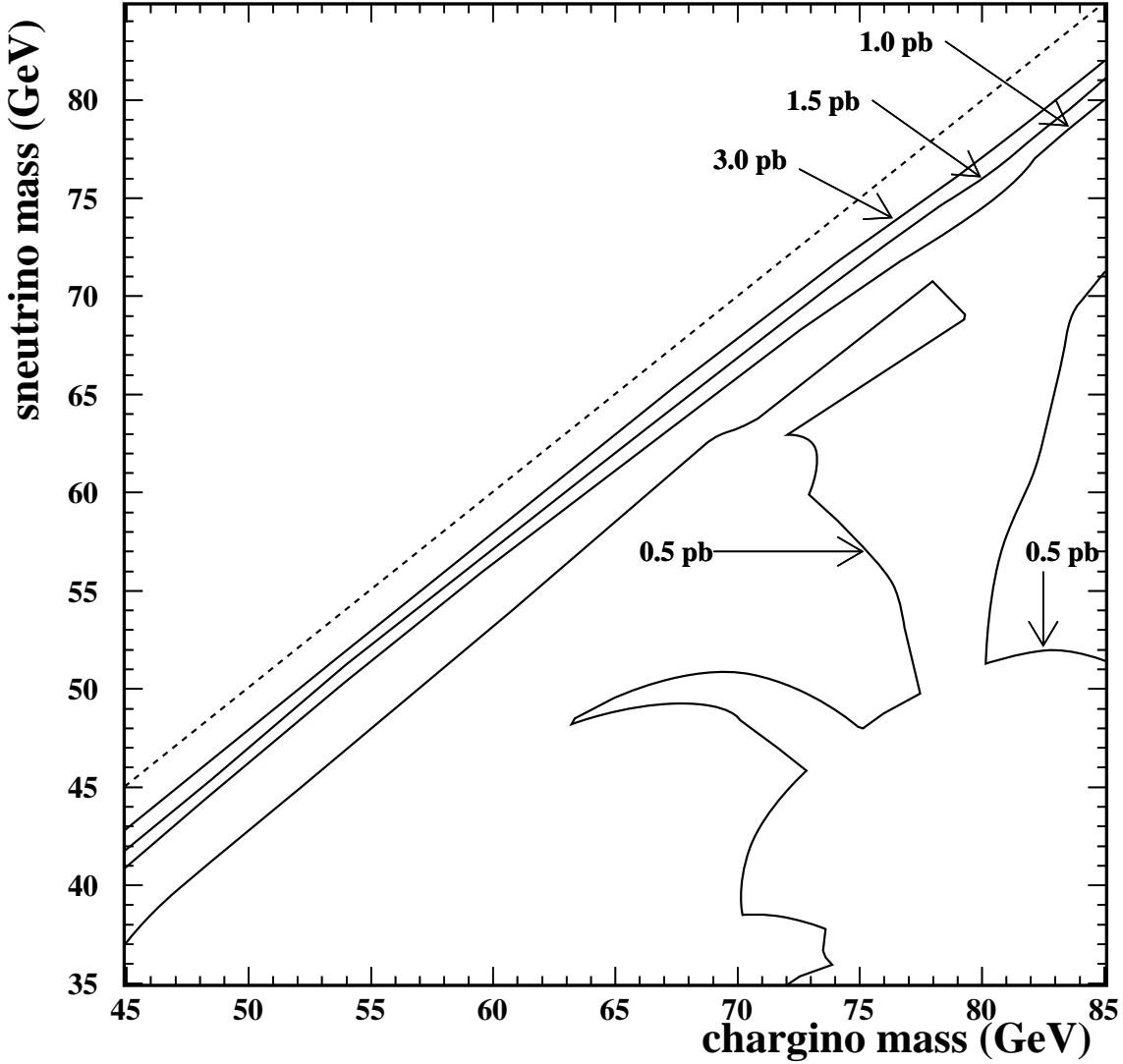}
 \caption{
Contours of the 95\% CL upper limits on the chargino pair
cross-section times branching ratio squared for 
$\chpm \rightarrow \ell^\pm \snu$ (2-body decay)
at $\protect\sqrt{s}$~=~172~GeV.
The  limits have been calculated for the 
case where the three sneutrino 
generations are mass degenerate.
The limit is obtained by combining the 161--172~GeV data-sets 
assuming a $\beta/s$ dependence of the cross-section.
This channel was not analysed in~\cite{ref:paperppe} 
at $\protect\sqrt{s}$~=~130--136~GeV.
The kinematically allowed region is indicated by the dashed lines.
} 
\label{fig:limit_8}
\end{figure}

The upper limits at 95\% CL on the chargino pair production 
cross-section times branching
ratio squared for the decay \dchtwo\  (2-body decay)
are shown in figure \ref{fig:limit_8}. 
The  limits have been calculated for the 
case where the three sneutrino 
generations are mass degenerate.
The data at the different centre-of-mass energies have been combined 
by weighting the integrated luminosity according to $\beta/s$, which 
corresponds to the approximate dependence
on $\beta$ and $s$ of the expected production cross-section for charginos.

\begin{figure}[htbp]
 \epsfxsize=\textwidth 
 \epsffile[55 230 486 673]{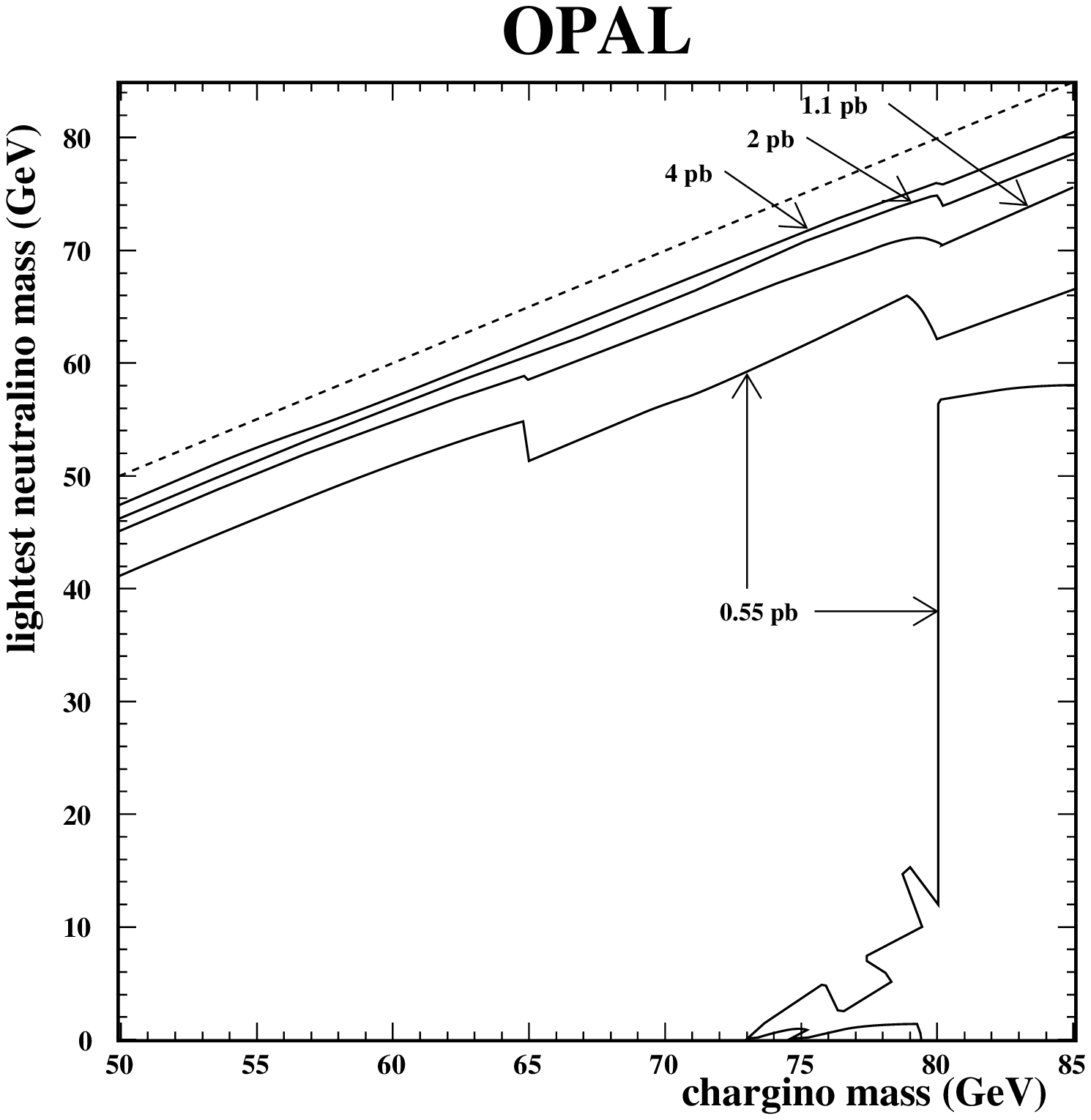}
 \caption{
Contours of the 95\% CL upper limits on the chargino pair
cross-section times branching ratio squared for 
$\chpm \rightarrow \ell^\pm \nu \chz$
 (3-body decay) at $\protect\sqrt{s}$~=~172~GeV,
The limit is obtained by combining the 130--172~GeV data-sets 
assuming a $\beta/s$ dependence of the cross-section.
The kinematically allowed region is indicated by the dashed lines.
} 
\label{fig:limit_4}
\end{figure}

The upper limits at 95\% CL on the chargino pair production 
cross-section times branching
ratio squared for the decay \dchthree\ (3-body decay)
are shown in figure \ref{fig:limit_4}. 
The data at the different centre-of-mass energies have been combined 
by weighting the integrated luminosity according to $\beta/s$.

\begin{figure}[htbp]
 \epsfxsize=\textwidth 
 \epsffile[0 0 580 600]{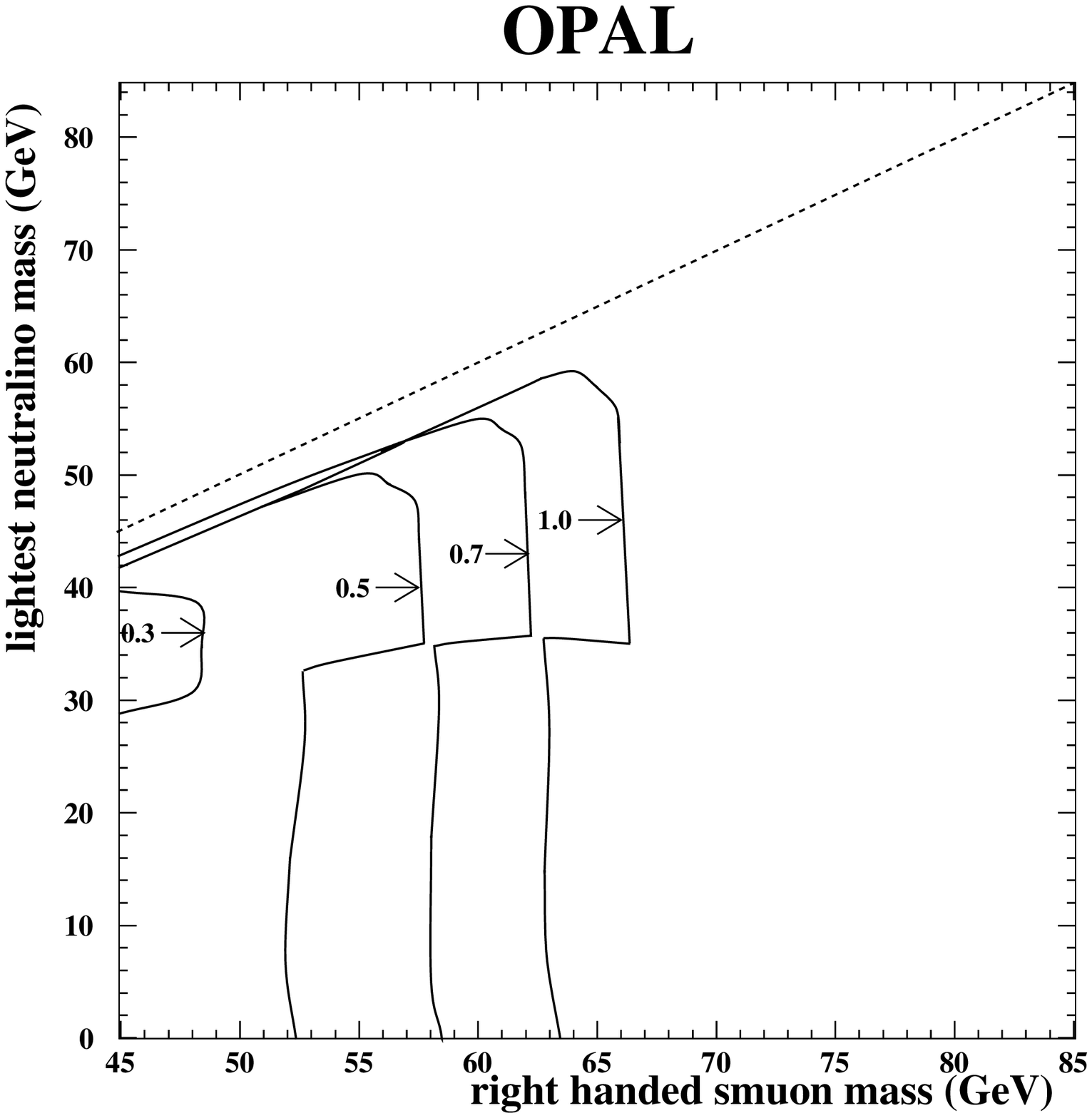}
 \caption{
95\% CL exclusion region for right-handed smuon pair 
production obtained by combining the 130--172~GeV data-sets.
The limits are calculated for several values of
the branching ratio squared for 
$\smu^\pm_R \rightarrow  {\mu^\pm} \nt_1$ that are indicated in the figure.
Otherwise they have no supersymmetry model assumptions.
The kinematically allowed region is indicated by the dashed line.
} 
\label{fig-mssm_2}
\end{figure}

\begin{figure}[htbp]
 \epsfxsize=\textwidth 
 \epsffile[0 0 580 600]{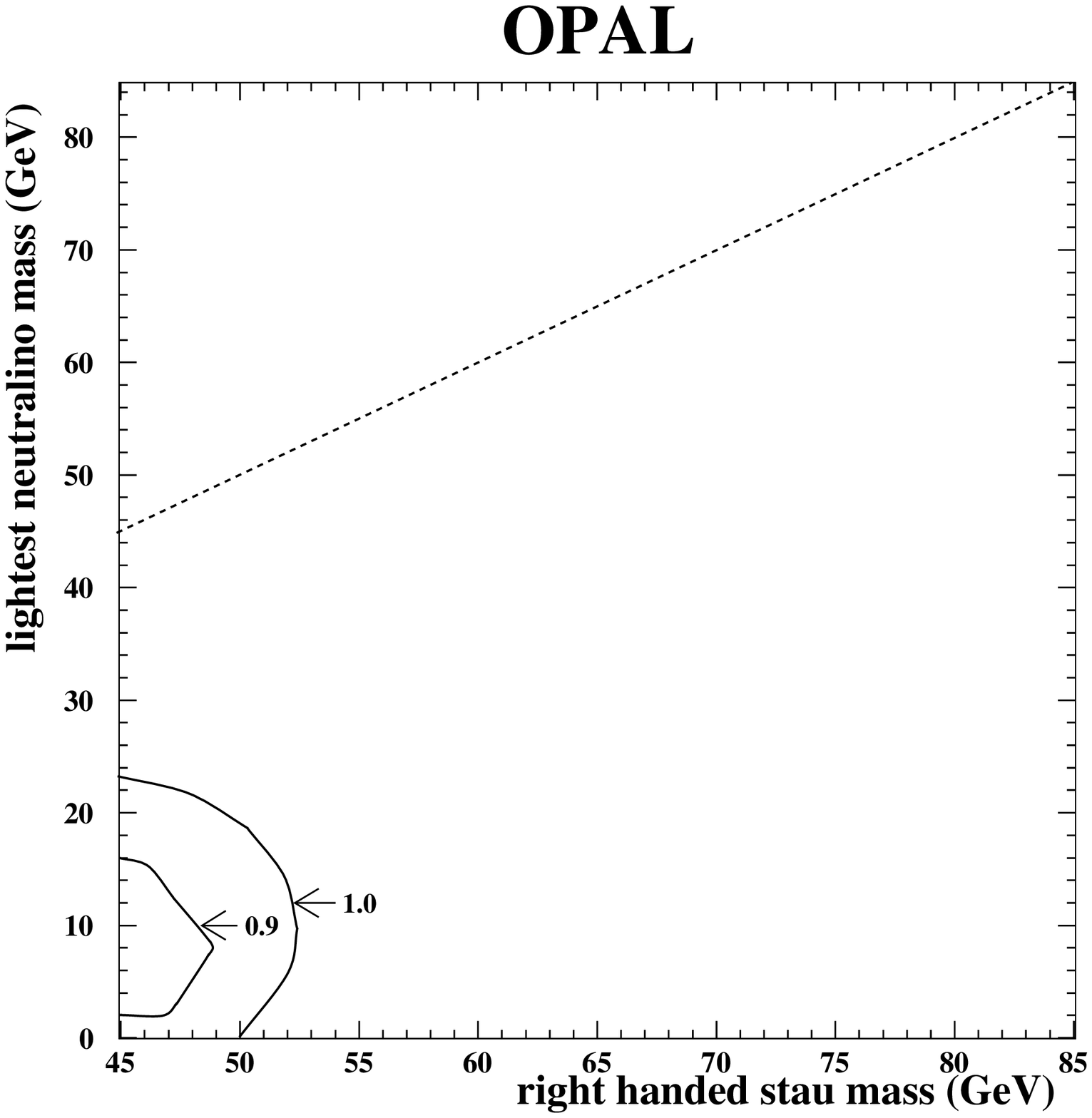}
 \caption{
95\% CL exclusion region for right-handed stau pair 
production obtained by combining the 130--172~GeV data-sets.
The limits are calculated for two values of
the branching ratio squared for 
$\stau^\pm_R \rightarrow  {\tau^\pm} \nt_1$.
Otherwise they have no supersymmetry model assumptions.
The kinematically allowed region is indicated by the dashed line.
} 
\label{fig-mssm_3}
\end{figure}

We can use our data to set limits on the masses of right-handed 
sleptons\footnote{
The right-handed slepton is expected to be lighter
than the left-handed slepton. The
right-handed one tends (not generally valid for selectrons)
to
have a lower pair production cross-section, and so
conventionally limits are given for this (usually) conservative case.}
based on the expected right-handed slepton pair cross-sections and
branching ratios.
The right-handed smuon pair production cross-section
can be calculated from the relevant photon- and Z-exchange
diagrams. The production cross-section depends simply on the
smuon mass. 
However, the branching ratio
is a function of the masses and couplings of the particles involved in 
open decay channels.
Of particular relevance here is the $\nt_2$, which may decay to
$\nt_1\gamma^{(*)}$ or $\nt_1$Z$^*$ and whose mass depends 
on the MSSM parameters $M_1$, $M_2$, $\mu$ and $\tan{\beta}$.
In particular 
 regions of parameter space, the square of the branching ratio for 
$\smu^\pm_R \rightarrow  {\mu^\pm} \nt_1$  
can be essentially zero\footnote{
For example, for $M_2=40$~GeV, $\mu=-25$~GeV, $\tan{\beta}=2$,
the branching ratio squared is calculated to be 0.02 for the 
case of ($m_{\smu_{\mathrm{R}}}$, $m_{\nt_1}$)$=$(70,20) GeV. 
All the quantitative predictions within the MSSM 
are obtained using {\sc Susygen}
and are calculated with the gauge unification relation,
$M_1 =  \frac{5}{3} \tan^2 \theta_W M_2$.}, and so 
 it is not possible to provide general limits within the MSSM
on smuon production 
 on the basis of this search alone.
In figure~\ref{fig-mssm_2} we show  limits on smuon pair
production as a function of smuon mass and lightest
neutralino mass for several assumed values of
the branching ratio squared for $\smu^\pm_R \rightarrow  {\mu^\pm} \nt_1$.
The limits depend on the neutralino mass 
as the efficiency, number of candidates and expected number of 
background events vary with  neutralino mass as well as the smuon mass.
For a branching ratio $\smu^\pm_R \rightarrow  {\mu^\pm} \nt_1$ of
100\% and for a smuon-neutralino mass difference exceeding 4~GeV,
right-handed smuon pair production is excluded at 95\% CL for 
smuon masses below 62.7~GeV.
The  95\% CL upper limit on the  production of
right-handed \staupair\ times
 branching ratio squared for $\stau^\pm_R \rightarrow  {\tau^\pm} \nt_1$
is shown in figure~\ref{fig-mssm_3}.
The present data-set slightly extends the limit established at LEP1.

\begin{figure}[htbp]
 \epsfxsize=\textwidth 
 \epsffile[0 0 580 600]{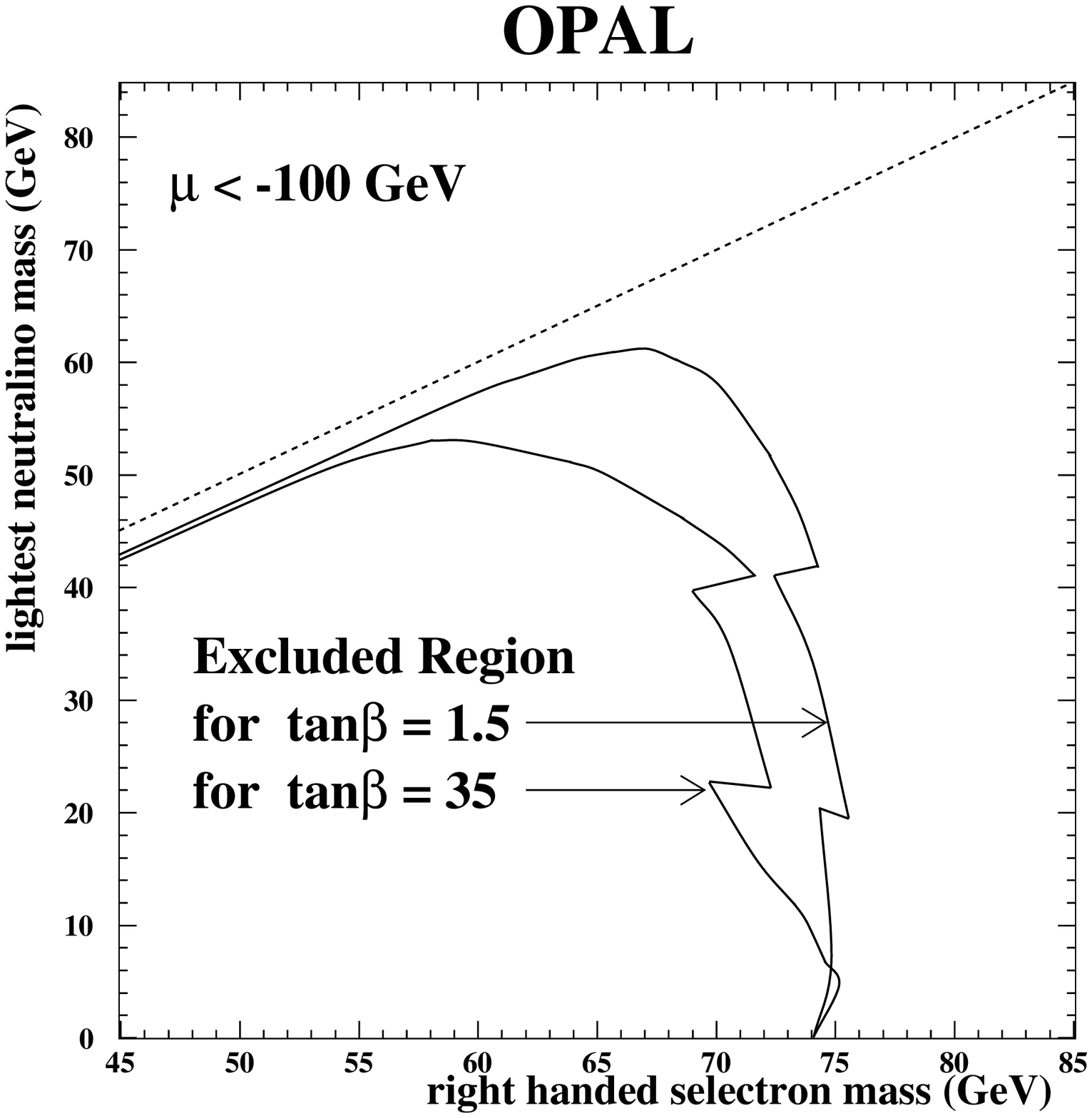}
 \caption{
For two values of $\tan{\beta}$ and $\mu < -100$~GeV,
95\% CL exclusion regions for right-handed selectron pairs 
obtained by combining the 130--172~GeV data-sets.
The excluded regions are calculated 
taking into account the 
predicted branching ratio for 
$\sele^\pm_R \rightarrow  {\mathrm{e}^\pm} \nt_1$.
The kinematically allowed region is indicated by the dashed line.
} 
\label{fig-mssm_1}
\end{figure}

An alternative approach is to set limits
taking into account the 
predicted branching ratio for $\smu^\pm_R \rightarrow  {\mu^\pm} \nt_1$
for specific choices of the MSSM parameters.
Figure~\ref{fig-mssm_1} shows 95\% CL exclusion regions 
for right-handed smuon pairs  
in the ($m_{\smu_{\mathrm{R}}}$, $m_{\nt_1}$) 
plane, for $\mu < -100$~GeV and for two
values of $\tan{\beta}$ (1.5 and 35).
The data at the different centre-of-mass energies 130--172~GeV 
have been combined 
by weighting the integrated luminosity according to
the MSSM-predicted  cross-section times branching ratio squared.
For $\mu < -100$~GeV and $\tan{\beta}=1.5$ and for 
smuon-neutralino mass differences exceeding 4~GeV, smuon masses  
below 55.6~GeV are excluded at 95\% CL.

The right-handed selectron pair production cross-section 
can be enhanced significantly by the t-channel neutralino exchange 
diagram when the neutralino mass is small and its coupling to electron and
right-handed selectron is high (gaugino-like). However due to possible
interference,
the cross-section may even be smaller than for smuon pair production. 
The presented limits are therefore quite model dependent. 
We have evaluated the expected cross-section times branching ratio
squared 
for $\mu < -100$~GeV  and for two
values of $\tan{\beta}$ (1.5 and 35)
in the ($m_{\sele_{\mathrm{R}}}$, $m_{\nt_1}$)
plane taking into 
account the production cross-section and the calculated branching ratio 
for $\sele^\pm_R \rightarrow  {\mathrm{e}^\pm} \nt_1$.
The 95\% CL exclusion region is shown in figure~\ref{fig-mssm_1}. 
For $\mu < -100$~GeV and $\tan{\beta}=1.5$ and for 
selectron-neutralino mass differences exceeding 5~GeV, selectron masses  
below 66.5~GeV are excluded at 95\% CL.

\section{Summary and Conclusions}

A selection of di-lepton events with significant missing transverse momentum 
has been performed using a total data sample of 20.6~pb$^{-1}$
at centre-of-mass energies of 161~GeV and 172~GeV.
Thirteen events are observed, which
is consistent with the $16.8 \pm 0.5$ events 
expected from Standard Model processes.

Further event selection criteria, in the form of kinematic cuts and lepton 
identification requirements, have been implemented in order to search
for scalar charged lepton pair, charged Higgs and chargino pair production.
The sensitivity to new physics has been maximised by using an 
algorithm to optimise the kinematic cut values
as functions of the masses of the
pair produced new particle and the neutral particle to which it is
assumed to decay. 

No evidence for new phenomena is apparent and limits 
on the pair production cross-section times branching ratio squared 
are presented for 
selectrons, smuons, staus, leptonically decaying charged Higgs
 and charginos that decay to produce a charged lepton and invisible particles.
In addition, 95\%\ CL exclusion regions 
in the ($\mathrm{m_{\sell_R}}$, $\mathrm{m_{\nt_1}}$)
plane for  selectrons, smuons
and staus are presented.
The limits are computed combining  
data collected at $\sqrt{s} = 161$ and $172$~GeV, described in
this paper, together 
with data previously collected at 
$\sqrt{s} = 130-136$~GeV~\cite{ref:paperppe}. 

With respect to our previous
publications~\cite{ref:paperppe,ref:crap161}, the analysis presented
here has an improved sensitivity to new physics sources
of di-lepton events with significant missing transverse momentum.
The results given here supersede those given 
in~\cite{ref:paperppe,ref:crap161}.
Model dependent limits on charged slepton pair production at LEP2
energies have been presented by the ALEPH
collaboration~\cite{ref:otherlep}. 

\newpage
\noindent {\Large\bf Appendices}

\appendix
\renewcommand{\thesection}{\Roman{section}}
\section{Event Selection I}
\label{sec-selI}

\subsection{Overview}
\label{sec-introcuts}
 
In designing the first
selection particular emphasis has been placed on retaining efficiency 
for events with very low visible energy, but nevertheless significant 
missing transverse momentum.
The event selection requires evidence that a pair of leptons has been
produced in association with an invisible system that carries away
significant missing energy and momentum.
The remaining cuts reduce the probability that the signature of 
missing momentum is faked by \sm\ processes containing, for example,
secondary neutrinos from tau decays  or
particles that strike regions of the detector where they are
undetected or poorly measured.

At least one track in the central detector must satisfy requirements
on lepton identification, isolation and momentum in the plane perpendicular
to the beam axis (\ptpt ).
In order to maintain a high efficiency, especially in the region of
small \dm , very much looser requirements are made on the possible presence
of a second lepton in the event.
A significant missing momentum is required by applying cuts on the
quantities \stevt\ and \staxic\ and the angle to the beam direction of
the missing momentum vector (see section~\ref{sec-introcuts} for definitions).

The dominant background that survives these cuts arises from \tp\ \lp\
in which one of the electrons is scattered at a significant angle to
the beam direction.
Such events are suppressed by requiring no significant energy 
to be present in the
SW and FD detectors.
The inner edge of the SW calorimeter is at approximately 0.025~rad to
the beam direction.
A beam energy scattered electron or radiated photon can therefore
carry away a \ptevt\ of approximately 0.025\Ebeam\ and not be detected.
This sets the scale for the minimum \ptevt\ that must be required in
order to suppress the \tpb .
A larger \ptevt\ may occur without the production of a tag in SW 
if, for example, both beam electrons in a \tp\
event are scattered
at approximately the same azimuthal angle, $\phi$.
However, the probability for such an occurrence is rather small.
\mc\ simulations of the \smp\ are used to tune the event selection
cuts and to estimate the residual background.
Given the above discussion,
the cuts applied on \ptevt\ and \ptaxic\ are scaled with \Ebeam .
This allows the same cuts on these quantities
to be applied at \ssix\ and \sseven\ as employed
in the analysis of the data collected at \sthree~\cite{ref:paperppe}.

\subsection{Lepton Identification, Isolation, etc.}
\label{sec-detail}
Unless otherwise explicitly stated, tracks in the central
detector and clusters in the  electromagnetic calorimeter
are required to satisfy the normal quality criteria employed in the
analysis of \sm\ \lp~\cite{bib-LL}. 
These criteria are as follows.
Tracks must have: \ptpt~$>$~0.1~GeV,
\dz~$<$~1~cm, \zz~$<$~40~cm and a total of 
at least 20 measured points in the CV, CJ and CZ tracking chambers, 
the first of which is at a radius of less than 75~cm.
\dz\ is the point of closest approach of the track to the beam axis in the
transverse plane and \zz\ is the $z$ coordinate at this point.
In addition to the quality criteria of~\cite{bib-LL}, 
tracks are required either to be matched to an ECAL
cluster or to hits in CV, or to have at least 50 CJ hits. 

Barrel electromagnetic clusters are considered if they have a deposited energy
$E$~$>$~0.1~GeV.
Endcap electromagnetic clusters are considered if they have a deposited energy
$E$~$>$~0.2~GeV,
they contain at least two blocks, and the fraction of the total energy
of the cluster given by the most energetic block is less than 99\%.
Algorithms are adopted to
avoid double-counting ECAL energy deposits 
associated with charged particles.

Clusters in the FD calorimeter are considered if their energy is
at least 1~GeV.
Clusters in the GC are considered if their energy is
at least 5~GeV.
Clusters in the SW calorimeter are considered if their energy is
at least 1~GeV or if they are consistent with the passage of a single
minimum ionizing particle through the calorimeter.
An algorithm to detect short-lived noisy regions in the calorimeters
is used to suppress noise
clusters in EB, EE, SW, FD, GC and HCAL. 

In order to suppress `junk' events, such as those originating from
beam-gas or beam-wall collisions, it is required that
at least one track  in the event is matched to 6 or more
hits in the   axial sectors of the vertex drift chamber or to an
ECAL cluster 
with energy of at least 0.1~GeV.
Events arising from the passage of cosmic ray muons through the
detector  are rejected using an
algorithm similar to that employed in the
analysis of \sm\ muon pairs~\cite{bib-LL}.

A track is identified as a candidate lepton if it has $p$~$>$~1.5~GeV and
satisfies:
\begin{description}
\item[electron] any one of the following three criteria:
\begin{enumerate}
\item
The output of the neural network described in~\cite{bib-NN} is greater
than 0.8.
\item
0.8~$<$~$E/p$~$<$~1.3, where $p$ is the momentum of the track and $E$ is the
energy of the associated electromagnetic cluster.
\item
0.5~$<$~$E/p$~$<$~2.0 and \wdedx\ is not in the range 
$-$0.04~$<$~\wdedx~$<$~0.0,
where \wdedx\ is the \dedx\ weight~\cite{bib-dedx} for the track to be
an electron.
\end{enumerate}
\item[muon] either of the following two criteria:
\begin{enumerate}
\item
The track is identified as a muon according to the criteria employed in the
analysis of \sm\ muon pairs~\cite{bib-LL}.
That is, it
has associated activity in the muon chambers or hadron calorimeter strips
or it has a high momentum but is associated with only a small energy deposit
in the electromagnetic calorimeter.
\item
The track is identified as a muon according to the criteria employed in the
analysis of inclusive muons in multihadronic events given
in~\cite{bib-MU}, with no cut on the kaon \dedx\ weight.
\end{enumerate}
\end{description}
Leptonic tau decays are usually identified as \epm\  or \mupm .
Hadronic tau decays are identified as follows:
\begin{description}
\item[hadronic tau] both of the following criteria:
\begin{enumerate}
\item
Within a cone of half opening angle 35$^\circ$ there are no more than three
tracks in total.
\item
The invariant mass of all tracks and clusters within the cone is 
less than the tau mass (assuming the pion mass for each track).
\end{enumerate}
\end{description}

The isolation of electron and muon candidates is defined by considering
charged particles and e.m.~calorimeter  clusters 
within a cone of half opening angle 20$^\circ$ around the lepton direction.
The isolation of hadronic tau candidates is defined by considering
charged particles and e.m.~calorimeter  clusters 
within a cone of half opening
angle 60$^\circ$, but outside the cone of half opening angle 
35$^\circ$, described above.
In order for a lepton candidate to be considered isolated
both of the following criteria must be satisfied: 
\begin{enumerate}
\item
There are no more than two additional charged particles within the
isolation cone and the sum of their momenta is less than 2~GeV.
\item
There are no more than two clusters in the ECAL within  the
isolation cone and the sum of
their energies is less than 2~GeV.
(In order to minimise the sensitivity to final-state photon
bremsstrahlung,  if there is only one photon within the isolation cone
of an electron or muon candidate it is classified as isolated
irrespective of the energy of the photon.)
\end{enumerate}
If a track has been identified as an electron or muon candidate, but
fails the relevant isolation cuts then it is considered as a tau
candidate if it satisfies the tau identification and isolation 
requirements given above.
This is in order to retain efficiency for hadrons from tau decay that
are misidentified as  electron or muon candidates.

Converting photons are identified using an
algorithm similar to that employed in the
analysis of \sm\ muon pairs~\cite{bib-LL}.
The tracks and clusters associated to the conversion are
replaced by a single 4-vector representing the photon. 
Isolated photons are defined by requiring that there be no
charged tracks within a cone of half angle 20$^\circ$ around an
electromagnetic cluster or converting photon.

The 4-momenta of any tracks and clusters within the isolation cone
are added to the 4-momentum of the lepton candidate.

\subsection{Event Selection Cuts}
\label{sec-gen}

The event selection cuts fall logically into three groups.
The first group of cuts requires evidence that a pair of leptons has been
produced, at least one of which must satisfy requirements
on lepton identification, isolation and \ptpt :

\begin{enumerate}
\item
The event must contain at least one isolated lepton
candidate with \ptpt~$>$~1.5~GeV.
\item
If the event contains a second isolated lepton candidate then there
must be no charged tracks other than those associated to the two
lepton candidates.

If any additional photons not associated to either of the leptons 
are present then the measured
acoplanarity and acollinearity\footnote{
The acollinearity angle (\acol ) is defined as 
180$^{\circ}$ minus 
the
three dimensional angle
between the two lepton candidates.} 
 are corrected by adding the 4-momentum
of the photon to the lepton to which it is nearest in $\phi$.
The values \stevt\ and \staxic\ are also corrected for the presence
of additional clusters.
\item
If the event contains only one isolated lepton candidate and this
lepton is identified as an electron or muon,
the tracks and clusters not associated to the lepton are
considered as a possible second lepton candidate if they satisfy the following
requirements: 
\begin{enumerate}
\item
There is at least one additional charged track.
\item
At least one   track has $\ptpt > 0.3$~GeV.
\item
The total number of additional tracks and clusters does not exceed 4. 
\item
The invariant mass of the additional tracks and clusters must not
exceed the tau mass.
The highest energy isolated photon in the event is excluded 
from this mass calculation;
this is in order to retain efficiency for radiative events in which
one of the leptons fails to be identified and
reduces the sensitivity of the calculated efficiency to possible
deficiencies in the simulation of photon bremsstrahlung. 
\end{enumerate}
If the event contains only one isolated lepton candidate 
and these requirements on the additional tracks and clusters are not 
satisfied then the event is rejected.
\begin{enumerate}
\setcounter{enumii}{4}
\item
If the event contains only one isolated lepton candidate and this
lepton is identified as a tau then the event is rejected.
\end{enumerate}
From now on the phrase ``lepton candidate'' will normally refer to
either an isolated electron, muon or tau candidate, or the rest of the
event if it satisfies the criteria 3~(a)--(d).
The majority of lepton candidates identified as ``tau'' or ``rest of
event'' arise from hadronic tau decays and are referred to by the
symbol ``$h$''.
\item
Both lepton candidates must satisfy $\mcost < 0.95$.
\end{enumerate}

The second group of cuts requires evidence for the production of
an invisible system that carries away
significant missing energy and momentum.
\begin{enumerate}
\setcounter{enumi}{4}
\item
Different cuts on the missing momentum and its 
direction are applied in the regions of small and large acoplanarity:
\begin{description}
\item[small acoplanarity] \acopc~$<$~1.2~rad:
\begin{enumerate}
\item
\stevt~$>$~0.035.
\item
\staxic~$>$~0.025.
\item
\axicos~$<$~0.99, where the direction of the missing momentum vector is
calculated using the missing momentum perpendicular to the event axis
in the transverse plane, \athet~$=$~tan$^{-1}$(\ptaxic/\pz ) and \pz\ 
is the total momentum of the observed particles in the z direction.
\end{enumerate}
\item[large acoplanarity] \acopc~$>$~1.2~rad:
\begin{enumerate}
\setcounter{enumii}{3}
\item
\stevt~$>$~0.045.
\item
\cosevt~$<$~0.90, where the direction of the missing momentum vector
is given by
 \pthet~$=$~tan$^{-1}$(\ptevt/\pz ).
\end{enumerate}
\end{description}
\end{enumerate}

The remaining cuts ensure that the signature of missing momentum
could not have been produced by \sm\ processes containing, for example,
secondary neutrinos from tau decays  or
particles that strike regions of the detector where they are
undetected or poorly measured.
In a number of these cuts we look for evidence of a particle that is
\btb\ in the transverse plane with the total momentum vector of 
the observed central detector tracks and electromagnetic clusters.
The idea is that the majority of the background events have no 
real missing momentum in the transverse plane.
If the  observed tracks and clusters in such a background event
appear to have a net momentum in
this plane then this is due to an additional particle recoiling \btb\
to them.
\begin{enumerate}
\setcounter{enumi}{5}
\item
Events with clusters in GC, FD or SW are
rejected if the scaled energy, $x_{\rm FDSW} = E$/$E_{\rm beam}$
exceeds 0.15.
Low energy, \efdsw~$<$~0.15, GC, FD or SW clusters are used to
 veto the event only if they
are \btb\ to within 1.2~rad 
with the total momentum vector of the observed tracks
and clusters in the central part of OPAL.
This is in order to reduce the potential loss in signal efficiency that might
arise from random low energy
clusters caused by possible detector noise or off-momentum electrons.
\item
A new feature of the OPAL detector
 in 1996 is the tungsten shield that was installed to
protect the central detector from radiation from the synchrotron mask.
This has the potentially serious consequence of creating a hole in the
SW acceptance in the angular region 0.028--0.031~rad.
However, events in which a beam energy electron strikes the shield tend
to yield a large number of low energy clusters in the inner edge of
the endcap electromagnetic calorimeter (EE)
and these can be used to provide an effective veto.
For the purpose of this cut the requirements on clusters in EE are
modified.
Clusters are considered if they
have a deposited energy of greater than 0.1~GeV and
they contain at least two blocks, they are not associated to any 
charged track, and they satisfy $\mcost > 0.964$.

An event is rejected if it contains 3 or more such EE clusters and
either of the
following criteria is satisfied:
\begin{enumerate}
\item
\acope~$<$~1.2~rad, where \acope\ is the acoplanarity angle between the
vector sum of the momenta of the above EE clusters and the total momentum
vector of the event.
\item
The sum of the energy of all such clusters does not exceed 4~GeV.
\end{enumerate}
If an event  contains exactly 2 such  EE clusters and both of the
above criteria are satisfied the event is rejected.
\item
Events containing high energy isolated photons form a potentially
serious source of background, because
quantities such as \ptevt\ and \ptaxic\ may be poorly measured.
However, the potential signal events may also contain  isolated photons and so
some care is needed in designing the selection.
Events are rejected if they satisfy any of the following three criteria:
\begin{enumerate}
\item
The energy of an isolated photon is greater than 25~GeV.
\item
The number of blocks in an isolated ECAL cluster is greater than 12.
If a photon strikes a part of the detector where there is a very large
amount of material in front of the calorimeter the observed energy
is particularly seriously degraded. 
In such cases the observed showers tend to be very broad.
\item
Events with medium energy ($E<25$~GeV) photons are examined to see
whether or not the observed \ptaxic\ could have been caused by the
observed photon\footnote{
This cut is necessary to remove a large potential background from
radiative lepton pairs, in particular radiative tau pairs.}.
The thrust axis in the plane perpendicular to the beam direction is
calculated without including the isolated photon.
If the photon lies on the opposite side of this axis to the two lepton
candidates and if the \ptpt\ of the photon \wrt\ this axis in the
transverse plane\footnote{
That is, the value of \at\ of the photon.}
 is greater than that of one of the leptons then the
event is rejected.
In the
regions 0.71~$<$~\mcost~$<$~0.83 and \mcost~$>$~0.965 the
resolution of the calorimeter is degraded due to upstream material or
poor containment.
If the photon lies in one of these regions stricter requirements are applied.
\end{enumerate}
\item
The number of tracks passing the
quality cuts given above divided by the total number of tracks
reconstructed in the central detector
is required to be greater than 0.2.
This is in order to reject `junk' events, such as those originating from
beam-gas or beam-wall collisions.
\item
The majority of \sm\ \lpair\ events are coplanar and collinear.
The most important cut in this analysis to remove such events is that
on the value of \staxic .
However, in order to provide additional protection against such events,
very loose cuts on acoplanarity and acollinearity 
are applied:
\begin{enumerate}
\item
The values of \acopc\ and \acolc\ are required to exceed 0.1~rad. 
\item
If the event contains an unassociated ECAL cluster of $E>15$~GeV then
\acopc\ and \acolc\ are required  to exceed 0.3~rad.
\end{enumerate}
A small residual background may arise from events in which the leptons
are perfectly coplanar, but that contain a low energy
cluster whose energy is overestimated by a large factor thus leading
to a spuriously high \staxic .
In order to reject such events the cuts given above on \ptaxic , 
acoplanarity and acollinearity are applied also to those quantities
calculated using the two leptons alone, i.e., without correction for the 
presence of unassociated photons.

\item
A particularly difficult potential background originates from \eemumu\
events in which one of the electrons and one of the muons is observed
and the second muon is scattered at an angle to the beam direction
of less than about cos$^{-1}$(0.965). 
At such an angle the muon can carry away a significant \ptevt\ and may
thus cause the event to be selected.
However, the muon is unlikely to 
lead to a track that passes the standard quality cuts 
 or a cluster in FD or SW that could be used to veto the event.
In order to reduce the potential background from this source,  events
are examined for any evidence in the muon chambers, hadron calorimeter
or central detector
of a muon escaping in the very forward region, \btb\ with the observed
lepton pair.
\item
If the charge of both lepton candidates is well measured (curvature
differs from zero by more than three sigma of the measurement error) then
events are rejected if both lepton candidates have a charge of +1 or 
both candidates have a charge of $-$1.
This removes half of the remaining e~--~$\mu$ background from the \eemumu\
events discussed in the previous point and also other events in which
one or more tracks have not been reconstructed.
\end{enumerate}

Cuts 4, 6, 7, 11 and 12 are designed primarily to remove background
from 2-photon processes, which produce events with relatively small
missing transverse momentum.
These cuts are applied only if the event has low missing
transverse momentum ($\staxic < 0.20$ for $\acopc < 1.2$~rad and
$\stevt < 0.25$ for $\acopc > 1.2$~rad). 
In addition to increasing the Monte Carlo calculated signal
efficiencies, this modification also reduces the sensitivity to noise
not simulated in the Monte Carlo.

\begin{figure}[htbp]
 \epsfxsize=\textwidth 
 \epsffile{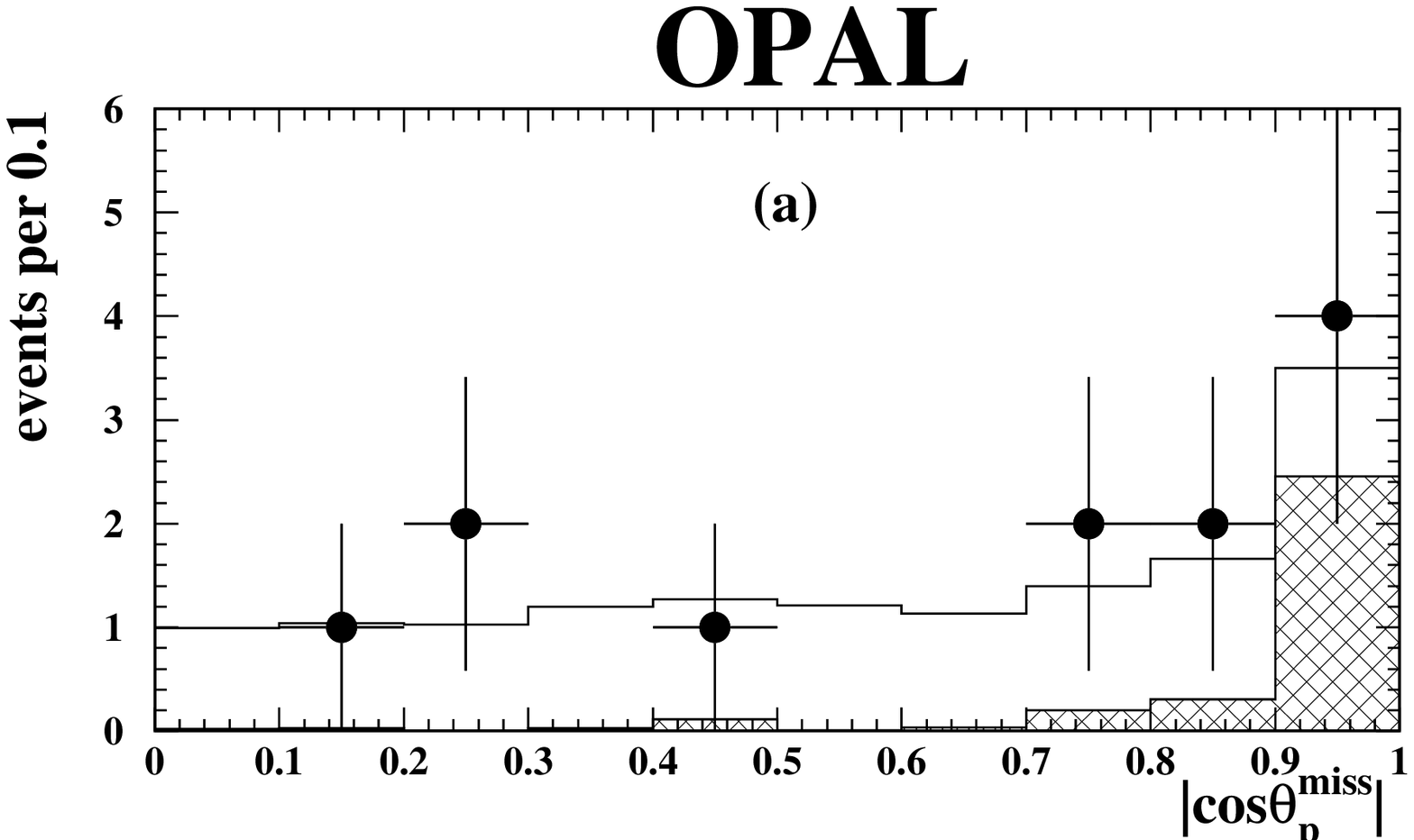}
 \epsfxsize=\textwidth 
 \epsffile{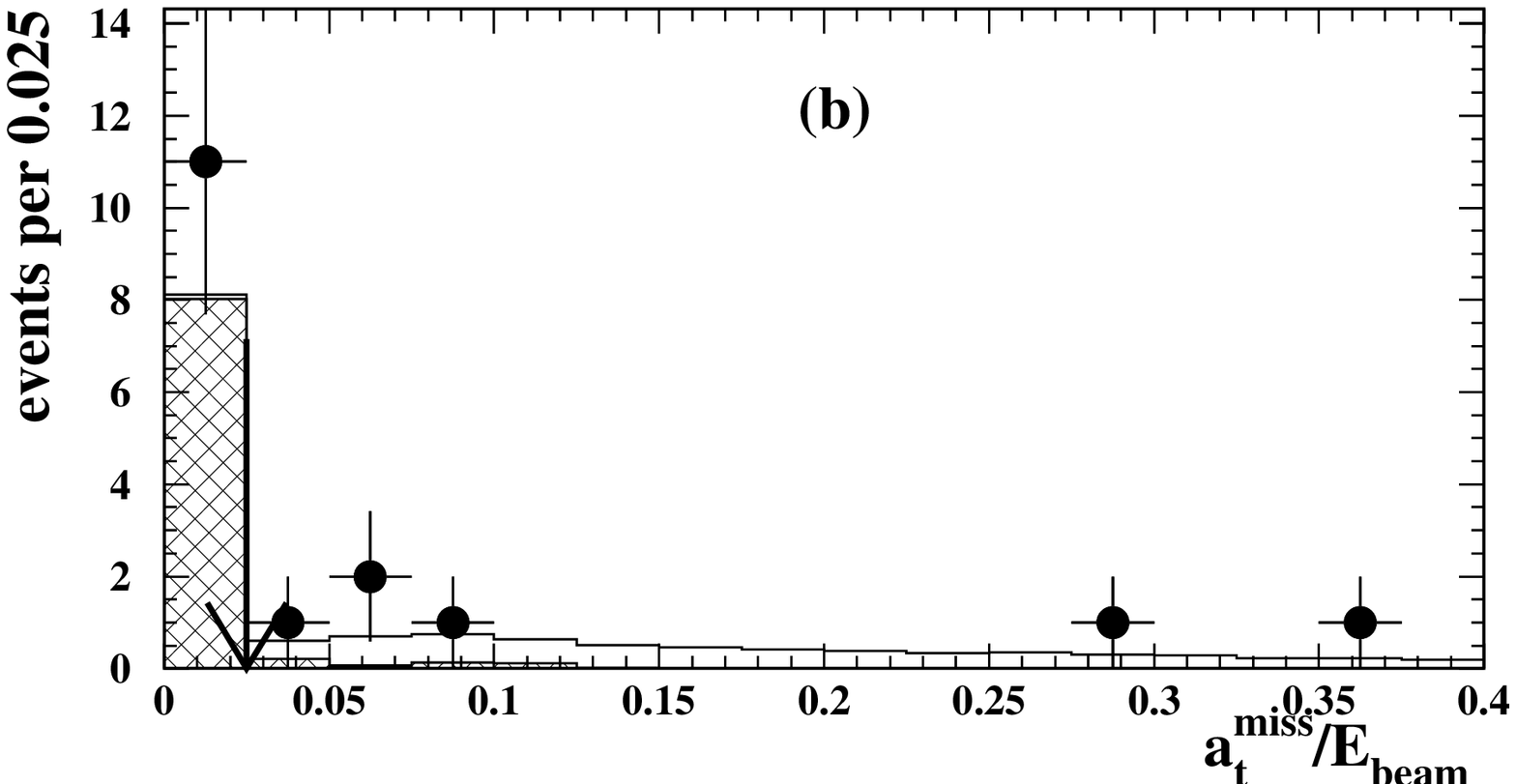}
 \caption{
(a) Distribution  of \cosevt\ after all selection cuts have been
applied except for those on the direction of the missing momentum
vector (cuts 5(c) and (e)).
(b) Distribution  of \staxic\ for events with $\acop < 1.2$~rad 
after all selection cuts have been
applied except for that on  \staxic\ (cut 5(b)).
The position of the cut on  \staxic\ is shown as the arrow.
The data at $\protect\sqrt{s}$~=~172~GeV 
 are shown as the points with error bars.
The \mc\ prediction for  4-fermion processes with
genuine prompt missing energy and momentum (\llnunu ) is shown as the
open histogram and the
background, coming mainly from processes with four charged leptons in
the final state, is shown as the shaded histogram.
} 
\label{fig-cosevt}
\end{figure}

We now describe some additional checks of the degree to which the
\smc\ describes the observed data.
These checks are particularly relevant to the \eell\ background.
Figure~\ref{fig-cosevt}~(a) shows the distribution  of \cosevt\ 
after all selection cuts have been
applied except for those on the direction of the missing momentum
vector (cuts 5(c) and (e)).
Figure~\ref{fig-cosevt}~(b) shows the distribution 
of \staxic\ for events with $\acop < 1.2$~rad 
after all selection cuts have been
applied except for that on  \staxic\ (cut 5(b)).

As a further test of the degree to which the \smc\ describes the data
we relax some of the selection cuts.
Cut 2 is relaxed to requiring no more than one charged track not
associated to either lepton pair.
Cuts 3b and 3c are relaxed to
requiring no more than 6 tracks and clusters with a mass of 3~GeV.
Cuts 4, 11 and 12 are removed.
Cut 5  on the magnitude and direction of the 
missing momentum is relaxed as follows:
\begin{description}
\item[small acoplanarity] \acopc~$<$~1.2~rad:
\begin{enumerate}
\item
\stevt~$>$~0.025.
\item
\staxic~$>$~0.015.
\item
no cut on \axicos . 
\end{enumerate}
\item[large acoplanarity] \acopc~$>$~1.2~rad:
\begin{enumerate}
\setcounter{enumii}{3}
\item
\stevt~$>$~0.035.
\item
no cut on \cosevt . 
\end{enumerate}
\end{description}

\begin{figure}[htbp]
 \epsfxsize=\textwidth 
 \epsffile{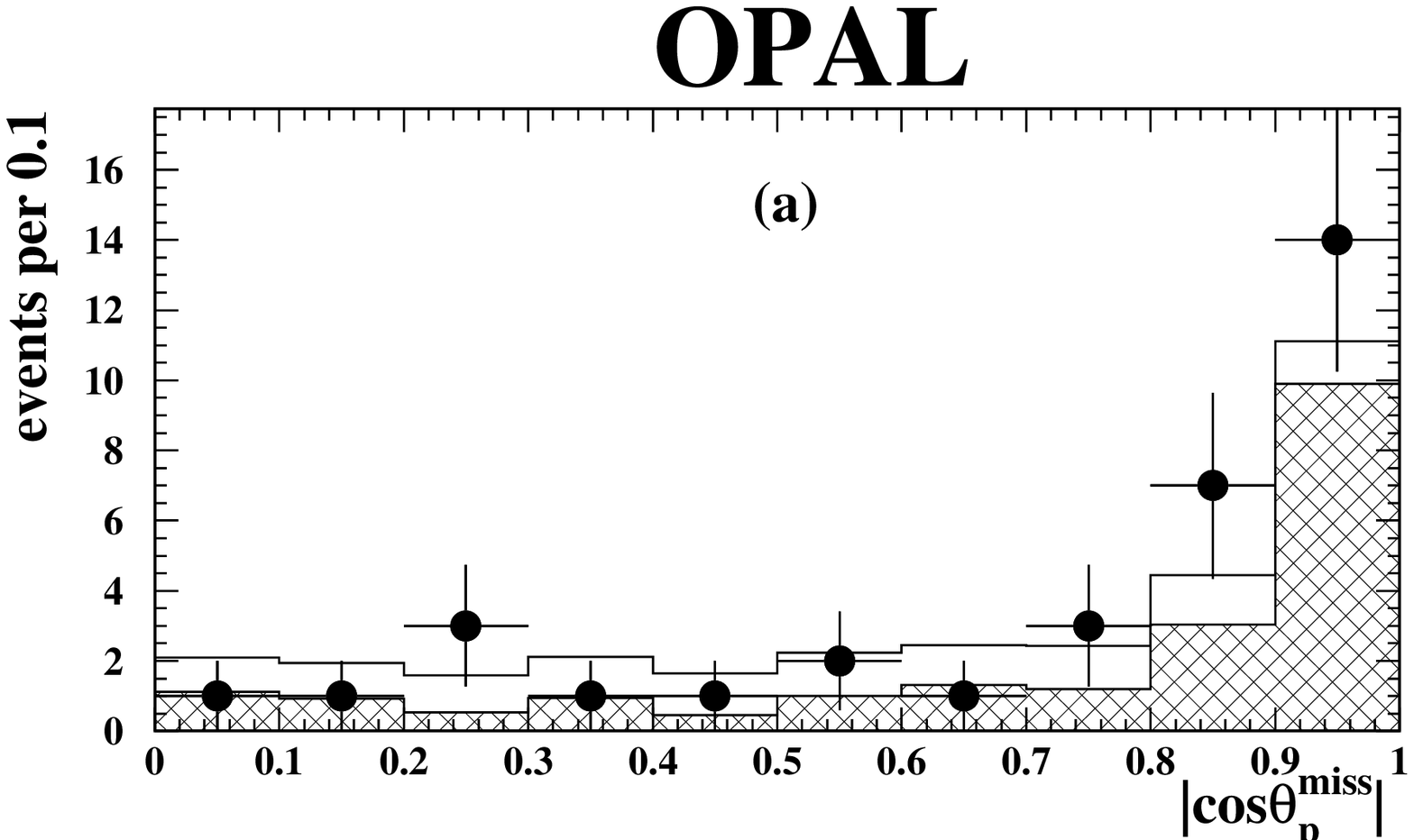}
 \epsfxsize=\textwidth 
 \epsffile{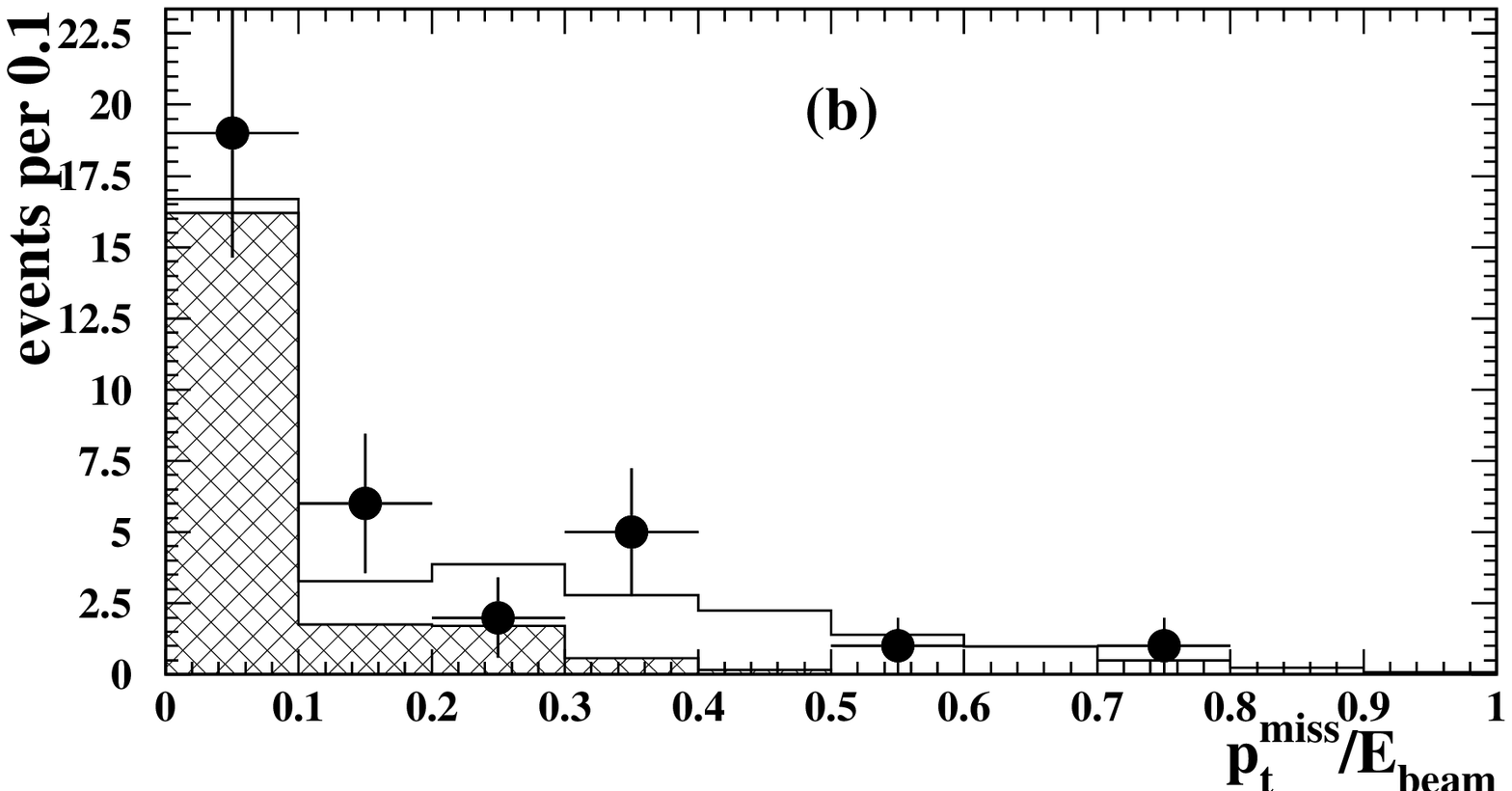}
 \caption{
Distributions  at $\protect\sqrt{s}$~=~172~GeV of
(a) \cosevt\ and (b)  \stevt\ of the events 
selected with the relaxed cuts described in the text 
compared with the \smc .
The data are shown as the points with error bars.
The \mc\ prediction for  4-fermion processes with
genuine prompt missing energy and momentum (\llnunu ) is shown as the
open histogram and the
background, coming mainly from processes with four charged leptons in
the final state, is shown as the shaded histogram.
} 
\label{fig-loose}
\end{figure}

The number of candidate events selected with these relaxed cuts at
161~GeV is 26.
The number  predicted by the \smc\ is $30.4 \pm 1.6$~(stat).
The number of candidate events selected with these relaxed cuts at
172~GeV is 34 and 
the number  predicted by the \smc\ is $32.1 \pm 1.5$~(stat).
A breakdown of the \mc\ predicted event samples into the
contributions from individual \sm\ processes is given in table~\ref{tab-loose}.
 Figure~\ref{fig-loose} shows the distributions of
(a)  \stevt\ and (b) \cosevt\ of the events at \sseven\ 
selected with these relaxed cuts  compared with the \smc .
In all of the above checks the data and \smc\ are in agreement. 

\begin{table}
\centering
\begin{tabular}{||c||c|c||c|c|c|c|c||}
\hline \hline
$\sqrt{s}$ & data & SM total & \lpair & \eell & \eeqq & \zee\ & \llnunu \\
\hline
  161      &  26  & 30.4 &  0.76  & 20.2  &  3.7  & 1.9   &  3.7    \\
  172      &  34  & 32.1 &  0.76  & 16.5  &  1.7  & 1.6   & 11.6    \\
\hline \hline
\end{tabular}
\caption[hbt]{\sl Comparison between data and \mc\ of the number of
  events selected with the relaxed cuts given in the text.
The total number of events predicted by the  \sm\ is given, 
together with a breakdown into the
contributions from individual processes.
The \sm\ \mc s are normalised to the experimental integrated  luminosity. 
\label{tab-loose}
}
\end{table}

\subsection{\wpair\  Event Selection Cuts}
\label{sec-wwcuts}

In order to suppress the contribution of \smp\ other than 
\wpair\ events some additional kinematic cuts are applied.
Firstly, a  loose cut is applied to suppress the remaining
background from two-photon processes and other \sm\ four-fermion
events containing four charged leptons in the final state,
e.g., \zee~$\rightarrow$~\eemumu :
events are rejected if \mrecoil~$>$~(140~GeV+\llmass ).
\llmass\ is the reconstructed mass in~GeV of the observed lepton pair
and \mrecoil\ 
is the mass in GeV of the invisible system recoiling against the lepton pair.

The remaining cuts depend on the flavour of the observed leptons:
\begin{enumerate}
\item
In events containing one electron and one muon (i.e., e$\mu$ events)
a somewhat tighter cut against  \sm\ four-fermion
events containing four charged leptons is applied.
If $\xmax<0.325$ events are rejected if \mrecoil~$>$~(100~GeV+\llmass ).
\item
In events containing two electrons or two muons (i.e., \epair , \mupair\
 events) with $xmin<0.325$  the recoil mass is required not to be 
consistent with the Z mass:
\begin{enumerate}
\item
If \llmass~$<$~35~GeV events are rejected if \mrecoil\ is within
10~GeV of the Z mass.
\item
If \llmass~$>$~35~GeV events are rejected if \mrecoil\ is within
5~GeV of the Z mass.
\end{enumerate}
\end{enumerate}

\section{Event Selection II}
\label{sec-selII}

The second selection
has been  optimised to select and measure the rate of
high visible energy events such as those from \wpair\ events in which both
 W's decay leptonically. 
Events are selected from a low multiplicity preselection~\cite{bib-LL}.
It is further required that the charged track multiplicity be 
at least one and
no more than six, and that the total number of charged tracks and
clusters does not exceed twelve. 
Low multiplicity ``jets'' of charged tracks and 
clusters in the ECAL, GC, FD and SW calorimeters 
are defined using
a cone algorithm~\cite{bib:cone} with a minimum energy of 2.5~GeV and a
cone half angle of 20$^{\circ}$.
A separate selection is defined for jet multiplicities of one, two and three.
For \wwllnunu\ events the fraction of preselected events per jet multiplicity
are 6\% 1-jet, 89\% 2-jet and 5\% 3-jet.
The former correspond mostly to events in which the decay products 
of one of the W's are either not reconstructed or partially reconstructed.
Events naturally fall most of the time in the 2-jet category, while events
with significant photon radiation observed in the detector may be classified 
as 3-jet events.

Electron and muon identification is applied to  
the most energetic track in each jet.
Electrons are identified using the ratio of 
the energy in the ECAL
to the track momentum (\Eovp).  Tracks
which are consistent with originating from photon conversions are removed,
and the ionisation energy loss, \dedx, is required to be within three
standard deviations of the value expected for an electron, if there are at
least 20 hits on the track with charge information.  Residual background in
the electron sample is removed by requiring that the track has fewer than
two associated hits beyond the first two layers of the HCAL,
and by requiring that the $\phi$ measurements of the track and
ECAL cluster match to within $1^{\circ}$.
Muons are identified as a track in the central
detector which has associated hits in the muon chambers or the 
HCAL, and has only a small energy deposit in the ECAL.
Identification of jets as electrons or muons, or 
as neither, is used primarily to classify
events rather than as a tool for background rejection.

\subsection{2-jet Selection}
Firstly we describe the criteria applied to
all 2-jet events, and then three sets of kinematic criteria, each of
which is sufficient to select the event.
\subsubsection{General 2-jet Criteria}
\begin{itemize}
\item[D1]{Both jets must contain at least one charged track
and have $|\cos{\theta}|  < 0.96$. At least one jet 
must satisfy $|\cos{\theta}|  < 0.90$. }
\item[D2]{The event must also have significant activity in the ECAL
and in the vertex drift chamber or the silicon micro-vertex detector :
\subitem{There must be at least two ECAL clusters, one of which has 
corrected energy exceeding
0.5~GeV. At least one of the jets must contain
an ECAL cluster, and for 
events where no electrons or muons are identified each jet is
required to contain an ECAL cluster.}
\subitem{Events must contain evidence for 
two distinct charged tracks originating from near the vertex.
Either the event must contain at least two tracks reconstructed in the
axial part of the vertex drift chamber or the highest energy charged 
track in each jet must be associated to hits in the silicon
microvertex detector.}}
\item[D3]{The acollinearity angle of the two jets must exceed 10$^{\circ}$.}
\item[D4]{For events with \ACOP $>$ 60$^{\circ}$ 
it is
    required that the direction of the missing momentum satisfies $\cosevt <
    0.95$.  For events with \ACOP $<$ 60$^{\circ}$ it is required
    that \staxic\ exceeds 0.025, and that the direction of
    missing momentum calculated using $a_t$  as described in
    section~\ref{sec-gen} 
    satisfies $|\cos{\theta_a}|<0.995$.}
\item[D5]{Events are rejected if there are any selected tracks which are not
    associated with either jet.}
\item[D6]{Events are rejected if a
 track segment in the muon chambers that is outside a 25$^{\circ}$
    cone around each jet is reconstructed with $|\cos{\theta}|>0.95$.}
\item[D7]{The event is rejected if 
the maximum azimuthal separation of any two vertex chamber axial tracks
    exceeds 177.5$^{\circ}$.}
\item[D8]{Events with 
acollinearity angle less than 20$^{\circ}$ are rejected if
the most energetic ECAL cluster has energy exceeding 80\% of the beam
    energy. This criterion is used for redundancy in the rejection of
Bhabha scattering events.
\item[D9]{
Events with kinematic properties and identified lepton flavour
similar to 
the \eemumu\ background
are subject to 
further requirements.
Events in which the two-jets are identified 
as $\mathrm{e}-\mu$, $h-\mu$ or $h-h$, where 
$h$ signifies that the jet is neither identified as an electron nor a muon,
and 
$\stevt < 0.25$ 
are considered candidates for rejection as consistent with this
background source.
Such events are rejected if any of the following conditions is satisfied :
\subitem[a]{The net charges of both jets with associated charged tracks are
non-zero and of the same sign.}
\subitem[b]{The least energetic jet 
has energy satisfying $\xmin < 0.08$ and
the electron candidate has $|\cos{\theta}| > 0.90$.}
\subitem[c]{There is a hadron calorimeter cluster 
reconstructed with $|\cos{\theta}|>0.95$ and
energy exceeding 1~GeV
outside a 25$^{\circ}$
cone around each jet.}}}
\end{itemize}
\subsubsection{Kinematic Criteria in Selection II}
Each 2-jet selection uses between two 
and four additional criteria based on the
following five variables.  Scaled variables are scaled by the beam energy.
The variables are \ACOP  ~, \XT ~,
\XONE ~, \XTWO ~ and \MLL .
Each event may be selected by at most three independent selections.  The
values for the cut variables are detailed in table~\ref{tab-lnln-cuts}.
\begin{itemize}
\item[A]{Kinematic selection without lepton identification.  This is designed to select with
    high efficiency events from the $\ee$, $\mm$ and $\emu$ classes based
on kinematics without requiring lepton identification.
}
\item[B]{Missing energy selection.  This is a rather inclusive selection of
    all di-lepton classes based simply on significant missing transverse
    energy.  
    It is required that there is at least
    one identified lepton (electron or muon). 
}
\item[C]{Di-lepton identification.  Four mutually exclusive selections are
    defined depending on the lepton identification results:}
\begin{itemize}
\item[C1]{Di-electron or di-muon.}
\item[C2]{Neither jet is identified as an electron or muon.}
\item[C3]{Electron-muon.}
\item[C4]{One jet is identified as electron or muon, the other is not
    identified as electron or muon.}
\end{itemize}
The selections with at least one identified lepton require that there is at
least one highly energetic jet (typically \XONE\ exceeding 0.30).
\end{itemize} 

\begin{table}
\centering
\begin{tabular}{|l|c|c|c|c|c|c|c|}
\hline
Cut      & A          & B      & C1   & C2      & C3  &  C4    \\
\hline
\ACOP~ ($^{\circ}$)    
& 5.0        &  15.0  & 5.0  &  15.0   & 5.0 &  5.0   \\
\XT      & 0.08       &  0.15  & 0.08 &  0.08   &  --  &  0.08  \\
\XONE    & 0.75$-$\XTWO &   --    & 0.30 &  0.20   & 0.30 & 0.30  \\
\XTWO    & 0.325      &   --    & --   &   --    &  --  &  --  \\
\MLL~(GeV)  & --         &  10.0  & 10.0 &  10.0   &  --  &   --  \\ \hline
\end{tabular}
\caption{
2-jet analysis kinematic cut values. 
The table lists the value of each cut quantity
for each selection.
Each variable is the minimum allowed value.
Cuts with ``--'' are not applied.
}
\label{tab-lnln-cuts}
\end{table}

\subsection{3-jet Selection}

Events passing the preselection and classified as 3-jet events are
selected if they satisfy:

\begin{itemize}
\item[T1]{The number of charged tracks should be between two and four, and 
exactly two of the jets should contain charged tracks.
\item[T2]{The sum of the opening angles among the three 2-jet pairings should be
    less than 357.5$^{\circ}$.}
\item[T3]{The energy of the jet without charged tracks is added to the energy of
    the closest charged jet.  At least one of the two jets must be identified
    as an electron or muon, and the total energy 
    must exceed 30\% of the beam energy.}
\item[T4]{For events in which the jets are not all in one half of the transverse
    plane, the following criteria designed to reject $\tau \tau \gamma$ events
    and tagged two-photon di-lepton events are applied.
    The event is required to have at
    least one charged jet with $|\cos{\theta}|<0.80$ and both must satisfy
    $|\cos{\theta}|<0.90$.  Events where the apparently neutral jet is below
    a polar angle at which it can be easily identified as neutral
    ($|\cos{\theta}| > 0.95$) are rejected if the energy of the neutral jet
    exceeds 15\% of the beam energy.  An axis in the transverse plane is
    defined using the most energetic charged jet and the event is rejected if
    the transverse momentum of the neutral jet with respect to this axis
    exceeds 80\% of the transverse momentum of the lowest energy charged jet.
    }
\item[T5]{$\stevt > 0.1$ and $\cosevt < 0.90$.
}
\item[T6]{The acoplanarity angle of the two charged jets should exceed
    10$^{\circ}$.}
\item[T7]{Analogously to the 2-jet selection, it is also required that 
there must be at least two ECAL clusters, one of which has energy exceeding
0.5~GeV, and at least two tracks reconstructed in the axial sectors of the
vertex chamber.}
\item[T8]{Background rejection: The 2-jet selection criteria D6
and D7, the Bhabha rejection cut D8, and the
like-sign rejection cut D9[a], are applied to all candidate 3-jet events.}}
\end{itemize}

\subsection{Single-jet Selection}

This selection is designed for 
events in which one high transverse momentum 
lepton
is observed at wide angle
with evidence for a partially reconstructed lepton at small polar angle.

The single jet selection requires the following :
\begin{itemize}
\item[S1]{The single jet which contains at least one charged track
should satisfy $|\cos{\theta}| < 0.82$, be associated
to an ECAL cluster 
and have scaled transverse momentum exceeding 0.25.}
\item[S2]{The jet must be identified as an electron or muon or
have ECAL energy exceeding 5~GeV.}
\item[S3]{A muon track segment or ECAL or HCAL
cluster 
should be present in the angular range $|\cos{\theta}| > 0.95$
and this forward activity should yield an acoplanarity angle
with respect to the jet of at least 10$^{\circ}$.}
\item[S4]{Remaining background contributions, particularly
from cosmic ray events
are reduced to a negligible level by 
requiring an in time time-of-flight hit, that a track in the 
jet be associated to a hit in the silicon micro-vertex detector, 
and if two or more tracks are found in the axial part of the 
vertex drift chamber, 
that the highest azimuthal opening angle
does not exceed 160$^{\circ}$. 
}
\end{itemize}

\newpage
\bigskip\bigskip\bigskip
\appendix
\noindent {\Large\bf  Acknowledgements}

\bigskip
\noindent 
We particularly wish to thank the SL Division for the efficient operation
of the LEP accelerator at all energies
 and for
their continuing close cooperation with
our experimental group.  We thank our colleagues from CEA, DAPNIA/SPP,
CE-Saclay for their efforts over the years on the time-of-flight and trigger
systems which we continue to use.  In addition to the support staff at our own
institutions we are pleased to acknowledge the  \\
Department of Energy, USA, \\
National Science Foundation, USA, \\
Particle Physics and Astronomy Research Council, UK, \\
Natural Sciences and Engineering Research Council, Canada, \\
Israel Science Foundation, administered by the Israel
Academy of Science and Humanities, \\
Minerva Gesellschaft, \\
Benoziyo Center for High Energy Physics,\\
Japanese Ministry of Education, Science and Culture (the
Monbusho) and a grant under the Monbusho International
Science Research Program,\\
German Israeli Bi-national Science Foundation (GIF), \\
Bundesministerium f\"ur Bildung, Wissenschaft,
Forschung und Technologie, Germany, \\
National Research Council of Canada, \\
Hungarian Foundation for Scientific Research, OTKA T-016660, 
T023793 and OTKA F-023259.\\


\newpage

\clearpage

\clearpage

\clearpage

\clearpage

\clearpage

\clearpage

\clearpage

\clearpage

\clearpage

\clearpage

\clearpage

\clearpage

\clearpage

\clearpage

\clearpage

\clearpage

\clearpage

\clearpage

\clearpage

\begin{figure}[htbp]
 \epsfxsize=\textwidth 
 \epsffile[0 0 580 600]{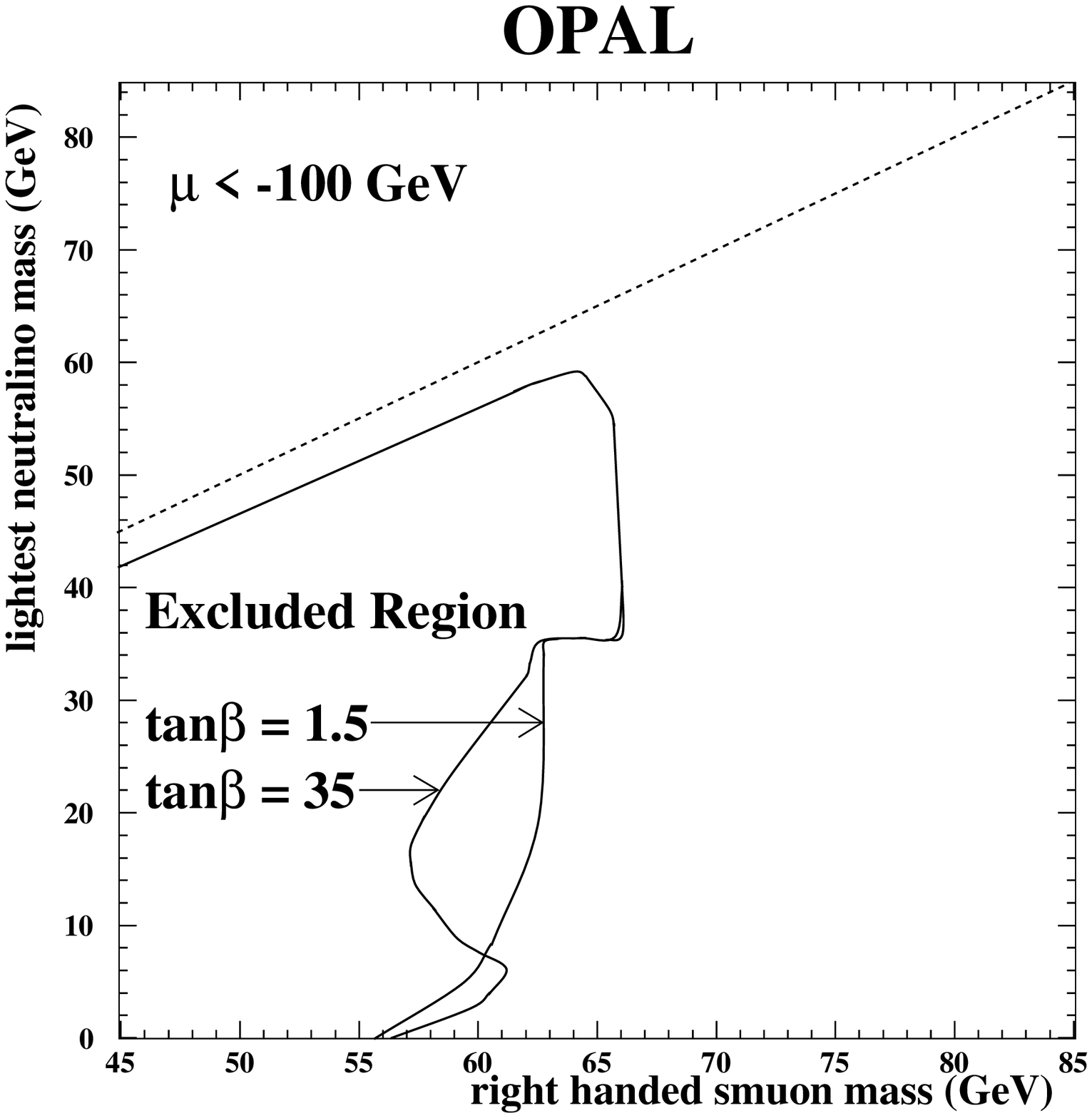}
 \caption{
For two values of $\tan{\beta}$ and $\mu < -100$~GeV,
 95\% CL exclusion regions for right-handed smuon pairs 
obtained by combining the 130--172~GeV data-sets.
The excluded regions are calculated 
taking into account the 
predicted branching ratio for $\smu^\pm_R \rightarrow  {\mu^\pm} \nt_1$.
The kinematically allowed region is indicated by the dashed line.
} 
\label{fig-mssm_2a}
\end{figure}
\clearpage

\clearpage

\clearpage

\clearpage

\end{document}